\documentclass[11pt, a4paper]{article}
\usepackage{jcappub}

\usepackage{booktabs}
\usepackage{multirow}
\usepackage{amssymb}
\usepackage[export]{adjustbox}
\usepackage{floatrow}
\newfloatcommand{capbtabbox}{table}[][3.5in]
\newfloatcommand{capbfigbox}{figure}[][2.3in]
\usepackage{blindtext}
\usepackage{amsmath}

\newcommand{\Fermi}{\textit{Fermi}}
\newcommand{\kpc}{\,\text{kpc}}
\newcommand{\PV}{\texttt{P6V11} }
\newcommand{\pc}{\,\text{pc}}
\newcommand{\GeV}{\,\text{GeV}}
\newcommand{\MeV}{\,\text{MeV}}
\newcommand{\s}{\,\text{s}}
\newcommand{\sr}{\,\text{sr}}
\newcommand{\cm}{\,\text{cm}}
\newcommand{\km}{\,\text{km}}
\newcommand{\fex}{e.g.~}
\newcommand{\cf}{cf.~}
\newcommand{\ie}{i.e.~}
\newcommand{\eg}{e.g.~}

\begin{document}

\hspace*{110mm}{\large \tt FERMILAB-PUB-14-289-A}

\vskip 0.2in

\title{Background model systematics for the Fermi GeV excess}

\author[a]{Francesca Calore,}
\emailAdd{f.calore@uva.nl}

\author[b]{Ilias Cholis}
\emailAdd{cholis@fnal.gov}

\author[a]{and Christoph Weniger}
\emailAdd{c.weniger@uva.nl}

\affiliation[a]{GRAPPA, University of Amsterdam, Science Park 904, 1090 GL
Amsterdam, Netherlands}
\affiliation[b]{Fermi National Accelerator Laboratory, Center for Particle
Astrophysics, Batavia, IL 60510, USA}

\abstract{
    The possible gamma-ray excess in the inner Galaxy and the Galactic center
    (GC) suggested by \Fermi-LAT observations has triggered a large number of
    studies. It has been interpreted as a variety of different phenomena such
    as a signal from WIMP dark matter annihilation, gamma-ray emission from a
    population of millisecond pulsars, or  emission from cosmic rays injected
    in a sequence of burst-like events or continuously at the GC.  We present
    the first comprehensive study of model systematics coming from the Galactic
    diffuse emission in the inner part of our Galaxy and their impact on the
    inferred properties of the excess emission at Galactic latitudes
    $2^\circ<|b|<20^\circ$ and 300 MeV to 500 GeV.  We study both theoretical
    and empirical model systematics, which we deduce from a large range of
    Galactic diffuse emission models and a principal component analysis of
    residuals in numerous test regions along the Galactic plane.  We show that
    the hypothesis of an extended spherical excess emission with a uniform
    energy spectrum is compatible with the \Fermi-LAT data in our region of
    interest at $95\%$ CL.  Assuming that this excess is the extended
    counterpart of the one seen in the inner few degrees of the Galaxy, we
    derive a lower limit of $10.0^\circ$ ($95\%$ CL) on its extension away from
    the GC.  We show that, in light of the large correlated uncertainties that
    affect the subtraction of the Galactic diffuse emission in the relevant
    regions, the energy spectrum of the excess is equally compatible with both
    a simple broken power-law of break energy $E_\text{break}=2.1\pm0.2\GeV$,
    and with spectra predicted by the self-annihilation of dark matter,
    implying in the case of $\bar{b}b$ final states a dark matter mass of
    $m_\chi=49^{+6.4}_{-5.4} \GeV$.
}

\maketitle


\section{Introduction}
\label{sec:Intro}

The study of Galactic cosmic rays (CRs) and gamma-ray physics has been a very
active field in the last few decades with the launch of PAMELA \cite{PAMELA,
PAMELA2}, AMS-02 \cite{AMS} aboard the International Space Station (ISS) and,
in particular, of the Large Area Telescope (LAT) aboard the Gamma-Ray Space
Telescope (\Fermi) \cite{1999APh....11..277G}. The latter has produced the most
detailed maps of the gamma-ray sky ever, in a wide energy range and with good
energy and angular resolution~\cite{Rando:2009yq}.  The interaction of Galactic
CRs with the interstellar medium (ISM) results in the diffuse emission from the
Milky Way, which is the brightest source of gamma rays seen with the
\Fermi-LAT. Thus, the \Fermi-LAT data provide an important handle for
understanding the origin and the propagation of CRs in our Galaxy. This is
complemented by measurements in microwaves \cite{Bennett:2012zja,
Planck:2006aa}, X-rays \cite{Eriksen:2011qf}, lower energy
\cite{Winkler:2003nn} and higher energy gamma rays \cite{Albert:2005kh,
Aharonian:2005jn, Acciari:2009xb}, and neutrinos \cite{Aartsen:2014gkd}. 

More specifically, focused studies of known Galactic sources in gamma rays by
both the \Fermi-LAT and Atmospheric Cherenkov Telescopes (ACTs), together with
measurements in X-rays, microwaves and now  neutrinos, have helped us in
studying and modeling the source properties and the primary CRs composition
injected \cite{Yuan:2010pg, Tanaka:2011vc, Castro:2012eca, Berezhko:2012mp,
Castro:2013fya, Saha:2014fga, Cardillo:2014kaa, Fukuda:2014oqa, HESS:2014dza}
by for example supernova remnants (SNRs) and pulsars.  Moreover, the analysis
of gamma-ray data has led to the discovery of the \Fermi~bubbles
\cite{Dobler:2009xz, Su:2010qj}, which are believed to be the inverse Compton
scattering (ICS) counterpart  \cite{Dobler:2009xz} of the WMAP haze
\cite{Finkbeiner:2003im, Dobler:2007wv},  recently observed also with Planck
\cite{Dobler:2012rta}.  

In addition, the unprecedented energy and angular resolution of \Fermi-LAT,
combined with an energy range of more than three orders of magnitude (30 MeV to
500 GeV) and the large field of view, have led to a revolution in the
measurement of the gamma-ray properties for specific classes of objects such as
active Galactic nuclei \cite{Collaboration:2010gqa, Ajello:2013lka},
star-forming galaxies \cite{Abdo:2009aa, Lacki:2010vs, Ackermann:2012vca} and
millisecond pulsars (MSPs) \cite{Abdo:2009ax, TheFermi-LAT:2013ssa,
Cholis:2014noa}.  For the latter class of objects very few detections and even
fewer spectral measurements had been made at gamma-ray energies.  Those same
properties of the LAT instrument have also facilitated an unprecedented
measurement of the isotropic gamma-ray background (IGRB) \cite{Abdo:2010nz,
AckermannEGB} and, for the first time, of its power-spectrum
\cite{Ackermann:2012uf}. These measurements have improved our understanding of
extragalactic astrophysical objects \cite{Cuoco:2012yf, AckermannEGB,
Ackermann:2012vca, Stecker:2010di, Makiya:2010zt} and  ultra-high-energy CRs
\cite{Berezinsky:2010xa, Ahlers:2010fw, Murase:2012df}.

\medskip

A wide range of cosmological and astrophysical observations have shown that
about 85\% of the matter content in the Universe is non-baryonic, dark and
cold~\cite{Ade:2013zuv}. The currently leading candidates for this \emph{dark
matter} (DM) are Weakly Interacting Massive Particles (WIMPs), which appear in
a large number of scenarios of beyond-the-Standard Model
physics~\cite{Jungman:1995df, Bertone:2004pz, Bergstrom:2012fi}.  The
`freeze-out mechanism' that sets their abundance in the early Universe provides
clear predictions for the self-annihilation rate of these particles today.
Their annihilation products could contribute to the cosmic- and gamma-ray
fluxes observed at Earth with rates that are tantalizingly close to the
sensitivity of current experiments.

\bigskip

Recently, a number of groups searching for DM signals have analyzed the
gamma-ray emission from the inner few degrees around the Galactic center (GC),
and either claimed~\cite{Goodenough:2009gk, Hooper:2010mq, Hooper:2011ti,
Abazajian:2012pn, Macias:2013vya, Abazajian:2014fta, Daylan:2014rsa,
Zhou:2014lva} or refuted~\cite{Boyarsky:2010dr} the existence of an extended
diffuse and spherical emission component on top of the standard astrophysical
backgrounds.  It was found that the excess emission is compatible with a radial
volume emissivity profile $\propto r^{-\Gamma}$, with $\Gamma$ in the range
2.2~\cite{Abazajian:2012pn} to 2.4~\cite{Macias:2013vya}, and with $r$ denoting
the Galacto-centric distance.  Furthermore, it was demonstrated that the excess
features an energy spectrum that peaks at a few GeV and is broadly consistent
with a signal from DM annihilation, although other source spectra (a
log-parabola or a power-law with exponential cutoff) can be accommodated by the
data, see \fex refs.~\cite{Hooper:2010mq, Macias:2013vya, Abazajian:2014fta}.

However, the modeling of the Galactic diffuse emission (GDE) is very uncertain
in the inner few hundred pc of the GC ($150\pc$ projected distance corresponds
to about $1^\circ$), and is strongly affected by systematics related to point
source subtraction and the modeling of diffuse
backgrounds~\cite{FermiLAT:2012aa, Abazajian:2014fta, Boyarsky:2010dr}.
Complementary observations, like local measurements of CRs, are not able to set
strong constraints on the CR propagation properties or gas densities in that
region. Moreover, the magnetic field intensity can be only indirectly
constrained by the synchrotron emission at microwaves (or the dispersion
measures) for given assumptions on high (low) energy electrons.  Indeed, given
the high density of astrophysical objects and CRs at the GC, and the
unprecedented sensitivity of \Fermi-LAT, it seems far from surprising that
residuals were found above the astrophysical emission that had been
anticipated~\cite{Strong:2004de, FermiLAT:2012aa}.

\smallskip

If the observed GC excess is indeed caused by the annihilation of DM particles
(which is a very exciting but also exotic possibility), it should visibly
extend tens of degrees above and below the GC~\cite{Serpico:2008ga,
Nezri:2012xu}.  The solid confirmation of an excess emission with the
\emph{same} spectral properties as observed in the inner few degrees around the
GC is hence a necessary (though not sufficient) criterion for the
interpretation of the excess at the GC in terms of DM annihilation, and as such
of utmost importance.  The first claim that such an extension to high
latitudes, up to tens of degrees, is indeed observed in the \Fermi-LAT data
(most notably in the \Fermi\ bubble regions~\cite{Su:2010qj}, which happen to
have a good signal-to-noise for DM signals), was put forward in
ref.~\cite{Hooper:2013rwa}, and reproduced in refs.~\cite{Huang:2013pda,
Daylan:2014rsa}.  Nonetheless, also this high latitude region is extremely
difficult to analyze, mainly because of the dominant background represented by
gamma rays originating from the interactions of CRs with the ISM and photon
fields.

\bigskip

Besides its status as a ``compelling case for annihilating
DM''~\cite{Daylan:2014rsa}, the GC excess emission has also been interpreted in
terms of various astrophysical processes.  Firstly, the excess gamma-ray
emission could originate from a population of gamma-ray MSPs associated with
the central stellar cluster and not yet detected by the \Fermi-LAT
\cite{Abazajian:2010zy, Abazajian:2014fta, Gordon:2013vta,Yuan:2014rca} (as
already suggested in ref.~\cite{Wang:2005av} years ago using EGRET data).
Recently, several arguments disfavor the MSPs hypothesis, suggesting that only
up to 5--10$\%$ of the observed emission can originate from MSPs
\cite{Hooper:2013nhl,Calore:2014oga,Cholis:2014lta}.

Another possible explanation of the GC excess relies on the interactions of CRs
with gas, for example non-thermal bremsstrahlung from a population of electrons
scattering off neutral molecular clouds in the inner $2^{\circ}$
\cite{2013ApJ...762...33Y}, or interactions between the gas and protons
accelerated by the super-massive black hole sitting at the
GC~\cite{Linden:2012iv}.  In general, those mechanisms would lead to an excess
emission correlated with the gas distribution itself.  In particular, the
former process could not only explain part of the GeV excess emission in the
Galactic ridge region, but also the excess at TeV energies as seen by
HESS~\cite{Aharonian:2006au, YusefZadeh:2012nh}.  However, in both cases an
extended signal up to a few kpc is excluded, unless a large amount of
unidentified spherically distributed gas is present at the GC.  It has also
been suggested that burst-like events during an active past of our Galaxy may
represent a viable mechanism for producing gamma rays in the inner Galaxy with
the observed spectrum and morphology. Hadronic \cite{Carlson:2014cwa} or
leptonic \cite{Petrovic:2014uda} scenarios might explain some of the observed
excess features at high latitudes.

\bigskip

In this paper, we will reanalyze the \Fermi-LAT gamma-ray data at high Galactic
latitudes.  We will consider a Region Of Interest (ROI) at Galactic latitudes
$2^\circ\leq |b|\leq 20^\circ$ and Galactic longitudes $|\ell|<20^\circ$, which
we refer to as the \emph{inner Galaxy} throughout.  This region avoids the
inner 2 degrees of the Galactic disk, which is the region of the sky that is
most contaminated with strong gamma-ray point sources and large uncertainties
in the diffuse emission.

\smallskip

The aims of this paper are the following:  {\it i}) to robustly confirm the
existence of an extended excess emission in the inner Galaxy on top of the
standard astrophysical background, henceforth the \emph{Galactic center excess}
(GCE); {\it ii}) to characterize its spectral and morphological properties and
its status as the extended counterpart of the excess seen at the GC; and {\it
iii}) to investigate in a comprehensive manner the systematic uncertainties
related to the modeling of diffuse backgrounds.   To this end, we study
\emph{theoretical} model systematics, which are related to variations in the
different possible models for the GDE, and \emph{empirical} model systematics,
which we estimate by an analysis of residuals from a large number of test
regions along the Galactic disk. 

\medskip

We adopt template-based multi-linear regression techniques to fit the
\Fermi-LAT data in our ROI.  To this end, we model the GDE from the Milky Way
based on physical models for the production and propagation of CRs in the
Galaxy.  By means of the \texttt{Galprop} code, we build a large set of GDE
models with the aim of spanning the full range of possible physical conditions
that affect the gamma-ray emission coming from direction of the inner Galaxy.
Each background model is then tested against \Fermi-LAT data during the fitting
procedure.  

Two criteria guide our search for GDE models for the inner Galaxy: {\it a}) a
\emph{good statistical fit}, \ie small $TS$ values, where $TS$ $\equiv - 2 \ln
\mathcal{L}$ throughout; and {\it b}) \emph{self-consistency}, \ie agreement
between the predicted and measured levels of the individual GDE components
(this is not guaranteed in a template fit, where spectra can vary arbitrarily).  

\bigskip

The paper is organized as follows:  In section~\ref{sec:methods} we discuss the
data reduction, the statistical methods and the modeling of some of the
backgrounds. Section~\ref{sec:diffuse} is dedicated to a detailed explanation
about how we model the main components of the Galactic diffuse emission.  In
section~\ref{sec:results} we then present the results on the spectrum and
morphology of the excess emission, and estimate theoretical and empirical model
systematics.  We perform parametric fits to the data in
section~\ref{sec:parametric}, and leave our discussion to
section~\ref{sec:discussions}.  Finally, in section~\ref{sec:conclusions} we
state our conclusions.


\section{Data analysis}
\label{sec:methods}

In this work, we study gamma rays collected by the \Fermi-LAT from the inner
Galaxy with template-based multi-linear regression techniques, see \fex
ref.~\cite{Dobler:2009xz, Su:2010qj}.  The main difference with respect to
previous studies is an extensive treatment and discussion of systematics
related to the modeling of the GDE, and an incorporation of these uncertainties
in the final spectral and morphological fits to the GCE.  In addition, we
introduce a few technical improvements, including a weighted adaptive masking
of point sources, the proper treatment of the point-spread-function (PSF) of
\Fermi-LAT, and the non-logarithmic binning of energies which facilitates flux
measurements at high energies.

\subsection{Data selection}
\label{sec:data}

In our analysis, we use 284 weeks of reprocessed \Fermi-LAT data (starting from
4 Aug 2008) with energies between 300 MeV and 500 GeV.\footnote{See
\url{http://fermi.gsfc.nasa.gov}.} We apply the standard zenith-angle cut
$\theta<100^\circ$ in order to avoid contamination from the Earth limb, and the
recommended \texttt{gtmktime} cut \texttt{(DATA\_QUAL>0) \&\&
(LAT\_CONFIG==1)}.  To maximize the available number of photons at high
energies we use front- and back-converted \texttt{P7REP\_CLEAN}
events.\footnote{We leave an analysis based on a subset of events with better
angular resolution at sub-GeV energies for future
work~\cite{Portillo:2014ena}.} The initial event selection and the calculation
of exposure maps is done using the standard \texttt{Fermi ScienceTools~v9r32p5}
with the instrument response functions (IRFS) \texttt{P7REP\_CLEAN}
\texttt{\_V15}.

We use the \texttt{healpix} projection for the spatial binning of
data~\cite{Gorski:2004by}.  The \texttt{healpix} grid is a hierarchical
equal-area iso-latitude pixelization of the sphere, and it is well suited for
full-sky analyses of astronomical data.  In our work, we adopt the resolution
parameter $n_\text{size}=256$, which corresponds to a pixel size of
$1.598\times10^{-5}\rm\ sr$ (roughly $0.23^\circ$ edge length for square
pixels).  All mask definitions in this paper refer to the pixel center as
defined in the \texttt{healpix} grid, leading to $\pm \sim 0.1^\circ$
variations at the mask edges.

\medskip

As noted above, the baseline ROI of this work, which we refer to as \emph{inner
Galaxy}, is defined as
\begin{equation}
    |\ell|\leq 20^\circ \quad \text{and} \quad 2^\circ\leq |b| \leq 20^\circ\;,
    \label{eqn:ROI}
\end{equation}
where $\ell$ and $b$ are the Galactic longitude and latitude, respectively.  As
discussed above, the lower latitude cut is applied to reduce contamination with
diffuse and point source emission from the Galactic disk.  The overall region
is kept small (compared to previous analyses~\cite{Hooper:2013rwa,
Daylan:2014rsa}) in order to avoid biasing our analysis results by potential
mis-modeling of the diffuse emission in regions of the sky that are irrelevant
for the GCE.  However, we will use a number of additional ROIs for validation
tests as well as for estimates of model systematics.

\medskip

In contrast to previous studies, we use energy bins that increase in
logarithmic size with energy.  This will partially counterbalance the reduced
photon statistics that usually complicate the analysis at energies above 10
GeV.  To this end, we define energy bins such that for a given photon flux with
a spectral index of $\Gamma$, each bin contains an equal number of expected
events.  For a given energy range $E_0 \equiv E_\text{min}$ to $E_\text{max}$,
and a given number of bins $n_\text{bins}$, we find that the boundaries of
these energy bins can be recursively determined by
\begin{equation}
    E_{j+1} = \left(E_j^{1-\Gamma} -
    \frac{E_\text{min}^{1-\Gamma} -
    E_\text{max}^{1-\Gamma}}{n_\text{bins}}
    \right)^\frac{1}{1-\Gamma}\;, \quad \text{with} \quad
    j=0, 1, \dots, n_\text{bins}\;.
\end{equation}

In the present analysis, we use $n_\text{bins}=20$ bins  in the range 500 MeV
to 500 GeV and adopt the value $\Gamma=1.45$, which is harder than the actual
spectrum.  This is a compromise between a loss in statistics at high energies
on the one hand, and unreasonably wide energy bins on the other hand.
Furthermore, we add four linearly spaced energy bins between 300 and 500 MeV.

\subsection{Statistical framework}
\label{sec:analysis}

We use the maximum likelihood technique for parameter inference and confidence
interval estimation.  To this end, we adopt a weighted Poisson likelihood
function for the photon data~\cite{Cash:1979vz},
\begin{equation}
    -2\ln\mathcal{L} = 2\sum_{i,j} w_{i,j} (\mu_{i,j} - k_{i,j} \ln \mu_{i,j})
    + \chi^2_\text{ext}\;,
    \label{eqn:Likelihood}
\end{equation}
where $\mu_{i,j}$ and $k_{i,j}$ are, respectively, the expected and observed
number of photons in the $i^\text{th}$ energy bin and $j^\text{th}$ pixel, and
$\chi^2_\text{ext}$ allows for external constraints on the model parameters.
The definition of the point source (PSC) mask weights $w_{i,j}\in[0,1]$, as
well as of the external constraints $\chi^2_\text{ext}$, will be discussed
below.  

\medskip

The expected number of photons is given by the sum of different space- and
sometimes energy-dependent templates that represent: \textit{i}) the GDE (tying
$\pi^0$ and bremsstrahlung components together, \cf section \ref{sec:diffuse}
for details), \textit{ii}) the \Fermi~bubbles, \textit{iii}) the IGRB,
\textit{iv}) the emission from the detected PSCs in the \Fermi-LAT Second
Source catalogue (2FGL)~\cite{Collaboration:2011bm}, and \textit{v}) a
spherically symmetric profile, centered at the GC that accounts for the GCE
excess emission (\cf section \ref{sec:simpleTemplates}).  We leave the
normalization of each model component free to float in each energy bin.  The
total model is defined by
\begin{equation}
    \mu_{i,j} = \sum_k \theta_{i,k} \mu_{i,j}^{(k)}\;,
\end{equation}
where $\theta_{i,k}$ is the normalization of component $k$ in energy bin $i$,
and $\mu_{i,j}^{(k)}$ is the predicted number of events of component $k$ when
$\theta_{i,k}=1$.

External constraints will be adopted both for the IGRB and for the
\Fermi~bubbles, which we leave free to float in our analysis, but which can be
more efficiently determined in regions outside of our ROI.  The constraints are
assumed to be Gaussian, such that $\chi^2_\text{ext}$ is of the form
\begin{equation}
    \label{eq:constrain}
    \chi^2_\text{ext} = \sum_{i,k}
    \left(\frac{\phi_{i,k}-\bar\phi_{i,k}}{\Delta\phi_{i,k}}\right)^2\;,
\end{equation}
where $\phi_{i,k}$ denotes the predicted flux of component $k$ in energy bin
$i$, and $\bar\phi_{i,k}$ and $\Delta \phi_{i,k}$ are the externally supplied
mean and standard deviation (see section~\ref{sec:simpleTemplates}).

\smallskip

We obtain the best-fit model parameters by minimizing $-2\ln\mathcal{L}$ with
respect to all parameters, using the minimizer \texttt{Minuit}.\footnote{See
\url{http://seal.web.cern.ch/seal/MathLibs/Minuit2/html/}.}  For the template
analysis, we derive error bars from the covariance matrix of
$-2\ln\mathcal{L}$, and checked that our results remain essentially unchanged
when using the more accurate (but much more time-consuming) \texttt{minos}
algorithm in \texttt{Minuit}.  Note that we allow negative values of
$\theta_{i,k}$ in our fits, which improves the stability of \texttt{Minuit}
without affecting our conclusions.

\bigskip

The finite angular resolution of the LAT is incorporated by following the
prescription described in ref.~\cite{FermiLAT:2012aa}.  Using \texttt{healpix},
we decompose the skymap into spherical harmonics, re-weight the moments
according to the decomposition of the \Fermi-LAT PSF (taking the GC as
reference position), and then transform them back to sky coordinates.  This
method is used for all diffuse emission components, except the GCE templates
(which remains non-smoothed for simplicity), and the PSC templates (which we
instead smooth using the more accurate \texttt{gtmodel} from the \texttt{Fermi
ScienceTools}).  We found that our results are not significantly affected by
the details of the smoothing (more specifically, neglecting smoothing of the
diffuse components entirely or smoothing them instead with a fixed-width
Gaussian with $2^\circ$ FWHM induces changes in the spectra at the \% level).

\bigskip

In order to minimize the impact of known point sources on our analysis, we use
a `soft' PSC mask, with values that can range from zero to one.  In practice,
this mask acts as an energy-dependent re-weighting of the exposure and the
associated count numbers in different regions of the sky.  The corresponding
reweighting factor, $w_{i,j}$, enters eq.~\eqref{eqn:Likelihood} as a simple
prefactor that multiplies the likelihood contributions from individual pixels.
In the Gaussian limit, this corresponds to increasing statistical errors by a
factor $w_{i,j}^{-1}$.

The weights are defined as follows. We infer the expected number of photons
from PSCs, $\mu_{i,j}^\text{PSC}$, in different pixels and energy bins by
creating a model map that contains all point sources of the
2FGL~\cite{Collaboration:2011bm}, fixing their fluxes to the best-fit
parameters of the catalogue.  We compare these numbers with the expected number
of background events from the GDE (including only the $\pi^0$, ICS and
Bremsstrahlung components at their nominal normalization),
$\mu_{i,j}^\text{BG}$.  For definiteness, we take the model P.\footnote{We
refer to appendix~\ref{app:60models} for the definition of model parameters. It
represents the reference model of ref.~\cite{FermiLAT:2012aa}.} The weights are
defined by
\begin{equation}
    w_{i,j} = \frac{1}{\left(\frac{\mu_{i,j}^\text{PSC}}{f_\text{PSC}\
    \mu_{i,j}^{BG}}\right)^{\alpha_\text{PSC}} + 1}\;,
\end{equation}
where $f_\text{PSC}$ denotes the threshold of the fraction of point source
contamination above which a pixel is masked, and $\alpha_\text{PSC}$
parameterizes the smoothness of the transition between $w=0$ and $w=1$.  As
default values, we select $\alpha_\text{PSC}=5$ and $f_\text{PSC}=0.1$.  We
discuss the systematics related to our choices concerning the PSC mask
definition in section~\ref{sec:spectrum}.

\medskip

We emphasize that we apply the PSC mask \emph{only} when performing fits and
calculating likelihood values.  The calculation of \emph{fluxes} integrated
over our ROI, as shown in various figures throughout the paper, remains
unaffected by the PSC mask,  in order to simplify comparison with other work.
Furthermore, as described below, we include as a default template in all
analysis steps the PSC flux derived from all of the 2FGL point sources with a
normalization that is fixed to one.  By construction, the 2FGL sources do not
affect the fits to the data, but they contribute to the overall flux in our ROI
and indicate where PSC are relevant and where not.


\subsection{Adopted templates}
\label{sec:simpleTemplates}

The PSC template is derived from the 2FGL~\cite{Collaboration:2011bm}, as
described above.  The GDE templates will be extensively discussed in
section~\ref{sec:diffuse}.  We describe here briefly the remaining spatial
templates that we use in our analysis and highlight their main characteristics.  

\begin{table}
    \centering
    \small
    \begin{tabular}{ccc}
        \toprule
        Name & Notes & Ref. \\\midrule
        PSC & Spectra fixed to 2FGL & \cite{Collaboration:2011bm} \\
        \Fermi~bubbles & Flat emission | Spectrum constrained & \cite{Su:2010qj, FranckowiakBubbl} \\
        IGRB & Constant emission | Spectrum constrained & \cite{AckermannEGB} \\
        GCE & Generalized NFW profile with inner slope $\gamma$ & -- \\
        \midrule
        Ackermann+ GDE models ($\times$13)& ($\pi^0$ + Bremss) + ICS & \cite{FermiLAT:2012aa} \\
        Additional GDE models ($\times$47) & ($\pi^0$ + Bremss) + ICS  & This work \\
        \bottomrule
    \end{tabular}
    \caption{Different templates and models used throughout the analysis.  We
    construct the models in the bottom row by using the \texttt{Galprop v54}
    code \cite{Strong:1998fr}.  As indicated, in our template analysis we tie
    both the bremsstrahlung and $\pi^0$ components together, ``$\pi_0$+Bremss",
    and vary them simultaneously by a common factor. Furthermore, we will split
    the GCE (ICS) templates in 10 (9) different segments during the morphology
    analysis in section~\ref{sec:morphology}.}
    \label{tab:templates}
\end{table}

\paragraph{\Fermi~bubbles.}  In our main analysis, we model the emission of the
\Fermi~bubbles as flat within the region defined in ref.~\cite{Su:2010qj}.  We
will discuss the impact of a possible latitude dependence of the \Fermi~bubbles
emission in appendix~\ref{sec:Bubbles}.

As mentioned in section~\ref{sec:analysis}, when performing fits in our main
ROI, which is relatively small and does not include the whole \Fermi~bubbles
template, we introduce an additional external constraint on the bubbles flux
implemented as in eq.~\eqref{eq:constrain}.  Mean and standard deviations are
taken from measurements in ref.~\cite{FranckowiakBubbl}, where the errors
incorporate systematics related to the uncertainties in the
GDE.\footnote{During the final stages of our work, the first \Fermi\ bubbles
analysis by  the \Fermi-LAT Collaboration has been released
\cite{Fermi-LAT:2014sfa}. Data above $10^{\circ}$ in latitude have been
analyzed and the spectrum and morphology of the \Fermi~bubbles have been
derived. Since we do use preliminary results of this analysis to constrain the
bubbles spectrum, we do not expect a large variation of our results due to the
imposed bubbles spectral shape.}

\paragraph{IGRB.} Fits within our main ROI do not well constrain the very
subdominant IGRB emission.  We hence constrain this component -- similarly to
what we do for the \Fermi~bubbles -- by introducing additional terms in the
likelihood function, with mean values and standard deviations (which include
statistical errors and systematical errors from the GDE and other sources) as
given in ref.~\cite{AckermannEGB}.

\paragraph{GCE template.} Throughout most of our analysis (except for the
morphology studies in sections~\ref{sec:results} and \ref{sec:parametric} and
appendix.~\ref{app:properties}), we will parametrize the volume emissivity of
the GCE in terms of the spherically symmetric\footnote{Although the DM profile
is generally assumed to be spherically symmetric, N-body simulations predict
some degree of tri-axiality of the halos in their inner part, see for example
ref.~\cite{Vera-Ciro:2014ita} for a recent discussion. In this case, an
elongation of the DM associated emission would be expected towards a specific
direction.  We discuss the elongation of the measured excess in
appendix~\ref{app:properties}.} generalized NFW profile, as given by
\begin{equation}
    \rho(r) = \rho_s \frac{(r/r_s)^{-\gamma}}{(1 + r/r_s)^{3 - \gamma}} \,,
    \label{eq:NFWgen}
\end{equation}
squared.  This is clearly motivated by the DM annihilation interpretation of
the GCE.  We will use the common normalization $\rho(r_\odot)=0.4\rm\ GeV/cm^3$
at the position of the Sun and a scale radius of $r_s = 20\kpc$  to ease of
comparison with previous results in the literature.  If not stated otherwise,
we will furthermore adopt the value $\gamma=1.2$, which is compatible with
previous (see \fex refs.~\cite{Hooper:2013rwa, Daylan:2014rsa}) and our own
findings below.

In general, the gamma-ray flux ($\GeV^{-1}\cm^{-2}\s^{-1}\sr^{-1}$) from
self-conjugate DM particles, $\chi$, annihilating in the Galactic DM halo is
given by
\begin{equation}
  \frac{dN}{dE}= \frac{\langle \sigma
  v\rangle}{8\pi \,m_\chi^2} \, \frac{dN_{\gamma}}{dE}
  \int_\text{l.o.s.}\!\!\!\!\! ds\;\rho^2(r(s, \psi))\,,
  \label{eqn:fluxADM}
\end{equation}
where $m_\chi$ is the DM mass, $\langle \sigma v\rangle$ the velocity averaged
total annihilation cross-section, $dN_{\gamma}/dE$ the averaged energy spectrum
of photons produced per annihilation, and $\rho(r)$ the radial DM energy
density distribution as function of the Galacto-centric distance $r$.  The
coordinate $s\geq0$ runs along the line-of-sight and is related to the distance
from the GC by $r(s,\psi) = \sqrt{(r_\odot-s\cos\psi)^2 + (s\sin\psi)^2}$,
where $\psi$ is the angular distance from the GC and $r_\odot=8.5\kpc$ denotes
the distance between Sun and GC.

\section{Galactic diffuse emission models}
\label{sec:diffuse}

The observed gamma rays with energies above 100 MeV typically originate from
CRs that propagate in the Galaxy. Firstly, CR nuclei produce via inelastic
nucleon-nucleon collisions with the interstellar gas neutral mesons (mainly
$\pi^{0}$s), which subsequently decay to gamma-ray pairs \cite{Strong:1998fr,
Kamae:2006bf, Kelner:2006tc}.  This is typically, and throughout this work,
referred to as the $\pi^{0}$ Galactic diffuse component.  Secondly, CR
electrons interact with the interstellar gas, giving rise to bremsstrahlung
emission \cite{Koch:1959zz, 1969PhRv..185...72G, Blumenthal:1970gc}, which can
be a significant diffuse gamma-ray component below a few GeV in the gas-rich
regions of our Galaxy.  Finally, CR electrons up-scatter low-energy photons via
ICS into the gamma-ray energy regime \cite{Jones:1968zza, Blumenthal:1970gc}.
ICS, bremsstrahlung and $\pi_0$ constitute the three components of the GDE.
Various ingredients enter the evaluation of these three GDE components, which
we will discuss in the remainder of this section.

\medskip

Firstly, the exact distribution of the CR sources is very important.  This is
especially true for electrons because they lose energy quickly, such that their
distribution in the Galaxy -- at higher energies -- is strongly correlated to
their initial injection (\ie acceleration and production) region.\footnote{CR
electrons are of different origin: They are diffusive shock accelerated ISM
electrons at SNRs, secondaries from inelastic pp and pHe collisions
(predominantly), and electrons produced in pulsars magnetospheres and further
accelerated at the termination shock of Pulsar Wind Nebulae (PWNe).  } In
addition, the assumptions on the diffusion scale height of the Galaxy, the
diffusion coefficient\footnote{Assuming that CRs diffuse isotropically in the
Galaxy, $\lambda$ is the diffusion length.} $D(=\frac{1}{2}\lambda c)$ and its
dependence on CR rigidity (which together set the time scale nuclei of a given
rigidity stay in the Galaxy before escaping to the intergalactic medium) can
strongly affect the diffuse components.  The large scale assumptions about
diffusion, convection and re-acceleration can strongly impact -- and are thus
constrained by -- the CR spectra \cite{Trotta:2010mx, Evoli:2008dv,
Cholis:2013lwa} that we measure locally. The same applies for the large scale
(angular and spatial) gamma-ray spectra \cite{Cholis:2011un, FermiLAT:2012aa,
Hryczuk:2014hpa}, which, in turn, impact any indirect DM search, either in
antiprotons \cite{Evoli:2011id, Cirelli:2013hv,Cirelli:2014lwa}, positrons
\cite{Cholis:2013psa} or gamma-rays \cite{Tavakoli:2013zva}.

\smallskip

Furthermore, the exact assumptions on the gas distribution in the inner Galaxy,
which is the target of CRs responsible for the $\pi^{0}$ and the bremsstrahlung
emission, affect in a direct manner the gamma-ray observations. The main
constituent of the interstellar gas is the atomic hydrogen (HI) traced by its
21 cm line emission \cite{Nakanishi:2003eb, Tavakoli:2012jx}, while the other
important component of the ISM gas is the molecular hydrogen (H2), which is
traced indirectly by the 2.6 mm line emission from CO \cite{Pohl:2007dz}.
Additionally, since CR electrons of energies $\mathcal{O}(1)$ GeV and above
suffer from fast energy losses due to synchrotron and ICS, the exact
assumptions on the magnetic field distribution and the interstellar radiation
field (ISRF) in the region of interest is of vital importance. In this context,
magnetic fields, which have a random and an ordered component at every position
of the Galaxy, can affect both the synchrotron electrons losses, which are
important above few GeV in energy, and the way CRs diffuse especially in
regions with strong ordered magnetic fields (see ref.~\cite{Dobler:2011mk} for
an example on the impact magnetic fields can have on causing anisotropic
diffusion of CRs and its impact on potential gamma-ray signals from DM).  The
ISRF assumptions matter for a second reason, namely the fact that those photons
are the target of CR electrons that up-scatter them to the observed gamma-ray
energy regime.

\smallskip

Emission directly correlated with the GC or the inner Galaxy comes only from
the inner 1--2 kpc of the Galaxy where the CRs propagation conditions can be
very different from those locally. For example, one may need to consider the
impact of potentially strong convective winds that could exist in that region
and could be associated to the \Fermi~bubbles.  Moreover, CR diffusive
re-acceleration may be different in that part of the Galaxy compared to the
local environment.  In fact, the propagation conditions in the inner Galaxy can
have only a very mild effect on the local CR spectra, which instead depend much
more on the assumptions regarding, for example, the spiral arms (see for
instance ref.~\cite{Gaggero:2013rya}), and are hence only weakly constraint.

\bigskip

In the following subsection we describe how the GDE can be modeled starting
from the fundamental ingredients of production and propagation of CRs in the
Galaxy. By using the \texttt{Galprop} code and varying the physical parameters
at stake, we build a set of GDE models in order to explore physical scenarios
for such a gamma-ray background. We discuss the parameter ranges that are
considered and the main assumptions of this approach. Finally, we will discuss
typical template variations that different assumptions on the GDE components
can lead to.

\subsection{Building GDE models with Galprop}
\label{sec:GDEGalprop} 

In our analysis, we make use of the \texttt{Galprop v54} code
\cite{Strong:2007nh, GALPROPSite, GalpropV54} to produce gamma-ray templates
for the GDE at various energies.  Additionally we download the \texttt{healpix}
GDE $\pi^{0}$, bremsstrahlung and ICS component maps from some of the models of
ref.~\cite{FermiLAT:2012aa}, which were also produced with \texttt{Galprop
v54}.  We refer for further details to ref.~\cite{FermiLAT:2012aa}.  We will
briefly summarize the properties of the adopted GDE models here, and provide
more details in appendix~\ref{app:60models}.

\medskip

CR sources are assumed to be distributed with a rotational symmetry on the
disk, with varying radial dependence. Their distribution can either follow the
distribution of SNRs, pulsars, OB stars (see ref.~\cite{FermiLAT:2012aa}) or
combinations of them. The primary CR electrons, protons and nuclei are then
injected into the interstellar medium with a power-law $dN/dE_{{\rm kin}} =
E_{{\rm kin}}^{-\alpha}$, which can vary between species and with kinetic
energy (per electron or nucleon) $E_{{\rm kin}}$.\footnote{We include all
species up to Carbon stable isotopes (Iron for the models of
ref.~\cite{FermiLAT:2012aa}) and take the CR protons and heavier CR nuclei to
have the same injection indices.} For the Galactic gasses, as we described,
there are two major neutral hydrogen components HI and H2 and a subdominant
contribution of ionized hydrogen HII.  These are all modeled as separate
components, with H2 suffering from the fact that a conversion from CO to H2 has
to be performed (see refs.~\cite{Pohl:2007dz, GALPROPSite}).\footnote{For the
GDE models that we built ourselves, we use the conversion factor $X_{CO}$
profile from ref.~\cite{Strong:2004td}, while the models from
ref.~\cite{FermiLAT:2012aa} come with their own $X_{CO}$ profiles. Given that
most of the H2 emission close to the inner Galaxy is actually masked by our low
latitude cut $|b|>2^\circ$, the precise radial dependence of $X_{CO}$ is of
little relevance here.}

Diffusion of CRs in the Galaxy is assumed to be homogeneous and isotropic
within a cylindrical volume of $r\leq r_{D}$ and $|z|\leq z_{D}$, where $z$ and
$r$ parametrize the position along the longitudinal and polar axes.  It is
described by a scalar diffusion coefficient depending on rigidity $R$ as
\begin{equation}
    D(R) \equiv D_{xx}(R) = D_{0} \left(\frac{R}{4 GV}\right)^{\delta}\,,
    \label{eqn:Diffusion}
\end{equation}
where $D_0$ is the diffusion coefficient at 4 GV and  $\delta$ is the diffusion
index with values between 0.3--0.6.\footnote{$\delta$ can vary with rigidity
but for our work we keep it fixed to 1/3.}  Diffusive re-acceleration in turn
is connected to spatial diffusion through a simple relation
\begin{equation}
    D_{pp}(R) =  \frac{4}{3 \delta (2-\delta)(4-\delta)(2+\delta)} \frac{R^{2} v_{A}^{2}}{D_{xx}(R)} \, ,
    \label{eqn:Reacceleration}
\end{equation}
with $v_{A}$ being the Alfv$\acute{\textrm{e}}$n speed.

Convection is considered to be taking place perpendicularly away from the
Galactic disk, with the convection velocity being zero on the disk but having a
gradient $dv/dz$.  The Galactic magnetic field responsible for synchrotron
losses of CR electrons is assumed to have a cylindrical symmetry with the
parametrization
\begin{equation}
    B(r,z) = B_{0}\, e^{(r_{\odot}-r)/r_c} \, e^{-|z|/z_{c}} \,,
    \label{eqn:B-field}
\end{equation}
where $B_0$ is the local magnetic field and $r_c$ and $z_c$ are the radial and
longitudinal extension, respectively ($r_\odot$ is 8.5 kpc).

Finally, the ISRF is built from the contribution of many stellar components and
includes the effects of absorption and re-emission from dust grains (see
ref.~\cite{GALPROPSite} for further details and ref.~\cite{Porter:2008ve} for a
description on how it is constructed). Within the code the ISRF is divided into
three basic components, related to the direct emission from stars, dust grains
and CMB.  The user is free to vary the normalization of each of these
components.

\bigskip

\begin{figure}
    \begin{center}
        \includegraphics[width=0.32\linewidth]{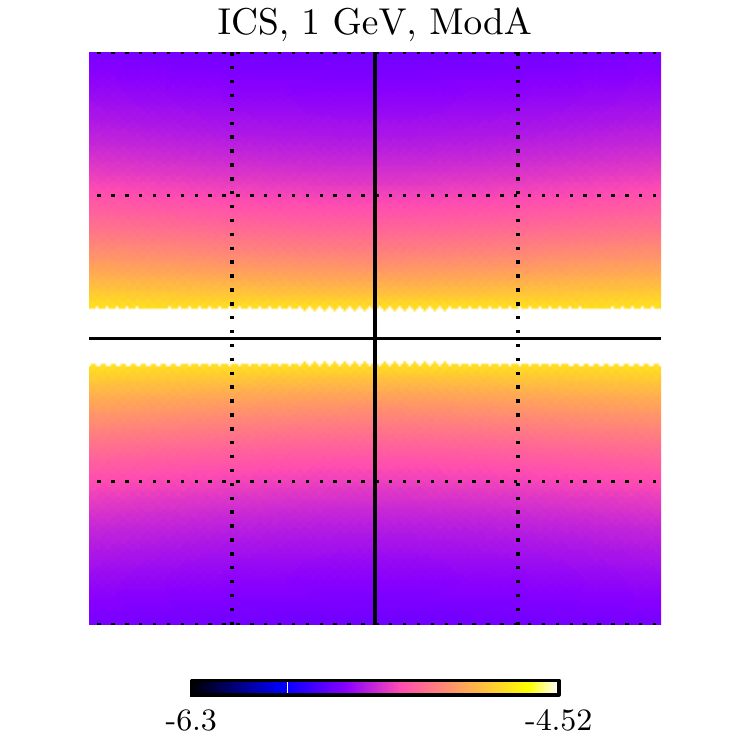}
        \includegraphics[width=0.32\linewidth]{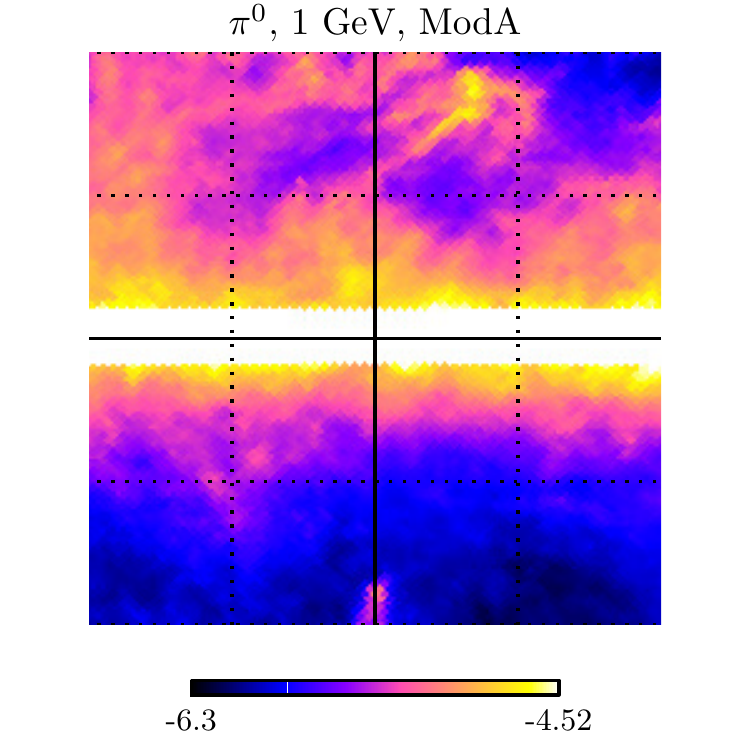}
        \includegraphics[width=0.32\linewidth]{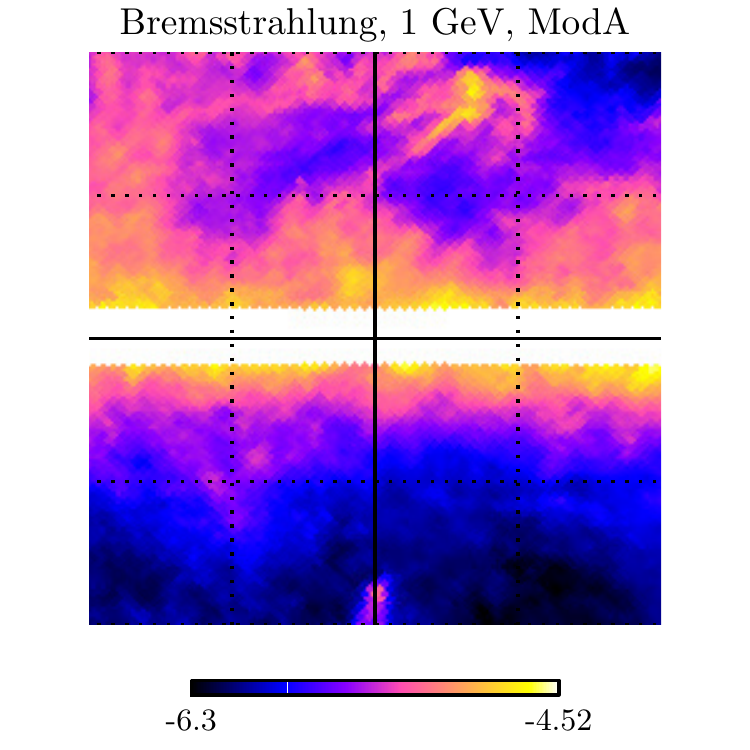}
    \end{center}
    \caption{Predicted emission for the GDE components of model A. From {\it
    left} to {\it right}: ICS,  $\pi^{0}$, and bremsstrahlung. The fluxes are
    shown in the $40^{\circ} \times 40^{\circ}$ sky-region, centered at the GC
    and masking out $|b| < 2^{\circ}$. The corresponding units are
    $\log_{10}(\rm GeV^{-1}\,cm^{-2}\,s^{-1}\,sr^{-1})$.}
    \label{fig:mapsDiff}
\end{figure}

\begin{figure}[th!]
    \begin{center}
        \includegraphics{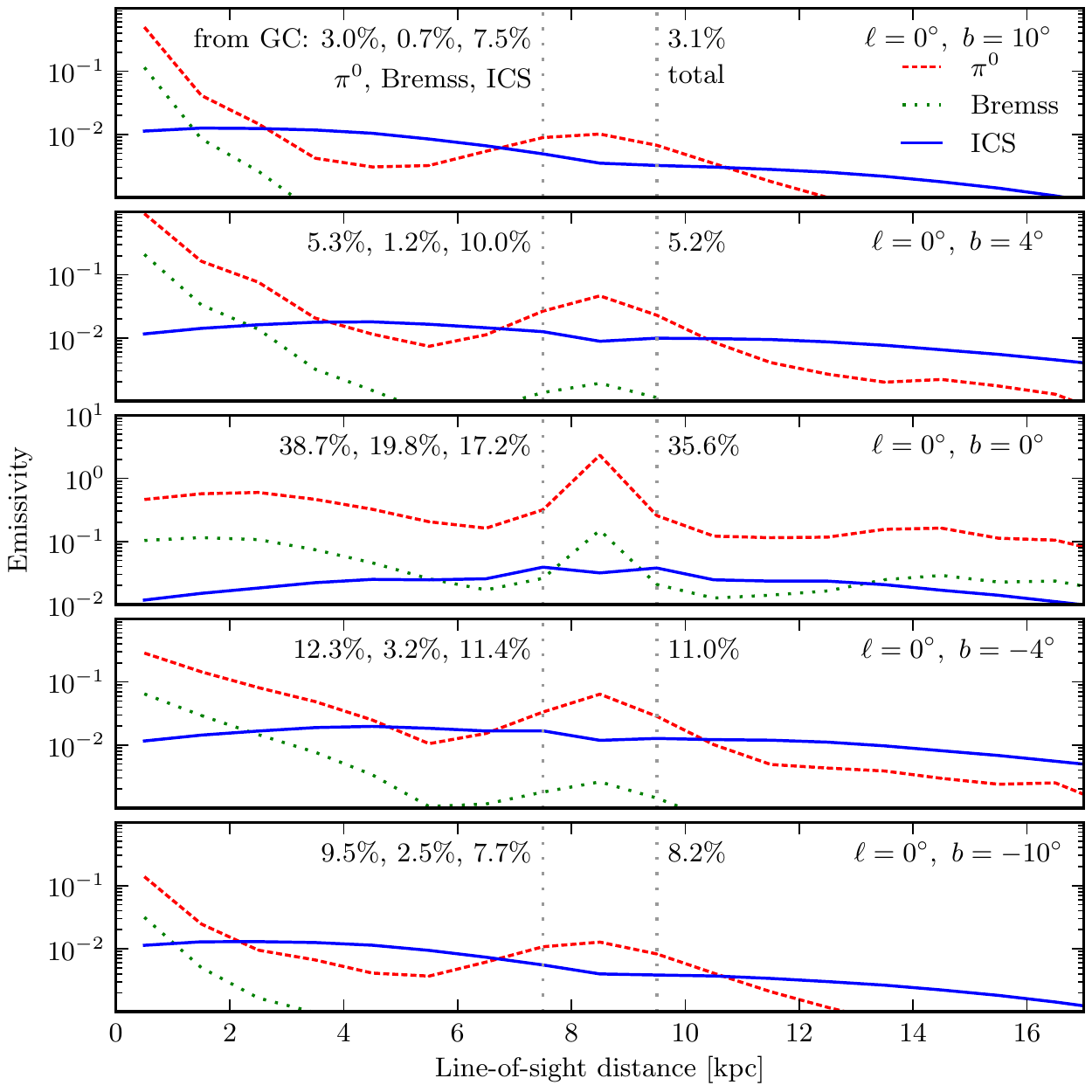}
    \end{center}
    \caption{Typical differential volume emissivity $dN/dV/dt/dE$ (in arbitrary
    units) of the three GDE components along the line-of-sight at five
    different Galactic latitudes, zero Galactic longitude, and gamma-ray
    energies of 2.6\GeV.  The numbers show the fraction of $\pi^0$,
    bremsstrahlung, ICS and total emission that comes from the distance range
    7.5--9.5\,kpc (with the GC being at 8.5\,kpc).  At latitudes above
    $|b|\geq4^\circ$, less than about 11\% of the total GDE comes from this
    central region.  Note that the reduced amount of \emph{local} gas in the
    southern hemisphere leads at negative latitudes to larger relative
    contributions from the Galactic center.}
    \label{fig:los}
\end{figure}

\begin{table}[t!]
    \centering
    \small
    \begin{tabular}{cccccccccc}
     \toprule
        Name & $z_{D}$ & $D_{0}$ & $v_{A}$ & $dv/dz$ & Source & $\alpha_{e}$($\alpha_{p}$) & $N_{e}$($N_{p}$) & $B$-field  & ISRF \\
     \midrule
        A        &  4  &  5.0  &   32.7    &   50  &  SNR  &  2.43(2.47)  & 2.03(5.8) & 090050020 & 1.36,1.36,1.0  \\ 
        B        &  4  &  28.0  &   31.0   &   0  &  SNR  &  2.43(2.39)  & 1.00(4.9) & 105050015 & 1.4,1.4,1.0  \\  
        C        &  4  &  5.0  &   32.7    &   0  &  SNR  &  2.43(2.39) & 0.40(4.9) & 250100020 & 1.0,1.0,1.0  \\  
        D        &  4  &  5.2  &   32.7    &   0  &  SNR  &  2.43(2.39)  & 0.40(4.9) & 050100020 & 0.5,0.5,1.0  \\  
        E        &  4  &  2.0  &  32.7    &   0  &  SNR  &  2.43(2.39)  & 0.40(4.9) & 050100020 & 1.0,1.0,1.0  \\  
        \bottomrule
    \end{tabular}
    \caption{The properties of GDE models A--E. Here, $z_{D}$ is in kpc, while
    $r_{D}$ is taken to be 20 kpc. $D_{0}$ is in units of $10^{28}$ $\cm^{2}
    \s^{-1} $, $v_{A}$ is in $\km \s^{-1} $ and $dv/dz$ in $\km \s^{-1}
    \kpc^{-1}$.  The CR electron and proton normalizations are $N_{e}$($N_{p}$)
    in units $ 10^{-9}$ $\cm^{-2} \sr^{-1} \s^{-1} \MeV^{-1}$ and refer to the
    differential flux at $E_{\rm kin}$ of 34.5 and 100 GeV.  $\alpha_{e}$ and
    $\alpha_{p}$ are the electron and proton injection indices above rigidities
    of 2.18 and 11.3 GV, respectively (and are respectively equal to 1.6 and
    1.89 below these rigidities). For the gas assumptions we take, $T_{S} =
    150$ K and an E(B-V) magnitude cut of 5 (see discussion in
    section~\ref{sec:Uncertainties}).  For model A the magnetic field
    ``090050020'' denotes in eq.~\eqref{eqn:B-field} $B_{0} = 9.0$ $\mu$G,
    $r_{c} = 5$ kpc and $z_{c} = 2$ kpc (similarly for the other models).
    Finally, the three numbers in the ``ISRF'' column refer to the
    multiplication factors of the ``optical'', ``IR'' and CMB components of the
    ISRF model used in \texttt{Galprop v54} webrun.}
    \label{tab:ModelA}
\end{table}

In figure \ref{fig:mapsDiff}, we show the typical morphology of the three
different diffuse emission components at 1 GeV from a model, model A, which has
parameters that are defined in table~\ref{tab:ModelA}.  We will refer to model
A as our reference model for further discussions, since as we will see below it
well describes the gamma-ray data and spectra in the inner Galaxy. The left
panel of figure \ref{fig:mapsDiff} refers to the ICS emission, which is smooth
and depends mainly on the electron distribution and the properties of the
$B$-field and the ISRF. On the other hand, $\pi_0$ (middle) and bremsstrahlung
(right) morphologies trace directly the distribution of the gas and depend
mainly on the proton and electron CR densities, as well as on the properties of
CR diffusion, re-acceleration and convection.

\medskip

The observed emission as shown in figure \ref{fig:mapsDiff} receives
contributions from all distances along the line-of-sight.  However, whether the
overall emission is dominated by locally produced gamma rays (within a few
kpc), or by gamma rays from the GC, is a strong function of the Galactic
latitude.  This is illustrated in figure \ref{fig:los}, where we show the
fractional contributions to the GDE components as function of the line-of-sight
for a typical GDE model.\footnote{To generate this figure, we used our own
modified version of \texttt{Galprop 54.1.984} where the line-of-sight
integration can be restricted.  We adopted a simple GDE model defined by the
galdef file \texttt{54\_01}.} We find that in the case of our baseline ROI,
less than 14\% (and for $|b|\geq4^\circ$ less than 11\%) of the GDE actually
comes from regions close to the GC.  The main challenge in extracting the GCE
in the inner Galaxy is hence to characterize the uncertainties and properties
of the \emph{local} gamma ray emissivity.

\bigskip

We close this subsection with a discussion of our model A, which we adopt as a
reference model throughout.  We \emph{tuned} model A to be ``self-consistent''
in the sense that, after the fit to the data that we will perform below, the
\emph{measured} and \emph{predicted} normalizations of the GDE template
components agree with high accuracy (see introduction, and see
section~\ref{sec:results} below).  The purpose of this model is to constitute a
proof-of-principle that shows that such a self-consistent model is indeed
possible with physically not unreasonable parameters.

However, we note that there are significant degeneracies between the physical
assumptions that can have a similar impact on the resulting diffuse gamma-ray
spectra. For instance, a harder ICS spectrum could be the result of a harder
injection spectrum for the CR electrons,  a lower energy loss rate (due to a
reduced $B$-field),  a different rigidity dependence of the CR diffusion or  a
different distribution in the energy density of the ISRF.  Thus, we can have
different combinations of physical properties leading to self-consistent
models.

Our model A adopts a CR electron normalization that is significantly higher (by
a factor of 5) than the typical $dN_{e}/dE \simeq$0.4$\times 10^{-9}$ $\cm^{-2}
\sr^{-1} \s^{-1} \MeV^{-1}$ at $\sim$30 GeV that is required to fit the
\emph{local} CR lepton data.  In particular, for the ICS emission at GeV scale
gamma-ray energies, the responsible CR electrons have typically energies
$\sim$50 GeV or higher. At these energies, the CR electron fluxes are dominated
by the contribution of sources within $\sim 1$kpc from the Sun's position and
are not very sensitive to higher normalizations in the electron flux at few kpc
distances.  Thus, a higher CR electrons normalization $\textit{along the
line-of-sight}$ and towards the GC is quite well possible.  Such a higher flux
may be due to some young or middle aged pulsars that lie along that direction,
either in the spiral arms or in the inner part of the Galaxy.  We note that
both the adopted $B$-field and ISRF have relatively high energy densities.
Given the uncertainties along the line-of-sight toward the GC, they are still
viable.  In fact model A suggests a $B$-field with an amplitude of 50$\mu$G at
the GC, which is in agreement with the lower limit of
ref.~\cite{Crocker:2010xc}.  All the other assumptions of model A are very
conventional ones.

\subsection{The explored parameter space}
\label{sec:Uncertainties}

Different assumptions on the source distribution, gas distribution, diffusion,
re-acceleration, convection, magnetic field distribution and the ISRF in the
inner part of the Galaxy (and along its relevant line-of-sight) will lead to
different GDE models.  The assumptions on all these factors and their
associated uncertainties need to be taken into consideration in any study of
gamma rays from the inner Galaxy.  

To conservatively estimate the impact of these uncertainties on the extraction
of the GCE, we systematically explore a large range of model parameters that go
beyond what is imposed by CRs measurements, allowing even for extreme
scenarios.  We are interested in testing uncertainties pertaining to the CR
source distribution and injection index, gas distribution, diffusion scale
height and coefficient, re-acceleration, convection and energy losses for the
CR electrons. As we will discuss in section~\ref{sec:diffuseTemplatesSpectra},
the uncertainties in the ICS are potentially the most important ones in the
search of diffuse excess emission towards the GC. For that reason, we change
both the ISRF energy density and the Galactic magnetic field amplitude and
profile in a large range.  In the present work, we present 60 different GDE
models to bracket the uncertainties related to what we discussed above.

\medskip

Our starting point are the 128 GDE models from ref.~\cite{FermiLAT:2012aa},
which were created using \texttt{Galprop v54} to probe different distributions
in sources, diffusion scale heights and radii, and different assumptions on the
Galactic gasses.  CR sources distribution are assumed to follow: {\it i}) the
SNRs distribution of ref.~\cite{Case:1998qg} (we will denote it as just
``SNR''), {\it ii}) the pulsar distribution of ref.~\cite{Lorimer:2006qs}
(``Pls$_{L}$" option), or {\it iii}) the one of ref.~\cite{Yusifov:2004fr}
(``Pls$_{Y}$" option) and {\it iv}) the distribution of OB stars as described
in ref.~\cite{Bronfman:2000tw} (``OB" option).  For the diffusion scale heights
and radii, ref.~\cite{FermiLAT:2012aa} considers values of $4 \leq z_{D}
\leq10$ kpc and $r_{D}$ being either equal to 20 or 30 kpc.  Gasses were
associated to assumptions on the spin temperature $T_{S}$ that corrects the
opacity of the 21 cm line of the HI gas:  a more conventional assumption of
$T_{S} = 150$ K and a more extreme one with $T_{S} = 10^{5}$ K. In addition,
dust can be taken as an alternative tracer of neutral hydrogen (HI $\&$ H2) and
allow to model out some of the GDE residuals from $\pi^{0}$ and bremsstrahlung
as was first done using EGRET data by
ref.~\cite{2005Sci...307.1292G}.\footnote{In ref.~\cite{FermiLAT:2012aa}, two
different magnitude cuts (2 or 5) were used on the E(B-V) reddening maps of
ref.~\cite{Schlegel:1997yv} when fitting to that (dust) map a linear
combination of the HI 21 cm line map and the CO line map, which is directly
related to the H2 map for a homogeneous conversion factor between the two
gasses ($X_{CO}=$const.).} We refer the reader to ref.~\cite{FermiLAT:2012aa}
for further details.  In ref.~\cite{FermiLAT:2012aa} each of these 128 models
was fitted to the 21-month full-sky \Fermi-LAT data as well as the most
relevant local CR measurements.  We note that the models from
ref.~\cite{FermiLAT:2012aa} were however not optimized for an inner Galaxy
study, but provide physical examples that give a good overall fit to gamma-ray
and CR data.

In order to avoid redundancies, and since we are interested in exploring more
extreme propagation scenarios as described above, we use only 13 of the models
from ref.~\cite{FermiLAT:2012aa} that probe the different choices in sources,
$r_{D}$, $z_{D}$ and gas (see appendix~\ref{app:60models} for further details).
Having added more of these models in the analysis would have not changed our
general results (see also ref.~\cite{Zhou:2014lva} for a discussion on all the
128 models of ref.~\cite{FermiLAT:2012aa}).

\medskip

In addition to our selection of 13 models from ref.~\cite{FermiLAT:2012aa}, we
generate our own models specifically for this study, using \texttt{Galprop v54}
(webrun version).  Those models explore remaining uncertainties, mainly related
to the diffusion coefficient, re-acceleration, convection, ISRF and $B$-field
distributions.  For the diffusion coefficient described in
eq.~\eqref{eqn:Diffusion}, a conventional value for $D_{0}$ in the range of
5--10$\times 10^{28}$ $\cm^{2} \s^{-1}$ (at 4GV) is used to fit the CR data and
the large scale gamma-ray data. As we discussed earlier, we probe significantly
larger ranges for the physical assumptions (which could well be realized in the
inner 2 kpc of the Galaxy without affecting much the CR data).  Thus for
$D_{0}$ we take a range between 2--60$\times 10^{28}$ $\cm^{2} \s^{-1}$. For
the re-acceleration, typical values coming from studying CRs are in the range
of 10--30 $\km \s^{-1} $ for the Alfv$\acute{\textrm{e}}$n speed, while we
consider in this work values between 0 and 100 $\km \s^{-1} $.  For the
gradient of convection velocity $dv/dz$, we allow values between 0 and 500 $\km
\s^{-1} \kpc^{-1}$, with CR antiprotons and large scale gamma-ray data not
showing any preference for values of $dv/dz > 50$ $\km \s^{-1} \kpc^{-1}$ for
the Galactic disk \cite{Evoli:2011id, Cholis:2011un}.  The standard assumption
for the ISRF model factors, is ``1.0, 1.0, 1.0'', see text in
section~\ref{sec:diffuseTemplatesSpectra}. We allow the ``optical'' and ``IR''
factors to span between 0.5--1.5 as our extreme options, given the level of
complexity of these models.  Finally, the magnetic field is known to have a
local value in the (generous) range of 3--10 $\mu$G with its $r_{c}$ and
$z_{c}$ scaling distances providing only a very rough description (see
ref.~\cite{Jansson:2009ip} for a discussion on the Galactic magnetic field
distribution and uncertainties). Typical \texttt{Galprop} assumptions include
$B_{0}=5$ $\mu$G, $r_{c} = 10$ kpc and $z_{c} = 2$ kpc as described in
eq.~\eqref{eqn:B-field}. Yet, the magnetic field close to the GC is expected to
have values as large as 50 $\mu$G \cite{Crocker:2010xc}. We take combinations
of $B_{0}$, $r_{c}$ and $z_{c}$ that allow values for the magnetic field at the
GC as low as 5.8 $\mu$G and as high as 117 $\mu$G with $5 \leq r_{c} \leq 10$
kpc and $1 \leq z_{c} \leq 2$ kpc.  Also, we include the possibility of a
significantly higher/lower CR electron population than what is measured
locally.

\medskip

We summarize the parameter ranges as follows:
\begin{itemize}
    \item geometry of the diffusion zone: $4 \leq z_{D} \leq10$ kpc and $r_{D}$
        = 20 or 30 kpc;

    \item source distributions: SNR, pulsars, OB stars;

    \item diffusion coefficient at 4 GV: $D_{0}  = 2-60 \times 10^{28}$
        $\cm^{2} \s^{-1}$;

    \item Alfv$\acute{\textrm{e}}$n speed: $v_{\rm A} = 0-100$ $\km \s^{-1} $;

    \item gradient of convection velocity:  $dv/dz$ = 0 -- 500 $\km \s^{-1} \kpc^{-1}$;

    \item ISRF model factors (for optical and infrared emission): 0.5 -- 1.5;

    \item  $B$-field parameters: $5 \leq r_{c} \leq 10$ kpc,  $1 \leq z_{c}
        \leq 2$ kpc, and $5.8 \leq B(r=0, z=0) \leq 117$ $\mu$G. 
\end{itemize}

It is evident that some (and in fact many) of our extreme models would be
completely ruled out by CR data and large scale diffuse gamma-ray data (or even
microwave data), if those options would describe the general Galactic
properties.  We include them with the attitude of testing whether the GCE
properties are significantly affected by \textit{extreme} Galactic diffuse
model assumptions.   

\bigskip

\paragraph{Limitations of our approach.}  When constructing the 60 GDE models
for the estimate of theoretical model uncertainties, we deliberately made a few
simplifying assumptions that we will summarize in the following.  This will be
a useful starting point for future attempts to explain the GCE in terms of
standard astrophysical processes.  However, the possible impact of those
limitations will be partially bracketed by the study of empirical model
uncertainties along the Galactic disk that we will present in
section~\ref{sec:ModelingUncertainties}.

\medskip

\texttt{Galprop} in its standard implementation solves the propagation equation
on a two dimensional spatial grid (assuming cylindrical symmetry), which is
prohibitive to the modeling of structures like the spiral arms and their impact
on the source distribution and CR propagation.  Any radial (and in the last two
cases also longitudinal) dependence in the description of convection,
re-acceleration or diffusion is absent from our GDE models.  Diffusion is taken
to be isotropic which can only be a rough approximation given the fact that
there are large scale magnetic fields in our Galaxy.\footnote{An example of a
modification of the \texttt{Galprop} code to account for inhomogeneous and
anisotropic diffusion of CRs due to the presence of large scale ordered
magnetic fields can be seen in ref.~\cite{Dobler:2011mk}. The anisotropic
diffusion of CRs strongly depends on the assumptions made about the ordered and
turbulent components of the Galactic magnetic field. Without including rotation
measures and also microwave data to probe the synchrotron total and polarized
intensity, we can not constrain those $B$-field components just from gamma-ray
observations. We let such questions for future work.} We summarize the main
limitations as follows:
\begin{itemize}
    \item assumption of homogeneity and isotropy of CR diffusion,
        eq.~\eqref{eqn:Diffusion};

    \item assumption of homogeneity of CR re-acceleration, described through a
        scalar quantity, eq.~\eqref{eqn:Reacceleration}; 

    \item lack of radial dependence of CR convection;

    \item assumption of radial symmetry of CR source distribution in the
        Galactic disk, not fully accounting for the spiral arms;

    \item assuming a steady state solution for the CRs, excluding transient
        phenomena;

    \item same spatial distribution of hadronic and leptonic CR sources;

    \item lack of a physical model for the \Fermi\ bubbles.
\end{itemize}

All these limitations will become important for future refined analyses of the
GCE, but are beyond the scope of the present work.

\subsection{Discussion about the template approach}
\label{sec:diffuseTemplatesSpectra}

\begin{figure*}
    \begin{center}
        \includegraphics[width=0.32\linewidth]{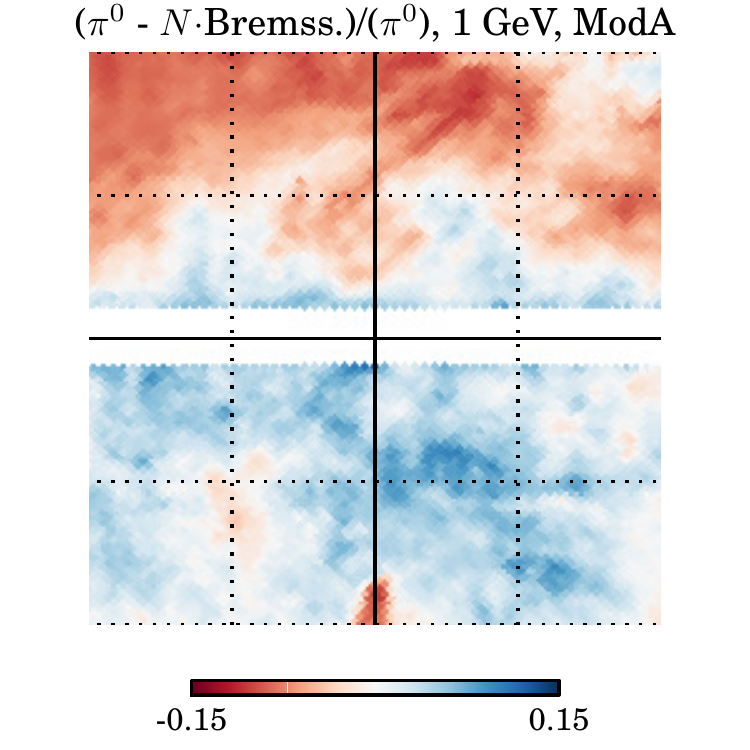}
        \includegraphics[width=0.32\linewidth]{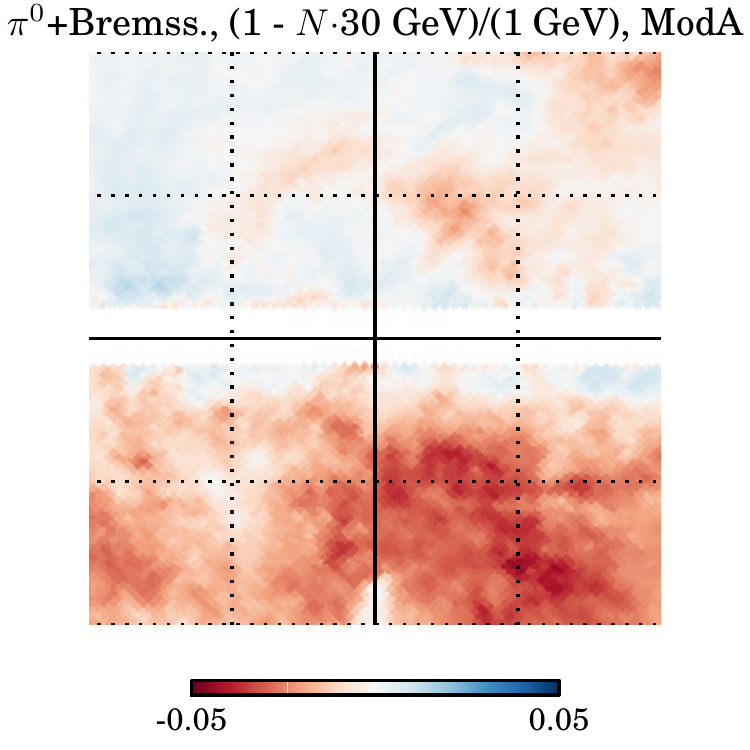}
        \includegraphics[width=0.32\linewidth]{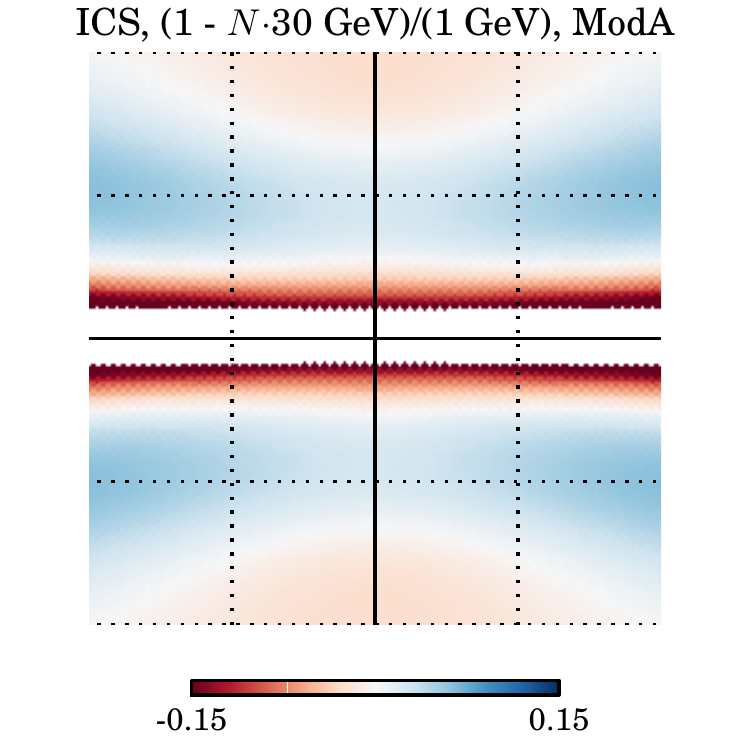}
    \end{center}
    \caption{Variations in the template morphologies within our baseline ROI,
    eq.~\eqref{eqn:ROI}.  \textit{Left panel}: Ratio of the $\pi^{0}$ map - $N
    \cdot$Bremss map over the $\pi^{0}$ map at 1 GeV for model A. The factor
    $N$ is fitted to minimize the sum of absolute residuals (we find $N=1.25$).
    The relative residuals exceed 5$\%$ (10\%) in 30$\%$ (0.36\%) of the
    pixels. The differential ratio map in combination with the fact that at 1
    GeV the two diffuse components are comparable makes it necessary to model
    the $\pi^{0}$ and bremsstrahlung emissions independently.  \textit{Central
    panel}: Ratio of the $\pi^{0}$+Bremss map at 1 GeV - $N
    \cdot$($\pi^{0}$+Bremss map) at 30 GeV over the $\pi^{0}$+Bremss map at 1
    GeV. The relative residuals never exceed a few $\%$.  Thus the
    $\pi^{0}$+Bremss map is morphologically the same in the energy range of
    interest and within the window of interest (at the $\%$ level accuracy).
    \textit{Right panel}: Ratio of the ICS map at 1 GeV - $N \cdot$ICS map at
    30 GeV over the ICS map at 1 GeV. The relative residuals exceed 5\% (10\%)
    close to the disk, in about 18\% (5.6\%) of the pixels. Thus there is an
    energy dependence of the ICS morphology that can lead to over/under
    subtracting the ICS emission at different energies if not taken into
    account.}
    \label{fig:TemplateComparisons1}
\end{figure*}

\begin{figure*}
    \begin{center}
        \includegraphics[width=0.32\linewidth]{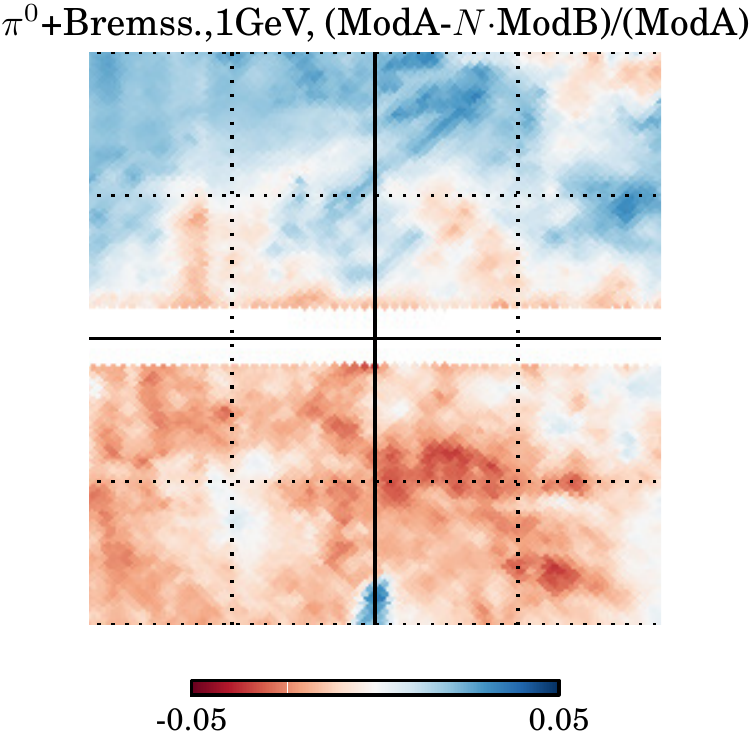}
        \includegraphics[width=0.32\linewidth]{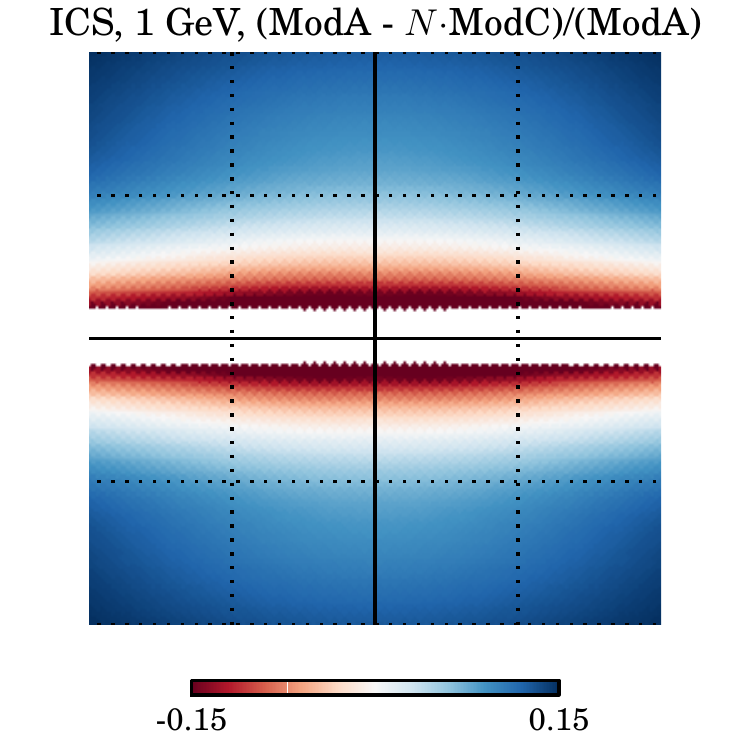}
        \includegraphics[width=0.32\linewidth]{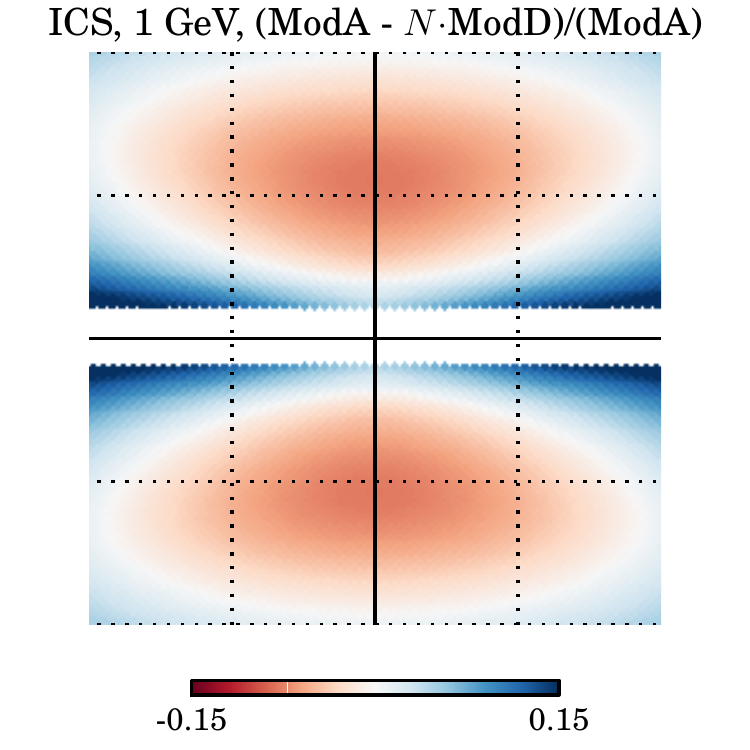}
    \end{center}
    \caption{We show in the \textit{left panel} the impact of faster CR
    diffusion on the $\pi^{0}$+Bremss map at 1 GeV. The differential ratio map
    does not exceed the few $\%$ level after including a normalization freedom
    on the combined $\pi^{0}$+Bremss map.  In the \textit{central} and
    \textit{right panels} we show the impact of different Galactic magnetic
    field and ISRF conditions on the ICS map. The differential ratio maps at 1
    GeV exceed the 5$\%$(10$\%$) level in 79$\%$(47$\%$) of the pixels when
    comparing model A and C, and 32$\%$(4.4$\%$) of the pixels when comparing
    models A and D. Disk-like, spherical and bubble-like morphologies can
    emerge in the differential ratio map, making it necessary to track down the
    exact physical assumptions towards the GC.}
    \label{fig:TemplateComparisons2}
\end{figure*}

\begin{figure}
    \begin{center}
        \includegraphics{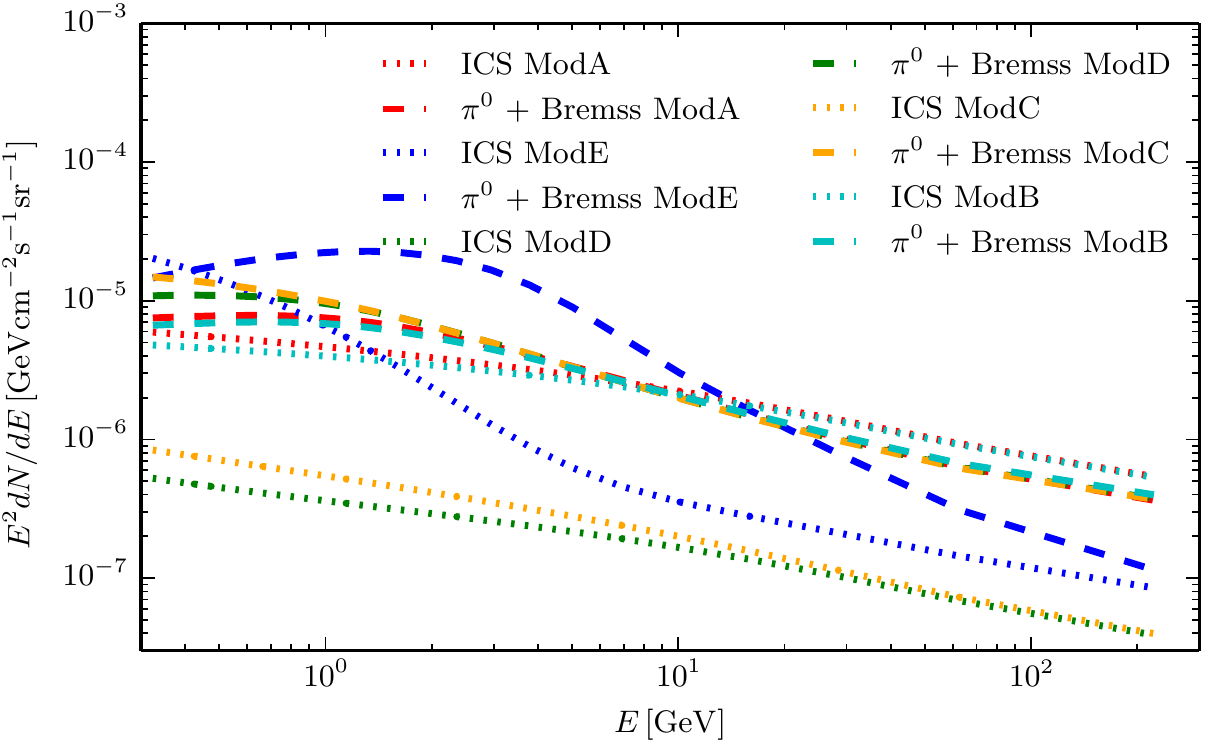}
    \end{center}
    \caption{Illustration of the predicted emission for the GDE components
    $\pi^{0} +$Bremss (dashed lines) and ICS (dotted lines) from five different
    models averaged over our baseline ROI.}
    \label{fig:spectrumDiff}
\end{figure}

\begin{figure}
    \begin{center}
        \includegraphics{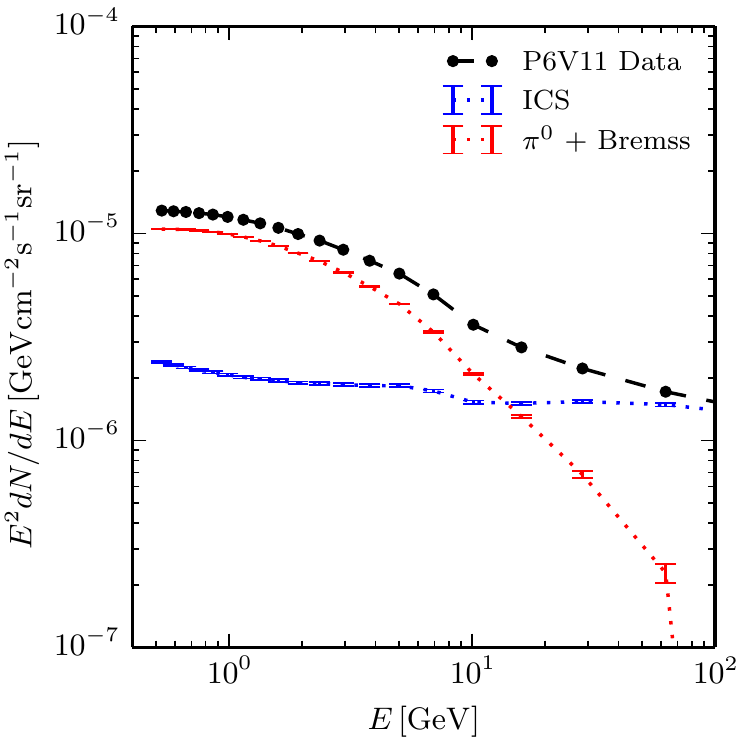}
        \includegraphics{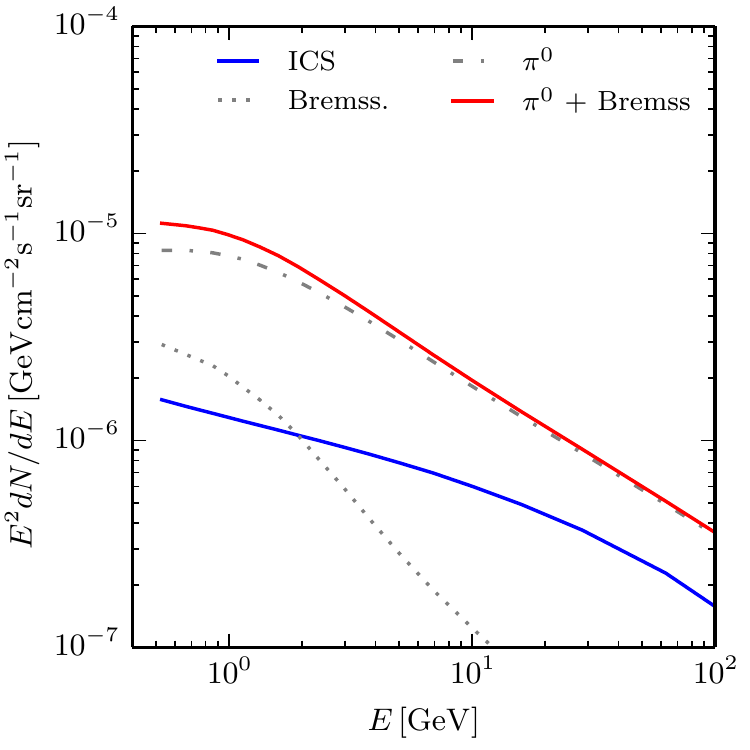}
    \end{center}
    \caption{\textit{Left panel:} Decomposition of the \PV background model
    into its contributions from ICS and $\pi^0$+Bremss.  The plot was generated
    by fitting simultaneously the ICS and $\pi^0$+Bremss components of the
    model P  to \PV  (see text for details). It does not vary much when other
    diffuse models are used instead. The extremely hard ICS emission at
    energies $>10\GeV$  is an \textit{intrinsic} property of the
    \texttt{P6V11}, which affects any analysis that employs it as GDE template.
    \textit{Right panel:} For comparison we show the actual spectra predicted
    by model P for ICS, $\pi^0$ and bremsstrahlung emission. Fluxes are
    displayed in the $40^{\circ} \times 40^{\circ}$ ROI, $|b|>2^{\circ}$.}
    \label{fig:P6V11}
\end{figure}

Before presenting our main results, we will illustrate and quantify the
limitations of the monolithic energy-independent templates for the individual
GDE components that were used in previous template-regression analyses. We will
also show the impact of variations of the astrophysical conditions (the
relevant reference models are summarized in table~\ref{tab:ModelA}).

Typically template analyses model both the $\pi^{0}$ and the bremsstrahlung
emission with a single template assuming that their morphological differences
are not significant. In figure~\ref{fig:TemplateComparisons1} left, we show
that such an assumption has limitations. We take the $\pi^{0}$ and the
bremsstrahlung maps at 1 GeV from the predictions of model A, and fit  the
bremsstrahlung map to the $\pi^{0}$ one within the shown region by minimizing
the residual template $|$$\pi^{0}$-$N \cdot$Bremss$|$ (summing over all
pixels). These two diffuse emission maps have generically a similar averaged
brightness at 1 GeV in that region of the sky (indeed, we find in the specific
case $N=1.25$ from the fit).  We then plot the ratio of the ($\pi^{0}$-$N
\cdot$Bremss)/$\pi^{0}$ maps at 1 GeV. As can be seen, in 30$\%$ of the ROI,
the pixels deviate from zero by more than 5$\%$.  Thus the disagreement between
the $\pi^{0}$ and the bremsstrahlung emission morphologies at 1 GeV can easily
exceed 5$\%$ toward the inner Galaxy.\footnote{While both the $\pi^{0}$ and
bremsstrahlung emission maps correlate with the same gasses, the fact that CR
electrons lose their energy much faster than CR protons as they propagate away
from the sources makes the bremsstrahlung maps brighter towards the disk at the
GeV energies.} This makes necessary to model the $\pi^{0}$ and the
bremsstrahlung emissions independently.  Therefore, for each considered GDE
model, we build separately  $\pi^{0}$ and the bremsstrahlung emission
templates.  However, as mentioned above, due to the large degeneracies between
both components we tie them together to a single $\pi^0$+Bremss template and
rescale them simultaneously in our template fit.

Given that the bremsstrahlung emission has generically a softer spectrum than
the $\pi^{0}$, one can understand that the  $\pi^{0}$+Bremss template will
change with energy. We show the impact of that in
figure~\ref{fig:TemplateComparisons1} middle plot, where we compare the
$\pi^{0}$+Bremss map at 1 GeV with the one at 30 GeV by producing the
($\pi^{0}$+Bremss(1GeV) - $N\cdot$ $\pi^{0}$+Bremss(30GeV))/
$\pi^{0}$+Bremss(1GeV) map. The change in the morphology is very smooth and at
the few $\%$ level.  For consistency we include the energy dependence of the
individual and the combined emission maps.  However, the ICS template does
change significantly faster with energy from 1 to 30 GeV (see right panel of
figure~\ref{fig:TemplateComparisons1}).  This is related to the fact that CR
electrons are solely responsible of the ICS emission, losing their energy much
faster than the CR protons responsible for the $\pi^{0}$ emission.  In about
18$\%$(5.6$\%$) of the pixels in the ROI shown we find relative absolute
residuals above 5$\%$ (10$\%$). Thus there is a sizeable energy dependence of
the ICS morphology that can lead to over/under subtracting the ICS emission at
different energies if not taken into account.

\bigskip

As discussed above, in the present work we are interested in \emph{extremal}
templates for the GDE emission, because we want to bracket the uncertainties in
extracting properties of the GCE.  Additionally, as stated earlier, the
physical conditions in the inner Galaxy can be quite different from what they
are locally, and we have fewer data to probe them than we do to probe the local
or the large scale averaged conditions.

In figure~\ref{fig:TemplateComparisons2} left we compare the $\pi^{0}$+Bremss
maps from models A and B. Their main difference is that model B assumes
significantly faster diffusion of CRs (factor of 5 in the diffusion
coefficient, see appendix~\ref{app:60models}).  The ratio of (Mod A
-$N\cdot$Mod B)/Mod A at 1 GeV remains below 5$\%$ (in absolute value) in all
pixels, which shows that after allowing for a free template normalization, this
effect remains rather small.

In the case of the ICS templates (shown in the central and right panels of
figure \ref{fig:TemplateComparisons2}) the assumptions on the physical
conditions, in particular the $B$-field and the ISRF distributions, can have a
size-able impact.  We find, again at 1 GeV, absolute residuals in the
differential ratio maps (Mod A -$N\cdot$Mod C)/Mod A and (Mod A -$N\cdot$Mod
D)/Mod A that exceed 5\% (10\%) in 79$\%$(47$\%$) and 32$\%$(4.4$\%$) of the
pixels.  This can be rather relevant for the GCE extraction, since the ICS
component is the one that is closest in morphology to the adopted GCE template.
In fact, in these differential ratio maps, disk-like, spherical and bubble-like
morphologies can potentially emerge, suggesting that the precise modeling of
ICS emission is of high importance for extracting information about the GCE.

\bigskip

Besides the morphology, also the spectral energy distribution of the different
components varies significantly from one model to another. To illustrate this
point, we show in figure \ref{fig:spectrumDiff} the energy spectra of the
combined $\pi^{0}$+Bremss component and the ICS component for five different
GDE models, averaged over our baseline ROI, eq.~\eqref{eqn:ROI}. Models B (E)
adopts a significantly faster (slower) CRs diffusion, while models C and D
refer to different assumptions on the ISRF and the $B$-field distribution in
the inner Galaxy. As can be seen in figure \ref{fig:spectrumDiff}, these models
predict very different $\pi^{0}$, bremsstrahlung and ICS spectra.  However, the
details of the spectra are not of relevance for the final results of this work,
since we allow a bin-by-bin refitting of the normalization of the individual
components, which, after the fit, leads to only small variations in the overall
flux of the template components.  Nevertheless, differences in the spectra are
relevant for the self-consistency check.

\bigskip

A simple model for the GDE that has been widely used in the literature is the
\texttt{P6V11} model by the LAT Collaboration.  This model was developed for
background subtraction in the study of point sources, and introduces
systematics into any analysis of the gamma-ray diffuse emission that are
basically unknown.

Since the \texttt{P6V11} Galactic diffuse model has just one free parameter
(its total normalization), the relative contributions from ICS, $\pi^0$ and
bremsstrahlung are fixed.  We show a \textit{decomposition} of this model into
its main components ($\pi^0$+Bremss and ICS) in figure~\ref{fig:P6V11} (left
panel).  To obtain this plot, we perform a full-sky ($|b|>2^{\circ}$) fit of
the $\pi^0$+Bremss and ICS components of the model P to the \texttt{P6V11}
model. The expected fluxes from the model in the region of interest are shown
in the right panel of figure~\ref{fig:P6V11}. We checked that our results do
not change qualitatively when using the $\pi^0$+Bremss and ICS of other diffuse
models, and can hence be considered as robust.  The plot shows that the
\texttt{P6V11} features an extremely hard ICS component at energies above 10
GeV, which in a template regression analysis can easily lead to
over-subtraction of other diffuse components in the data.\footnote{This
peculiarity of the \texttt{P6V11} is also contained in the \Fermi~documentation
at
\url{http://fermi.gsfc.nasa.gov/ssc/data/P6V11/access/lat/ring_for_FSSC_final4.pdf}.}
It is easily conceivable that this property of the \PV contributes to the
softening of the GCE emission above 10 GeV that was found \fex in
refs.~\cite{Hooper:2013rwa, Daylan:2014rsa}, but is absent in our analysis.

The reason for this hard ICS component at high energies is not obvious.  It
could potentially be related to neglecting the contribution from the \Fermi\
bubbles during its construction, which, in turn, could have been partially
absorbed into the ICS component.  Whatever the exact reason for this hard ICS
spectrum is, it shows that the \PV is not self-consistent in the above sense.


\section{Non-parametric analysis of the Galactic center excess}
\label{sec:results}

In this section we present our main results from the template-based
multi-linear regression analysis, and describe in detail how we estimate the
various model systematics for the GCE.  Parametric fits to the spectrum and
morphology of the GCE are kept for section~\ref{sec:parametric}. 

In subsection~\ref{sec:spectrum} we discuss the theoretical model systematics
of the GCE spectrum that we infer from a set of 60 GDE models.  In
subsection~\ref{sec:ModelingUncertainties} we estimate and discuss the
empirical model systematics by analyzing \Fermi-LAT data in 22 test regions
along the Galactic disk.  In subsection~\ref{sec:morphology} we finally present
a study of the GCE in different segments of our baseline ROI, which provides a
handle for a later study of the morphology and extension of the GCE.

\subsection{Theoretical model systematics}
\label{sec:spectrum}

\begin{figure}
    \begin{center}
        \includegraphics{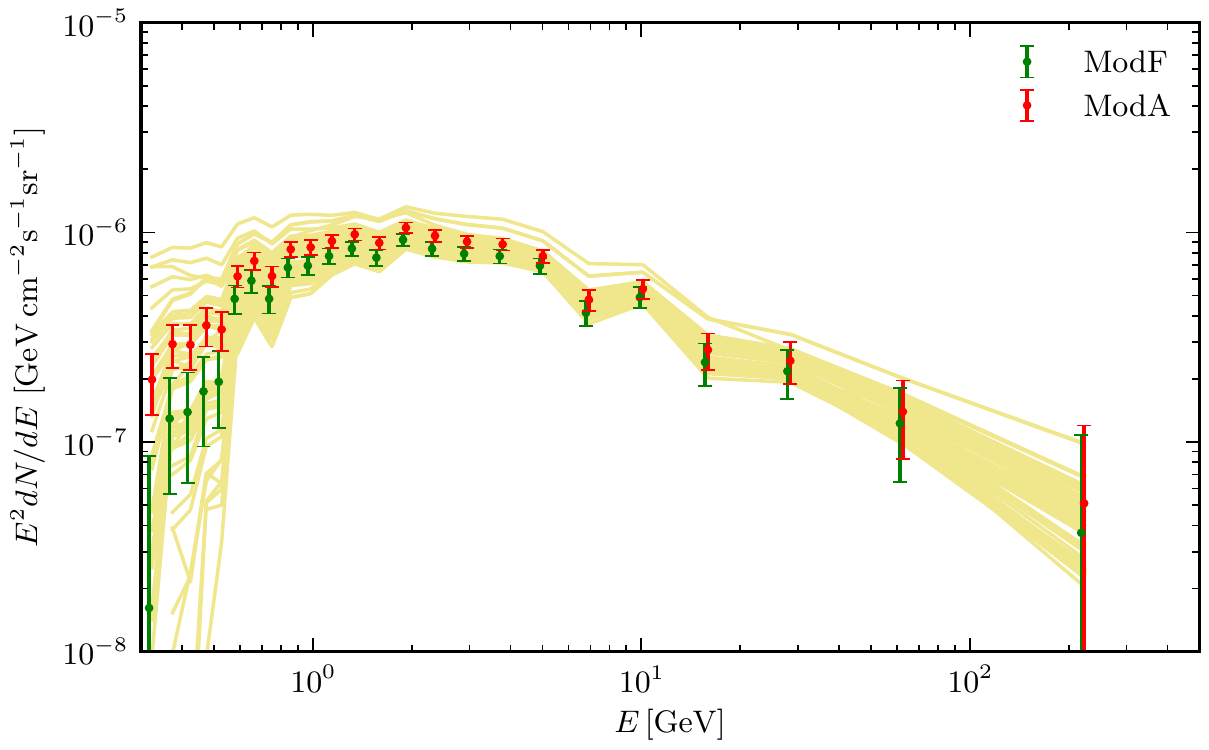}
    \end{center}
    \caption{Plain GCE energy spectrum as extracted from our baseline ROI,
    assuming a generalized NFW profile with an inner slope $\gamma=1.2$, for
    all of the 60 GDE models (\emph{yellow lines}).  We highlight the model
    that provides the best overall fit to the data (model F, \emph{green
    points}) and our reference model from the discussion in
    section~\ref{sec:diffuse} (model A, \emph{red points}), together with
    $\pm1\sigma$ statistical errors.  For all 60 GDE models, we find a
    pronounced excess that peaks at around 1--3 GeV, and follows a falling
    power-law at higher energies.}
    \label{fig:spectrum60models}
\end{figure}

In figure~\ref{fig:spectrum60models} we show the main result of this
subsection: the energy spectrum of the GCE that we find when adopting
\emph{any} of our GDE models.  The individual diffuse components that
contribute to the fit were discussed above and are summarized in
table~\ref{tab:templates}.  They include a set of 60 GDE models that span a
rather large range of physical conditions.  As described above, the
normalization of each emission component is left free to float in each energy
bin, but additional external constraints are applied to the IGRB and \Fermi\
bubbles templates.

We use our set of 60 GDE models to bracket the theoretical uncertainties that
affect the extraction of the GCE.  It is remarkable that the GCE emission
spectrum is rather stable.  At energies above 1 GeV, the overall flux varies by
less than a factor of 2--3, and features in all cases a pronounced peak at
energies around 1--3 GeV.  At higher energies, the spectrum appears to be well
described by a power-law with a spectral slope of $\sim -2.7$ (but see
discussion in section~\ref{sec:parametric}).  At lower energies we observe a
pronounced rise in the energy spectrum with a spectral index that is
significantly harder than $\sim2$ for all of the GDE models, though the exact
form of the spectrum is rather dependent on the adopted GDE model.  

In figure~\ref{fig:spectrum60models}, we also highlight the spectra that we
obtain for the GDE model F, which yields formally the best fit, and for the GDE
model A, which we used above in section~\ref{sec:diffuse} as reference model
(\cf parameters in table~\ref{tab:ModelsParameters1}).  These spectra are shown
together with their statistical errors, which are -- except at the highest
energies -- smaller than the width of the theoretical model systematics band.

\medskip

\begin{figure}
    \begin{center}
        \includegraphics{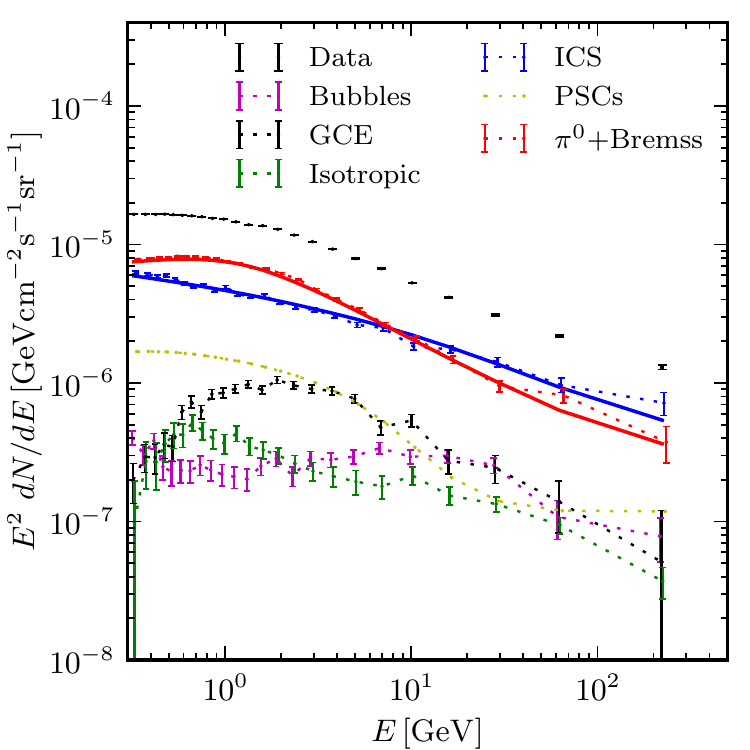}
        \includegraphics{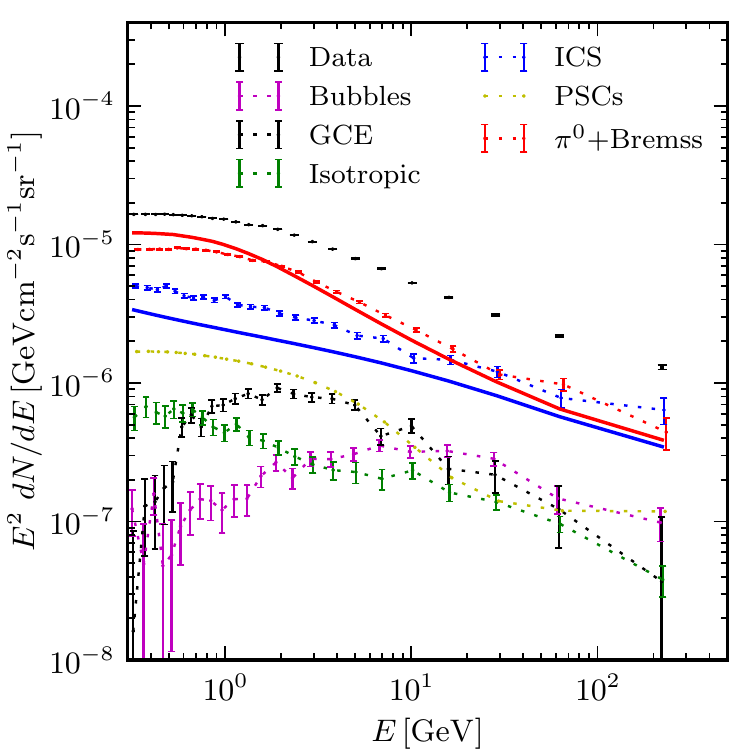}
    \end{center}
    \caption{Energy spectra of different components (\emph{dotted lines}) from
    a template fit to the data (\emph{black points}), compared to the predicted
    GDE model fluxes (\emph{solid lines}).  The reference model A is shown in
    the \emph{left panel}, while the GDE model that provides the best-fit to
    the data, model F, is shown in the \emph{right panel}.}
    \label{fig:consistency}
\end{figure}

In figure~\ref{fig:consistency}, we show the energy spectra of the different
diffuse and PSC components for model A and model F, averaged over the baseline
ROI and compared to the data. Since the normalization of all components is left
free to float, independently in each energy bin, it is not guaranteed that the
individual measured spectra actually correspond to a physical model.  However,
as already discussed above, we find that for model A (which was specifically
constructed for that purpose) the \emph{predicted} and the \emph{measured}
energy spectra of the GDE components agree very well at the level of
$5$--$10\%$.\footnote{We checked that this is also true when applying the
latitude cut $|b|\geq5^\circ$ instead and repeating the fits.}  This serves as
a proof-of-principle that the results obtained from the template fit can
actually correspond to a physical GDE model.  For model F, which yields the
best-fit, the fitted GDE fluxes deviate somewhat from the predicted ones, but
are still close to what we found for model A.  Below, we will use model A and F
as reference scenarios.\footnote{As can be seen in
figure~\ref{fig:consistency}, the spectrum of the IGRB (in the left panel) and
the \Fermi\  bubbles (in the right panel) is sometimes overly suppressed at
energies below 1 GeV, which suggests an over-subtraction of the GDE.  We
checked that this possible over-subtraction has only minor impact on our
results and decided to keep these energy ranges in our analysis, see discussion
in subsection~\ref{sec:ModelingUncertainties}.}

\medskip

\begin{figure}
    \begin{center}
        \includegraphics{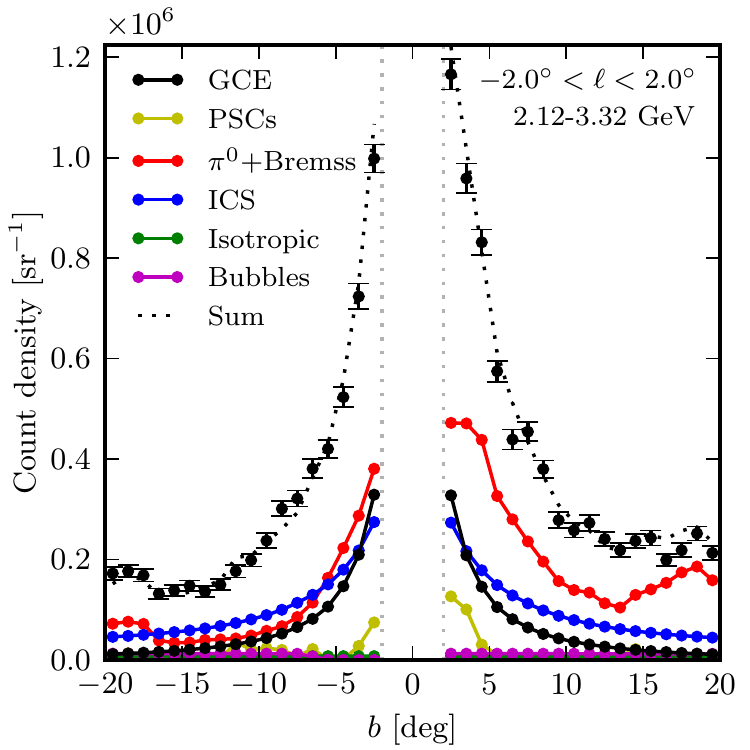}
        \includegraphics{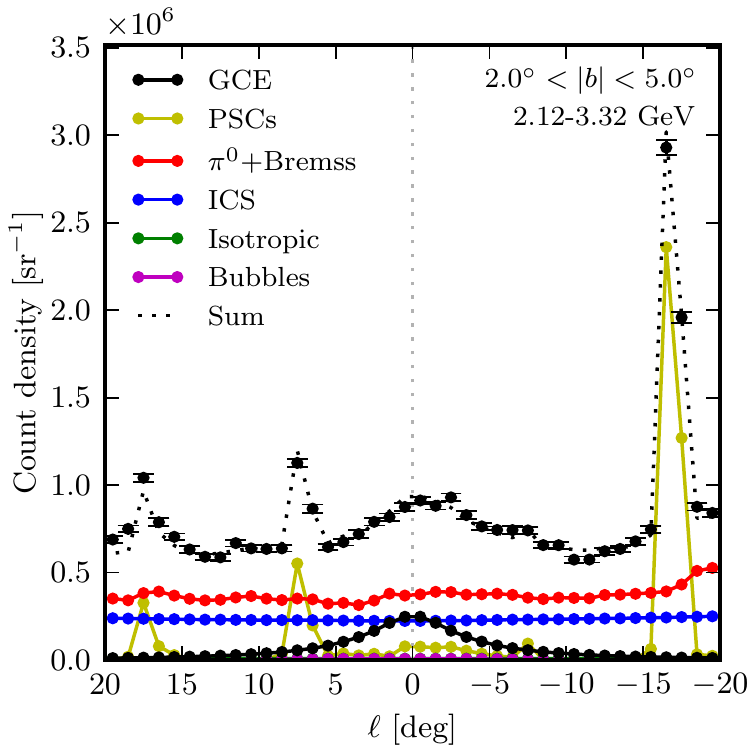}
        \includegraphics{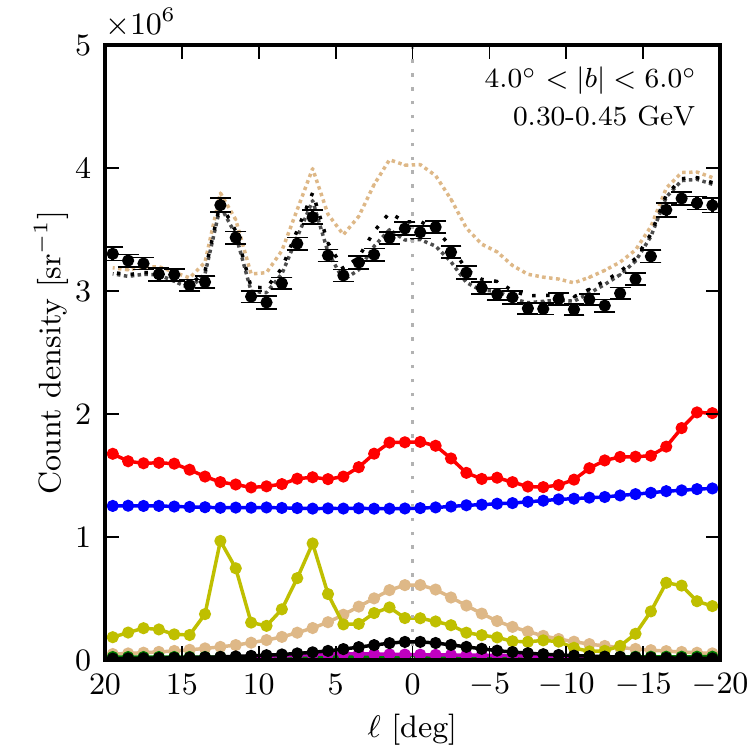}
        \includegraphics{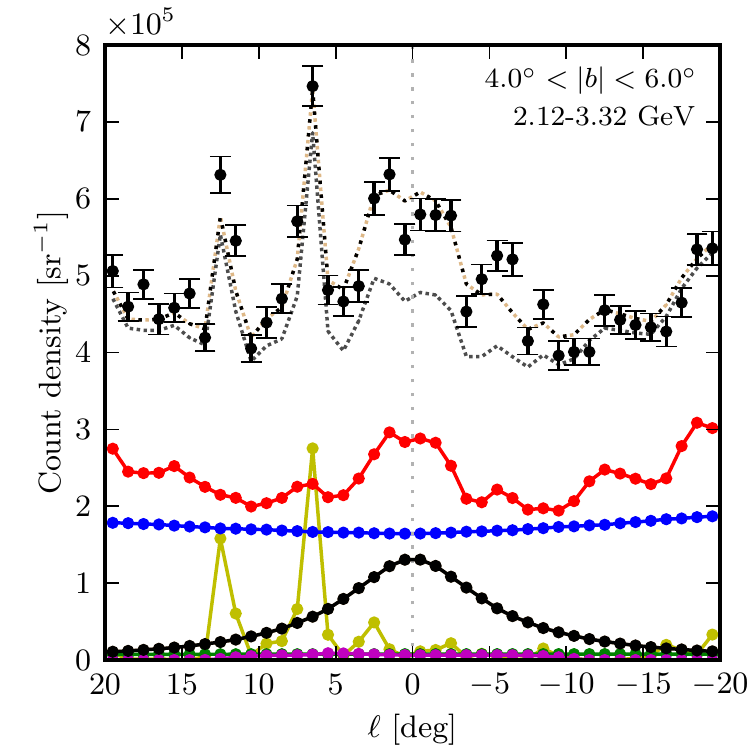}
    \end{center}
    \caption{\emph{Top panels:} Latitude (\emph{left}) and longitude
    (\emph{right}) dependence of the different components in narrow spatial
    strips, at energies around 3 GeV, for model A.  \emph{Bottom panels:}  Same
    as top right panel, but further away from the Galactic disk in the latitude
    range of $4^\circ<|b|<6^\circ$, at energies around 400 MeV (\emph{left})
    and 3 GeV (\emph{right}).  Furthermore, in the \emph{bottom left panel},
    the \emph{light pink line} illustrates  the situation of a GCE spectrum
    that is softer than the one that we find in our template fits. For
    comparison, we show here the case where the GCE component flux would follow
    a simple power-law with spectral index $2.0$ at energies below 2 GeV,
    keeping the normalization fixed to the one measured at 2 GeV.  The summed
    spectrum clearly overshoots the data in the inner few degrees.  Finally, in
    the \emph{bottom right panel}, the \emph{gray densly dotted} line shows in
    addition the sum of fluxes when the GCE component is neglected.  }
    \label{fig:profiles}
\end{figure}

\begin{figure}
    \begin{center}
        \includegraphics[width=0.93\linewidth]{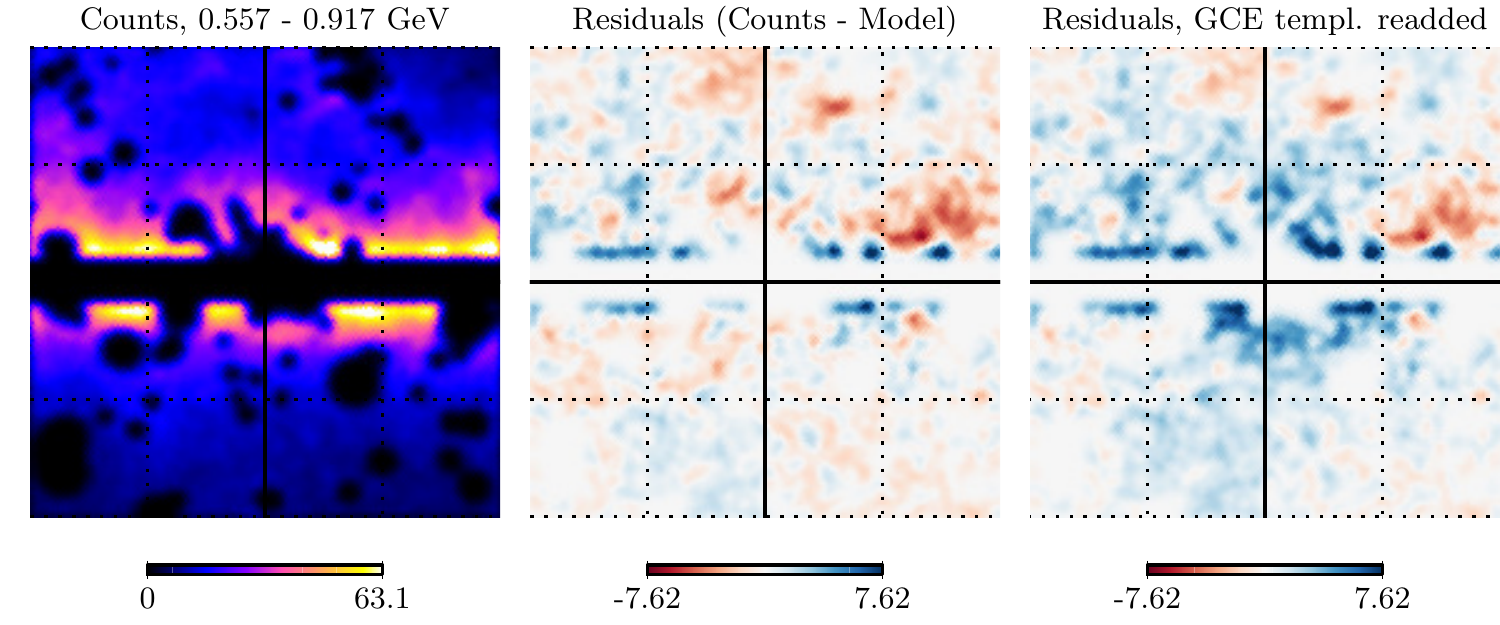}
        \includegraphics[width=0.93\linewidth]{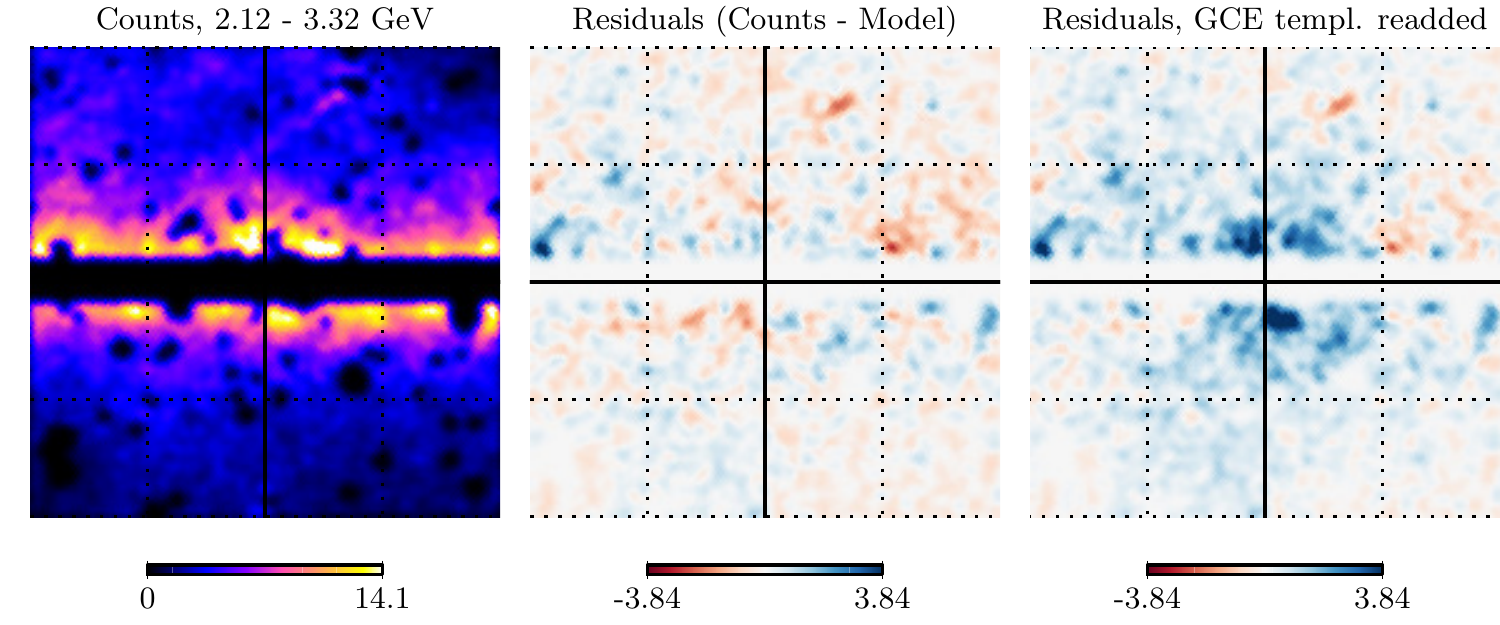}
        \includegraphics[width=0.93\linewidth]{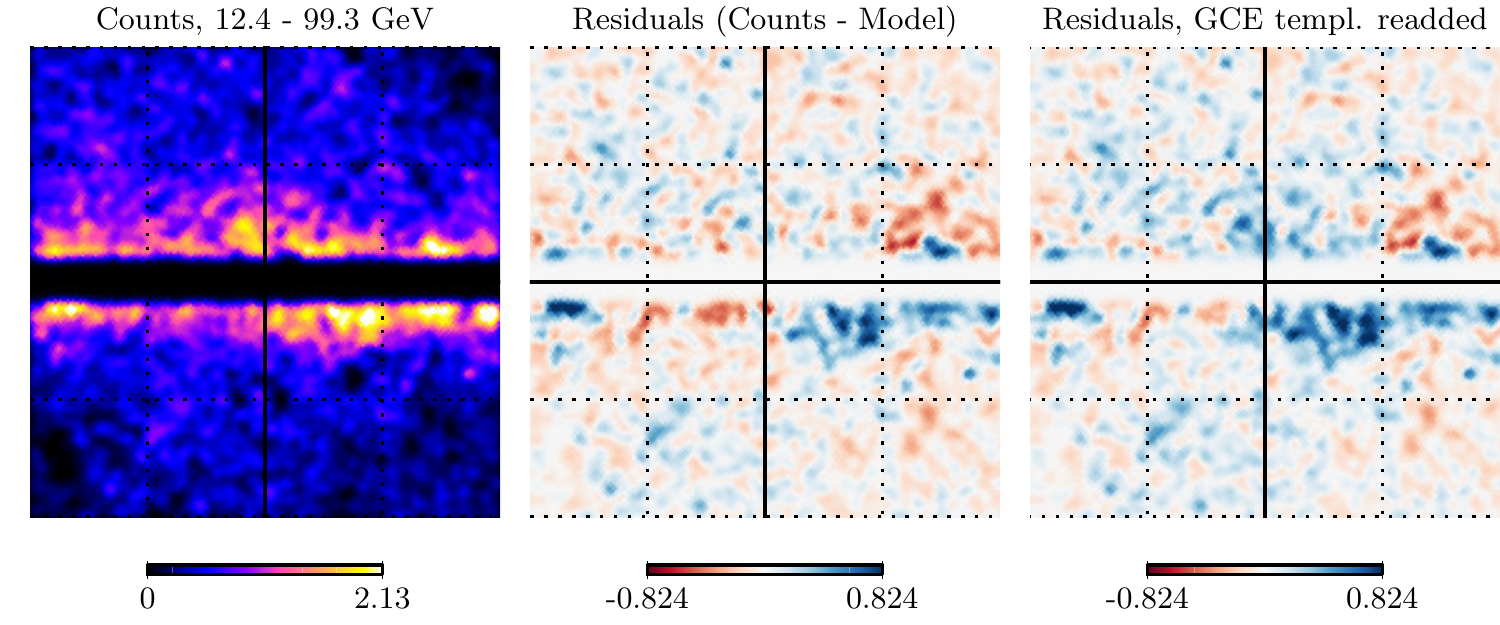}
    \end{center}
    \caption{\emph{Left panels:} Count maps at various energies (from
    \emph{top} to \emph{bottom}), with the disk cut $|b|>2^{\circ}$ and PSC
    mask applied.  \emph{Central panels:} Residuals after subtracting our
    self-consistent GDE model A.  \emph{Right panels:} Residuals after
    subtracting our self-consistent GDE model, but re-adding the GCE template
    associated to the model.  A Gaussian smoothing with $\sigma=0.4^\circ$ is
    applied to all plots.}
    \label{fig:residuals}
\end{figure}

In figure~\ref{fig:profiles}, we show the latitude (top left panel) and
longitude (remaining panels) profiles of the individual diffuse and PSC
components, and their sum compared to the actually measured fluxes for model A.
Data and model are in general in very good agreement.  We remind that, although
the contributions from point sources are shown in these plots for completeness,
the corresponding regions are actually masked during the fit.  However, we find
that even in the re-added masked regions the agreement between model and data
is reasonably good, except for regions that contain very bright sources.

One of the most striking, but also most critical aspect of the GCE spectrum is
the steep rise at sub-GeV energies.  It is instructive to see how the
predictions at low energies would change \emph{in absence} of such a rise.  To
this end, we show in the bottom left panel of figure~\ref{fig:profiles} the GCE
template and total emission for a GCE spectrum that is extrapolated down from
its value at 2 GeV with a spectral slope of -2.  In that case, the longitudinal
profile of the gamma-ray emission would be significantly enhanced close to the
GC, which is not observed.  Note that we selected for this plot the latitude
range at which this effect is best visible by eye.  In the right bottom panel
of figure~\ref{fig:profiles}, we show for completeness the flux that would be
predicted in absence of a GCE emission at energies around 3 GeV.

\medskip

In figure~\ref{fig:residuals}, we show exemplary count and residual maps for
model A.  The left panel of each row shows a map of photon counts in our
baseline ROI adopting also the PSC mask as described in
section~\ref{sec:analysis}.  The other panels show the total residuals (data -
model counts) when the emission associated with all templates is subtracted
from data (central panel) or when all templates but the GCE template are
subtracted (right panel). The comparison of the residuals with and without GCE
template gives a rough but useful idea about the intensity of the emission
associated with the GCE template and its spatial extension.  

In general, we find correlated residuals that are well above the level expected
from Poisson noise: about 10\% at  $E<1\GeV$ and about 25\% for  $E\sim3\GeV$
(including the GCE contribution).  The emission associated with the GC excess
is, after other components are subtracted, the most pronounced large-scale
excess in our ROI (and, as a matter of fact, in the entire Galactic plane with
$|\ell| \leq 70^\circ$).  However, we will discuss in the next subsection that
excesses of similar size are observed in other regions along the Galactic disk,
and we will characterize their properties and implications for the
interpretation of the GCE.

\bigskip

The residual plots in figure~\ref{fig:residuals} illustrates that it is hard to
construct an a priori model of the GDE that is in agreement with the
observations at the level of Poisson noise.  In fact, for our best-fit GDE
model F, the reduced $\chi^2$ in the energy range from 500 MeV to 3.31
GeV\footnote{At higher energies the low number of photons prohibits a simple
goodness-of-fit analysis.} is around $\chi^2/\text{dof}\simeq 295000/267000
\simeq 1.10$.  Although this value is close to one, thanks to the large number
of degrees of freedom the corresponding $p$-value is utterly small and around
$10^{-300}$.  Given the high quality and statistics of \Fermi-LAT data, and the
still rather rudimentary treatment of the GDE, this is hardly surprising.  

In any case, the comparison of $TS$ values remains an important tool for
selecting GDE models that provide gradual improvements when fitting the data.
However, the extremely small $p$-values that one obtains when fitting the data
suggest that it is mandatory to study the typical uncertainties of the GDE
modeling in light of the data before drawing strong conclusions from purely
statistical fits.  This is what we will do in the next subsection.

\subsection{Empirical model systematics}
\label{sec:ModelingUncertainties}

\begin{figure}
    \begin{center}
        \includegraphics[width=1.0\linewidth]{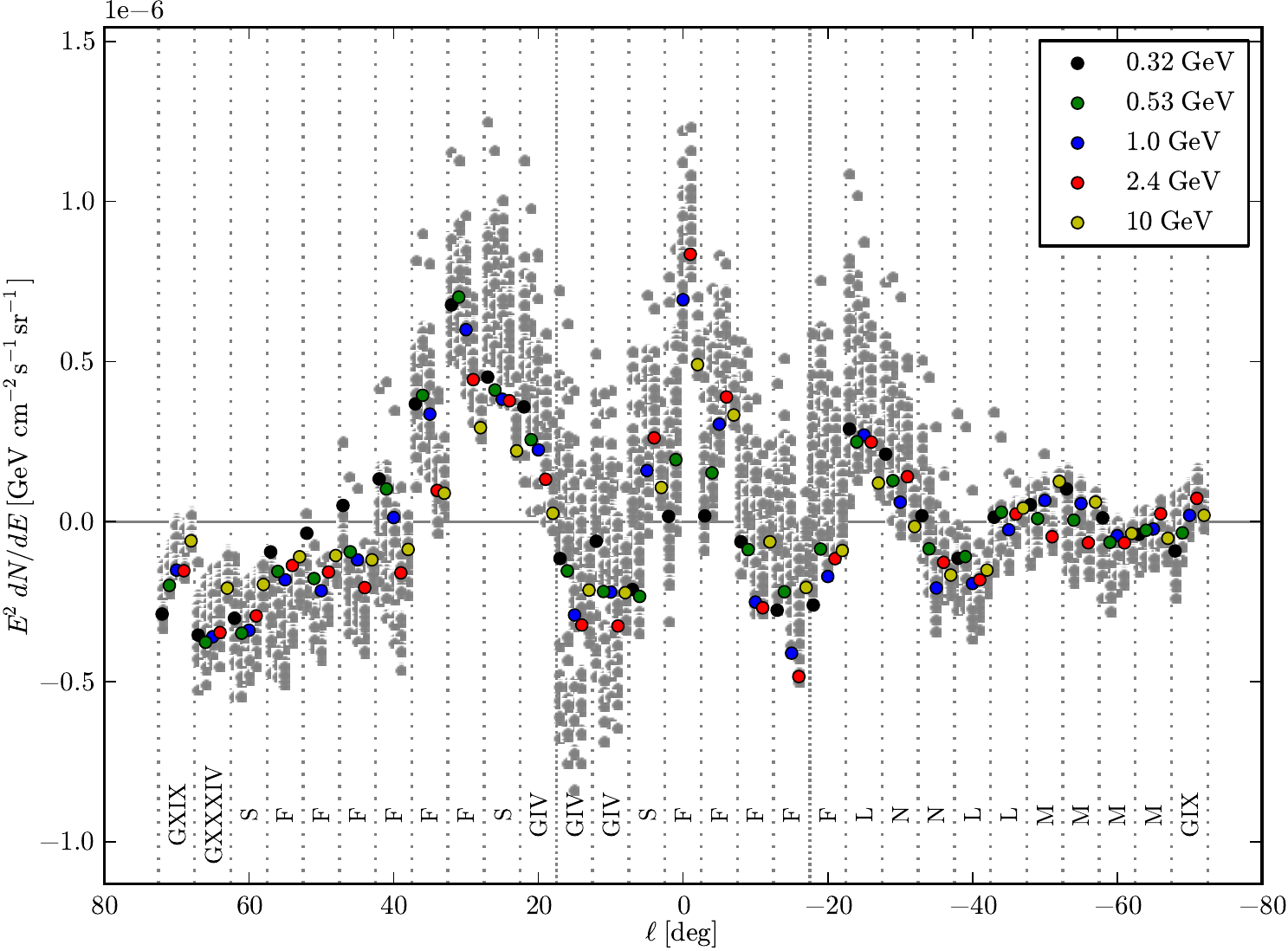}
    \end{center}
    \caption{Flux absorbed by the GCE template when moving it, as well as the
    ROI, along the Galactic disk in steps of $\Delta\ell=\pm 5^\circ$, for five
    different reference energies.  The \emph{colored dots} indicate the flux
    for the GDE model that gives {\it locally} the best-fit (these models are
    listed in the bottom of the plot), whereas the \emph{gray dots} indicate
    the fluxes for all other models.  The excess observed at the GC is -- at
    around 1--3 GeV -- clearly the largest in the considered region, although
    other excesses exist as well (see text for a discussion).  Regions with
    $|\ell| \gtrsim 20^\circ$ (indicated by the \emph{vertical dotted lines})
    will be used as test regions for estimates of the empirical model
    uncertainties of the adopted GDE models.}
    \label{fig:sliding}
\end{figure}

The modeling of the GDE in the present analysis is entirely based on the
numerical code \texttt{Galprop}.  The agreement between the GDE modeling and
actual data in the inner Galaxy is quite satisfactory, with typical residuals
that are significantly smaller than the GCE (see figures~\ref{fig:profiles}
and~\ref{fig:residuals}).  However, in order to increase the confidence in
these results and to study the robustness of the inferred GCE spectrum, we will
estimate typical residuals above the \texttt{Galprop} predicted GDE by
analyzing the diffuse emission from the Galactic disk, away from the GC, in a
systematic way.  

As discussed above, a significant part of observed the gamma-ray emission
towards the inner Galaxy is actually produced \emph{locally}, with typical
distances of a few kpc along the line-of-sight (\cf figure~\ref{fig:los}).  At
first approximation one can hence expect that the accuracy of GDE modeling, at
least at the $|b|>2^\circ$ latitudes we are interested in, is of a similar
level along the entire inner part of the Galactic disk, including the region
close to the GC.  We will here consider variations in the longitude range
$|\ell| < 90^\circ$, and use them to estimate the model systematics at the GC.

\subsubsection{Analysis of test regions along the Galactic disk}

We consider a number of ROIs along the Galactic disk, away from the GC, and
measure how much of the observed gamma-ray emission is absorbed in the
transposed GCE template centered in that ROI.  Namely, we consider gamma-ray
data in the region $|\ell|\leq 90^\circ$ and $|b|\leq 20^\circ$, in 29
overlapping ROIs defined by $-20^\circ + k\cdot 5^\circ \leq \ell \leq 20^\circ
+ k\cdot 5^\circ$ and $2^\circ\leq|b|\leq 20^\circ$, with $k = -14, -13, \dots,
14$. The GCE emission template moves along and is centered at $\ell = k\cdot
5^\circ$.  For all of our 60 GDE models and all of the ROIs we perform a
template analysis as above for the GC.

\medskip

The excess emission that we find along the Galactic disk is summarized in
figure~\ref{fig:sliding} in a rather compact way.  For all 60 GDE models we
show as gray dots the flux that is absorbed by the (transposed) GCE template,
as function of the template center in longitude steps of
$\Delta\ell=\pm5^\circ$, for five different reference energy bins centered at
0.32, 0.53, 1.0, 2.4, and 10 GeV.  For each longitude step, we determine the
\emph{locally} best-fit-model in that ROI.  We then highlight by colored points
the fluxes measured for that model at each of the five different energies.
This allows to read off a rudimentary estimate of the excess spectrum in each
region of the Galactic disk.

As can be seen in figure~\ref{fig:sliding}, we clearly reproduce the pronounced
excess at the GC ($\ell=0^\circ$), with a peak in the spectrum at energies
around 2.4 GeV (see the trend of the colored points at $\ell=0^\circ$).  At
these energies, the GC excess is the most pronounced excess in the entire test
region.  However, at Galactic longitudes around $\ell\sim\pm25^\circ$, we
observe residuals with almost identical size.  Further away from the GC, at
$|\ell|\geq40^\circ$, residuals are mostly consistent with zero, though
sometimes biased towards negative values.

\bigskip

The excesses along the Galactic disk might be on first sight discouraging,
since they show that uncertainties of the GDE as we model it in the present
analysis are almost of the same magnitude as the GCE itself.  This brings up
the question whether any reliable conclusions about the morphology, spectrum
and distinctiveness of the GCE can be drawn at all.

From figure \ref{fig:sliding} we find a number of differences between the GCE
and the excess emission away from the GC at $\ell\sim\pm25^\circ$.  The most
notable one is that the emission at GC has a different spectrum, with a peak at
energies around $2\rm\ GeV$, which is not present in any of the regions away
from the GC.  Furthermore, the GCE is strongly peaked at the GC, and falls off
rapidly as function of longitude, whereas the excesses away from the GC have a
smoother dependence on $\ell$.  Already this suggests that the GCE and excesses
away from the GC are of different physical origin.

At least one of the excesses away from the GC, at $\ell\sim 25^\circ$, appears
to be associated with a known structure, the Aquila Rift region, which is a
molecular cloud complex that is well identified in CO~\cite{Dame:2000sp}, and
has an enhanced star formation rate~\cite{Levshakov:2013dia}.  With a distance
from the Sun of a few hundred pc~\cite{Knude:2011js}, it is the closest
molecular complex towards the inner Galactic region.  Indeed, the observed
excess spectrum appears to be a featureless power-law with a spectral index of
very roughly $2.3$, compatible with the typical expectations for star forming
regions~\cite{Ackermann:2012vca}.  More generally, longitudes around
$\ell\sim\pm25^\circ$ (with a projected distance of at least $3.6\rm\ kpc$ from
the GC) coincide with the molecular ring with a high projected star formation
along the line-of-sight~\cite{Robitaille:2010xx}.

We tried to account for the observed excesses by allowing additional variations
in the HI and H2 maps or subsets thereof, however without much success.  The
\Fermi\ team encountered similar problems and used ad hoc templates to account
for the excess emission in models for the GDE.\footnote{See \texttt{P7V6}
discussion,
\url{http://fermi.gsfc.nasa.gov/ssc/data/access/lat/Model_details/Pass7_galactic.html}.}
In the present work, we will simply accept these excesses along the Galactic
disk as model uncertainties of state-of-the-art GDE models, and incorporate
them in our analysis below.

\begin{figure}
    \begin{center}
        \includegraphics{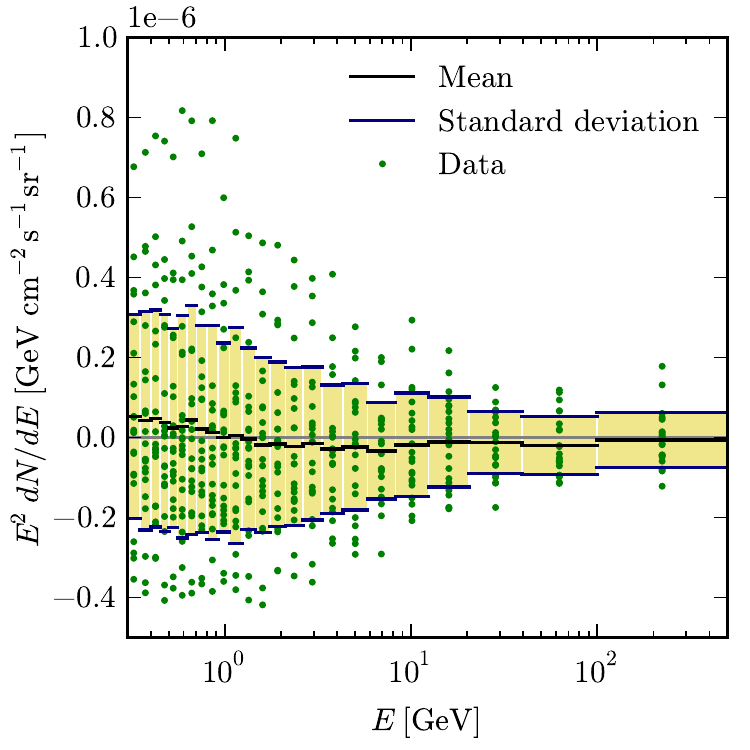}
        \includegraphics{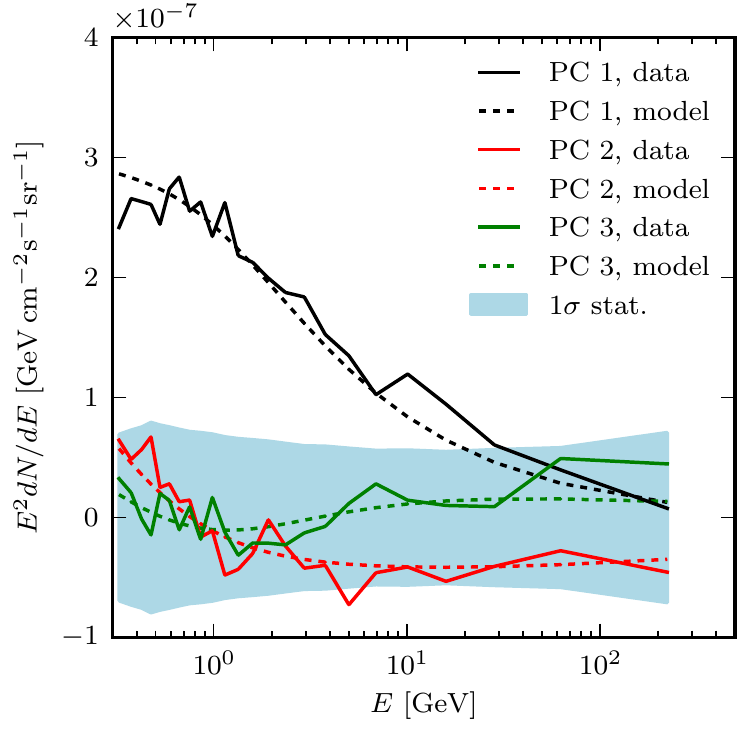}
    \end{center}
    \caption{\emph{Left panel:} Residuals absorbed by the transposed GCE
    template in 22 test regions along the Galactic disk (\emph{green points}),
    as shown in figure \ref{fig:sliding} by the colored dots;  the \emph{yellow
    boxes} indicate the mean and standard deviation.  \emph{Right panel:}
    Decomposition of the covariance matrix of the residuals in principal
    components.  We only show the three components with the largest standard
    deviation (\emph{solid lines}), and compare them to the statistical errors
    from the GCE fit at the GC (\emph{blue area}).  The \emph{dashed lines}
    show model predictions from a four parameter analytical model.  It provides
    a good fit and traces the observed variations back to uncertainties in the
    normalization and slope of the $\pi^0$ and ICS components (see
    appendix.~\ref{sec:systModeling} for details about the model).}
    \label{fig:variance}
\end{figure}

\subsubsection{The covariance matrix of empirical model systematics}

We will make use of the observed residuals in figure~\ref{fig:sliding} as an
estimate for empirical model systematics along the Galactic disk.  To this end,
we show in the left panel of figure~\ref{fig:variance} the flux that is
absorbed by the transposed GCE templates in different energy bins.  Each dot
represents one of the 22 test ROIs in the longitude range $20^\circ\leq |\ell|
\leq 70^\circ$.  We also indicate the mean values, which remain close to zero,
and the standard deviations.

\medskip
   
The observed variations along the disk are correlated in energy.  We quantify
this by analyzing the covariance matrix of the fluctuations, which is given by
\begin{equation}
    \Sigma_{ij,\rm\,mod} = 
    \left\langle \frac{dN}{dE_i}\frac{dN}{dE_j} \right\rangle -
    \left\langle \frac{dN}{dE_i}\right\rangle\left\langle\frac{dN}{dE_j}
    \right\rangle\;,
    \label{eqn:SigmaEstimate}
\end{equation}
where the average runs over the 22 test ROIs, and $dN/dE_i$ denotes the
GCE-like residual flux in energy bin $i$ as shown in the left panel of
figure~\ref{fig:variance}.

\medskip

We analyse the properties of this covariance matrix in terms of its principal
components.  These are simply the eigenvectors of this matrix, normalized to
the corresponding variance.  The three principal components with the largest
standard deviation are shown in figure \ref{fig:variance}, compared with the
$\pm1\sigma$ statistical error of the GCE at the GC.  We find that only the
first principal component is at almost all energies significantly larger than
the statistical errors.  It follows very roughly a power-law with a spectra
slope of 2.2, which is reminiscent of ICS emission.

\smallskip

The origin of the observed empirical model systematics can be understand in
terms of a simple analytical model that we discuss in
appendix~\ref{sec:systModeling}.  It takes as four parameters only the
uncertainties in the normalization and slope of the $\pi^0$+Bremss and ICS
components.  Fitting these four parameters to the three largest principal
components of the covariance matrix gives rise to a \emph{modeled} covariance
matrix with principal components as shown by the dashed lines in the right
panel of figure~\ref{fig:variance}.  The agreement is rather satisfactory,
except at the very lowest energies below 600 MeV where the modeled first
principal component overshoots slightly the observed one.  We hence conclude
that the empirically derived model systematics can be understood in terms of
variations in the normalization and spectral slopes of the primary diffuse
background components.

\medskip

Below, we will use the empirical covariance matrix when performing fits to the
GCE spectrum instead of the analytical model.  However, in order to avoid a
double-counting of statistical errors, we will truncate the principal
components that enter the empirical covariance matrix and restrict them to the
first three.  We will refer to this truncated matrix as $\Sigma_{ij,\,
\text{mod}}^\text{trunc}$.

\begin{figure}
    \begin{center}
        \includegraphics{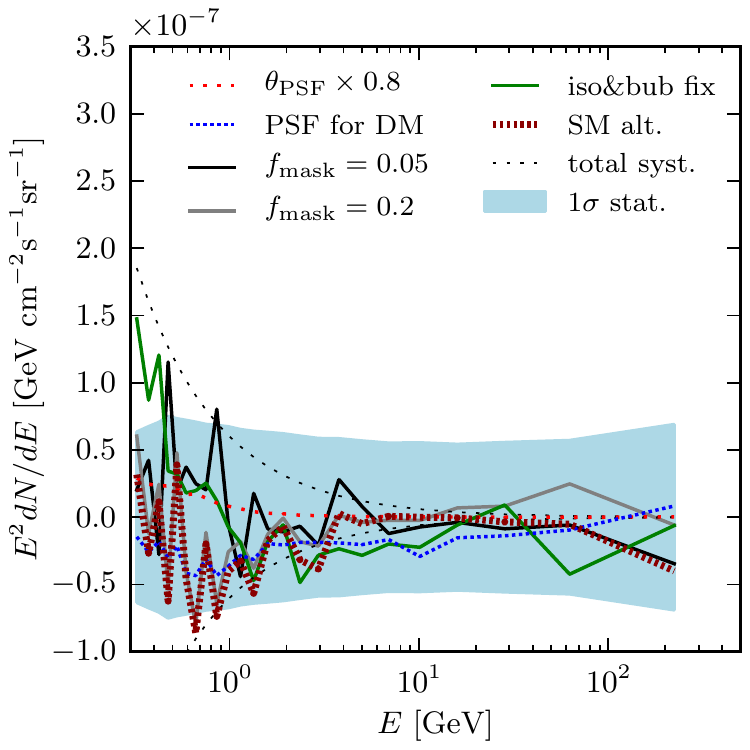}
    \end{center}
    \caption{Summary of different analysis systematics.  The impact of choices
    made in the PSC mask definition is shown by the \emph{solid black} and
    \emph{gray lines} (for variations in $f_\text{mask}$), and by the
    \emph{dotted dark red line} (for adopting an alternative GDE model in the
    mask definition).  We also show the impact of smoothing the GCE template
    with the \Fermi-LAT PSF (\emph{blue dotted line}), fixing the isotropic and
    bubbles normalization to their external constraints (\emph{green solid
    line}), rescaling the \Fermi-LAT PSF by a factor close to one (\emph{red
    dotted line}). The \emph{light blue region}) represents the statistical
    errors.  Results are shown for the reference model A, but should be rather
    similar for other GDE models.  The \emph{thin dotted line} shows our
    estimate for the overall analysis systematics, which at energies below 1
    GeV clearly exceed the statistical errors of the GCE.  See text for details
    on the implementation in spectral fits.}
    \label{fig:otherSyst}
\end{figure}

\subsubsection{Other systematics and the GCE spectrum}
\label{sec:otherSystematics}

Before showing the GCE spectrum with empirical model uncertainties, we
summarize further systematics that enter our analysis in figure
\ref{fig:otherSyst}. Namely, we display the impact on the flux absorbed in the
GCE template when {\it a}) decreasing the width of the PSF by a reference
factor of 0.8, {\it b}) including PSF smoothing also for the GCE template, {\it
c}) changing the definition of the PSC mask by varying $f_\text{mask}$ as
indicated or using model E instead of model P for the PSC mask definition, {\it
d}) fixing the flux of the IGRB and the \Fermi\  bubbles to their external
constraints rather then leaving them free to vary, and {\it e}) using a
different GDE model to define the PSC mask.  At energies above 1 GeV, all of
these variations are well below the statistical error and negligible.  However,
we take the variations below 1 GeV into account by modeling them as
$dN/dE^\text{res} =
6\times10^{-8}\rm\,GeV^{-1}\,cm^{-2}\,s^{-1}\,sr^{-1}\,(E/1\rm\,GeV)^{-3}$.  We
include this uncertainty in the final fit \emph{twice}: once as uncorrelated
error and once as fully correlated errors with a free normalization.  This
gives rise to a covariance matrix $\Sigma_{ij,\rm\,res} = dN/dE_i^\text{res}
dN/dE^\text{res}_j + \delta_{ij}dN/dE_i^\text{res} dN/dE_i^\text{res}$.

\medskip

\begin{figure}
    \begin{center}
        \includegraphics{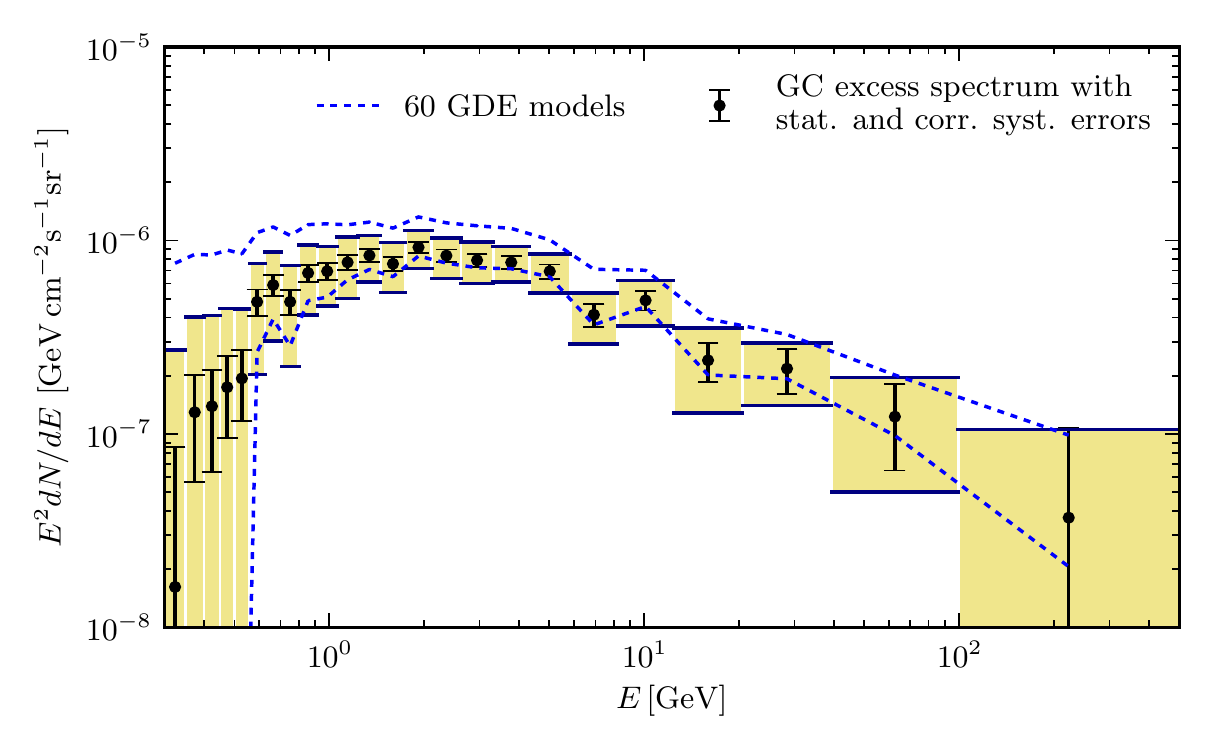}
    \end{center}
    \caption{Spectrum of the GCE emission for model F (\emph{black dots})
    together with statistical and systematical (\emph{yellow boxes}, \cf figure
    \ref{fig:variance}) errors. We also show the envelope of the GCE spectrum
    for all 60 GDE models (\emph{blue dashed line}, \cf figure
    \ref{fig:spectrum60models}).}
    \label{fig:spectrumSysStat}
\end{figure}

Finally, figure \ref{fig:spectrumSysStat} shows, as one of the main result of
this subsection, the spectrum of the GC excess emission for model F (\cf figure
\ref{fig:spectrum60models}), together with statistical errors and the (diagonal
part of the) empirical model systematics as inferred above (note that we omit
the sub-dominant method uncertainties $\Sigma_{ij,\rm\,res}$ in this plot).
Statistical and systematic errors are shown at $\pm1\sigma$.  For comparison,
we overlaid the envelope of the GCE spectra associated with the 60 GDE models
as shown of figure \ref{fig:spectrum60models}, to indicate the theoretical
model systematics.  We find that empirical and theoretical systematics are
roughly of the same order in the considered energy range and ROI.

\subsection{The morphology from ten sky segments}
\label{sec:morphology}

We will now set the stage for an investigation of the morphology of the GCE in
section~\ref{sec:parametric} below.  We are interested in studying \emph{a})
the symmetry of the excess emission around the GC, and, more importantly,
\emph{b}) how far from the disk the GCE extends.  To this end, we split the GCE
template in ten segments and repeat the analysis of the previous two
subsections.  Furthermore, we allow additional freedom in the ICS templates, as
we explain below.  We present additional morphological studies of the GCE,
which mostly reconfirm findings from previous works, in
appendix~\ref{app:properties}.

\medskip

\begin{figure}
    \begin{floatrow}
        \capbfigbox{%
            \includegraphics[width=2.3in]{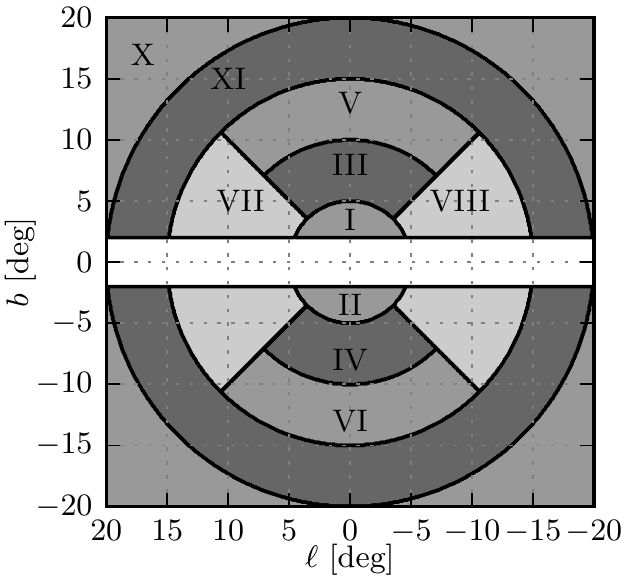}
        }{%
            \caption{Geometry of the ten GCE segments used in our morphology
            analysis, see table~\ref{tab:ROIdefinitions}.}
            \label{fig:ROIsplits}
        }
        \capbtabbox{%
            \footnotesize
            \begin{tabular}{ccc}
                \toprule
                \#ROI & Definition & $\Omega_\text{ROI}\rm\ [sr]$ \\
                \midrule
                I, II & $\sqrt{\ell^2 + b^2} < 5^\circ$, $\pm b>|\ell|$ & $6.0\times10^{-3}$ \\
                III, IV & $5^\circ<\sqrt{\ell^2 + b^2} < 10^\circ$, $\pm b>|\ell|$ & $1.78\times10^{-2}$ \\
                V, VI & $10^\circ<\sqrt{\ell^2 + b^2} < 15^\circ$, $\pm b>|\ell|$ & $2.93\times10^{-2}$\\
                VII, VIII & $5^\circ<\sqrt{\ell^2 + b^2} < 15^\circ$, $\pm\ell>|b|$ & $3.54\times10^{-2}$ \\
                IX & $15^\circ<\sqrt{\ell^2 + b^2} < 20^\circ$ & $1.51\times10^{-1}$ \\
                X & $20^\circ<\sqrt{\ell^2 + b^2}$ & $1.01\times10^{-1}$ \\
                \bottomrule
                \vspace{0.10in}
            \end{tabular}
        }{%
            \caption{Definition of the ten GCE segments that are shown in
            figure~\ref{fig:ROIsplits}, as function of Galactic latitude $b$
            and longitude $\ell$, together with their angular size
            $\Omega_\text{ROI}$.}
            \label{tab:ROIdefinitions}
        }
    \end{floatrow}
\end{figure}

We divide the GCE template within our main ROI, see eq.~\eqref{eqn:ROI}, into
\textit{ten GCE segments} as shown in figure \ref{fig:ROIsplits} and defined in
table~\ref{tab:ROIdefinitions}.  Each of the ten segments is zero outside of
its boundaries, and equals the standard GCE template (generalized NFW with
$\gamma=1.2$) inside its boundaries.  The normalization of each of the ten
templates is allowed to float freely in the fit.  The definition of the
segments aims at studying the symmetries of the GCE around the GC: Allowing
regions in the North (I, III, and V) and South (II, IV, and VI) hemisphere, as
well as in the West (VII) and East (VIII) ones, to vary independently, we can
test the spectrum absorbed by the GCE template in the different regions of the
sky.  Moreover, with the same segments, we can investigate its the extension in
latitude.

\smallskip

To facilitate the study of morphological properties of the excess, we
furthermore allow additional latitudinal variations in the ICS components of
the individual GDE models.  We split our ICS component into \textit{nine ICS
segments}, corresponding to 9 latitude strips with boundaries at $|b| =
2.0^\circ$, $2.6^\circ$, $3.3^\circ$, $4.3^\circ$, $5.6^\circ$, $7.2^\circ$,
$9.3^\circ$, $12.0^\circ$, $15.5^\circ$ and $20^\circ$.  We then allow the
normalization of the ICS strips to vary independently, though we keep the
normalization of strips that are symmetric under $b\to-b$ bound to the same
value; this gives nine free parameters for the ICS emission.\footnote{We note
that the spectral shape of the GC excess, as well as the envelope discussed in
section~\ref{sec:spectrum} and shown in figure \ref{fig:spectrum60models},
remains basically the same when applying this additional degrees of freedom in
these fits. See further discussion in appendix \ref{app:properties}.} The
advantage of these additional degrees of freedom is that they allow to mitigate
the effects of over- or under-subtraction of data at different latitudes.  The
GCE component has a significant correlation with the ICS emission.  An unbiased
treatment of the ICS is hence important.  Physically, these variability makes
sense since it accounts in a simple but efficient way for uncertainties in the
CR electron density and the ISRF at different latitudes along the
line-of-sight.  Practically, it reduces the scatter between the 60 GDE models
and removes outliers, without much affecting results for the best-fit models.

\bigskip

\begin{figure}
    \begin{center}
        \includegraphics[width=0.92\linewidth]{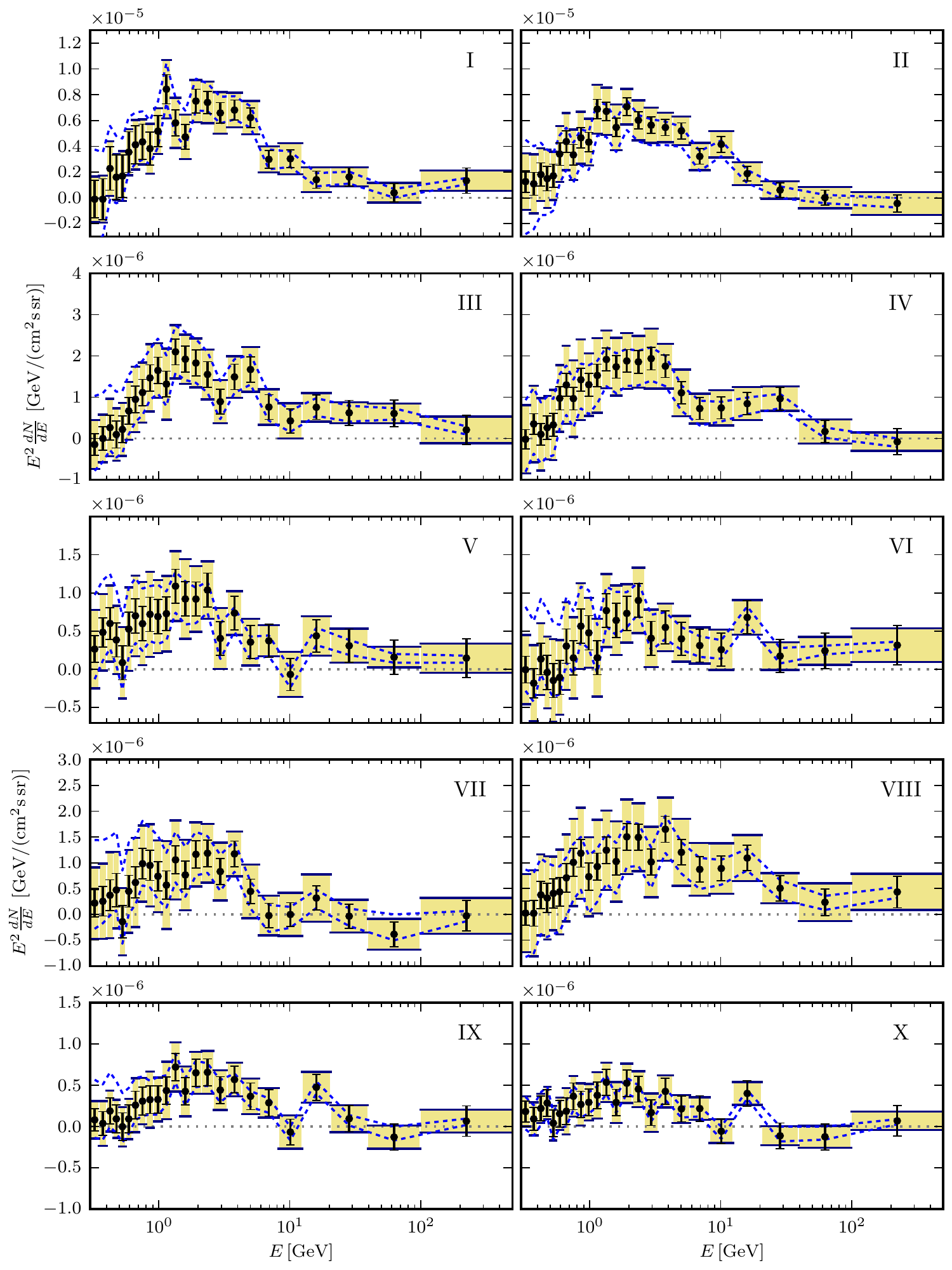}
    \end{center}
    \caption{Same as figure \ref{fig:spectrumSysStat}, but from a fit with the
    segmented GCE template as illustrated in figure \ref{fig:ROIsplits}.  We
    show results for GDE model F (\emph{black dots}), as well as the envelope
    for all 60 GDE models (\emph{blue dotted lines}) and the systematic errors
    that we derived from fits in 22 test regions along the Galactic disk
    (\emph{yellow boxes}, in analogy to figure \ref{fig:variance}).  See figure
    \ref{fig:DMsplitBG} below for the spectra of all components.}
    \label{fig:DMsplitSpectraSyst}
\end{figure}

With the above setup, namely ten GCE segments and nine ICS templates plus all
remaining components from table~\ref{tab:templates}, we analyze the \Fermi-LAT
data within our main ROI as well as the test regions along the Galactic disk
for all of our 60 GDE models (for the test regions we actually use only one
combined ICS template for efficiency, which might lead to a slight
overestimation of the empirical model systematics).  In analogy to what we did
above, we estimate the variance in each of the ten GCE template segments from
fits in ROIs that are centered at $20^\circ\leq|\ell|\leq 70^\circ$ and
$b=0^\circ$.  

\smallskip

The results for the extracted GCE template fluxes are shown in figure
\ref{fig:DMsplitSpectraSyst}.  The error bars correspond to the model that
yields the best $TS$ value (which is still model F), but we also indicate the
envelope from all models, as well as the model systematics that we found from
the Galactic disk analysis (\cf figure~\ref{fig:spectrumSysStat}).  In the
regions I--IV, the GC excess is clearly visible with a peak at energies around
1--3 GeV, and with a drop at energies above and below.  Similar excesses are
observed in the other template segments, with spectra that suggest
compatibility with what is observed in I--IV.  Such a compatibility of spectra
of the different segmented templates indicates that there is no clear asymmetry
with respect to the GC---neither in the North vs South hemispheres, nor in the
West vs East hemispheres.  We present a quantitative discussion about the
actual extension of the excess to higher latitudes and the compatibility of
spectra in the different sky regions in section~\ref{sec:parametric}.

\bigskip

We end with a few words about the covariance matrix of empirical model
uncertainties that we find for the segmented GCE template.  We analyze the
principal components of this matrix for each of the ten segments independently.
In general, we find that the observed fluctuations are significantly larger
than the ones that would be predicted from the analytical model in
appendix~\ref{sec:systModeling} when using the same model parameters that we
found in the previous subsection for the single GCE template.  This suggests
that there are, not surprisingly, variations in the backgrounds of the
individual GCE segments that average out when considering the single GCE
template.  We take these additional variations automatically into account in
the subsequent analysis by making use of the (truncated) covariance matrix in
our spectral and morphological fits.  However, we will neglect here possible
segment-to-segment correlations.


\section{Parametric fits to the Galactic center excess}
\label{sec:parametric}

Equipped with information about the GCE spectrum and morphology, and with
estimates for the theoretical and empirical model uncertainties, we perform in
this section a number of parametric fits to the data.

In the previous section, we found that theoretical and empirical model
uncertainties affect the GCE spectrum at a similar level (see
figure~\ref{fig:spectrumSysStat}).  However, theoretical model uncertainties in
the way we discussed them here are difficult to interpret in a purely
statistical sense, since the $TS$ values that we find for fits with our 60 GDE
models differ typically by $>\mathcal{O}(100)$ values (see
appendix~\ref{app:60models}), and even our best-fit model for the GDE gives
formally a poor fit to the data.  This is a generic problem of modeling the
GDE~\cite{FermiLAT:2012aa}, as we discussed at the end of
section~\ref{sec:spectrum}.  On the other hand, the empirical model
uncertainties are simple to interpret statistically and give by construction a
realistic account for typical systematics of state-of-the-art GDE modeling.

\medskip

We will hence adopt the following \emph{strategy}:  We will use the GCE
spectrum and associated statistical errors from model F only, which gives
formally the best-fit to the \Fermi-LAT data in our ROI.  In fits to the GCE
spectrum we then only consider the \emph{empirical} model systematics, and
neglect the theoretical ones.  Given the small scatter for the GCE spectrum
that we find for different GDE models, this is well justified.  We checked
explicitly that using different GDE model as starting point in the spectral
fits would not alter our results significantly (see
appendix~\ref{sec:otherModeling}).  Hence, we consider our approach as
statistically sound and sufficiently robust to derive meaningful results.

\medskip

We will introduce general aspects of fits with correlated errors in
subsection~\ref{sec:correlatedFits}, and then test the most common
interpretations of the GCE emission in terms of a number of DM and
astrophysical toy models in subsection~\ref{sec:DMfits}
and~\ref{sec:AstroFits}.

\begin{figure}[t]
    \begin{center}
        \includegraphics{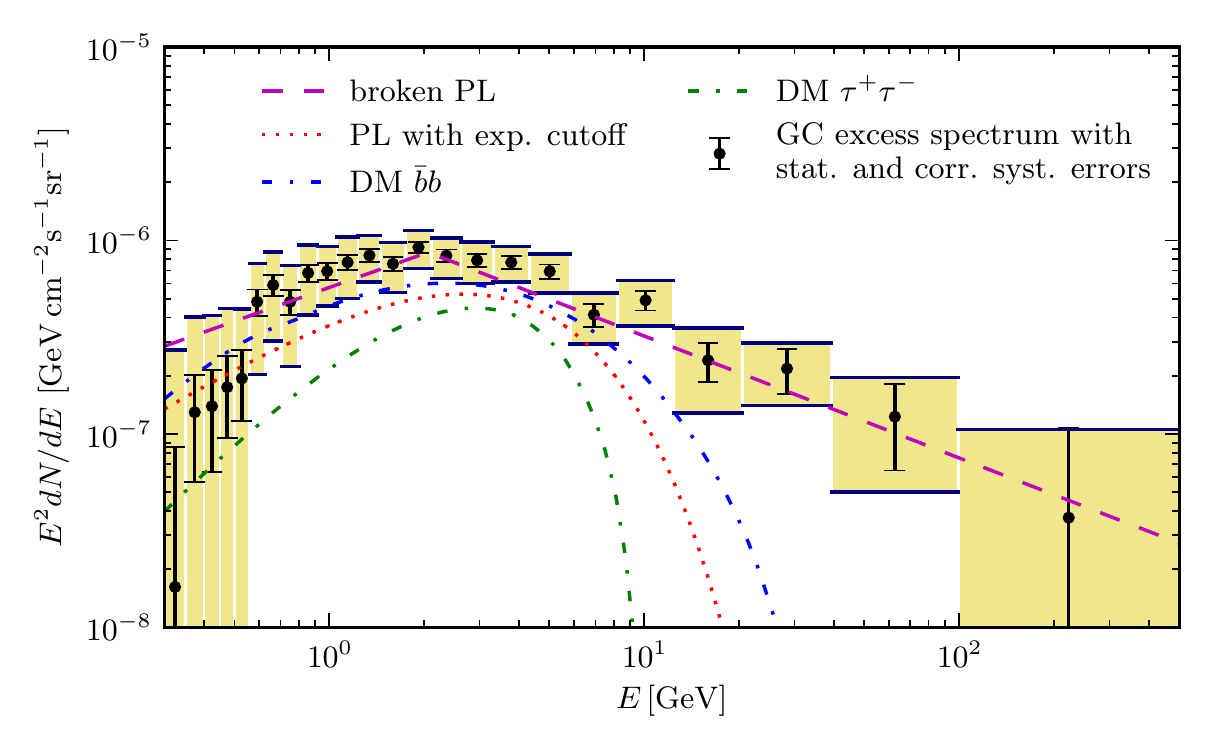}
    \end{center}
    \caption{Spectrum of the GCE emission, together with statistical and
    systematical errors, for model F (\cf figure \ref{fig:spectrumSysStat}).
    We show fits to the GCE with various spectral models.  We emphasize that
    the shown systematic errors are \emph{correlated}, and that the spectral
    models actually \emph{do} provide a good fit to the data in most cases.  We
    show the best-fit model parameters, along with indicators for the fit
    quality, in table~\ref{tab:fitResults} (\cf figures~\ref{fig:DMfits1}
    and~\ref{fig:AstroFits}).  See text for details on the fitting procedure.}
    \label{fig:spectrumSysStat_fits}
\end{figure}

\subsection{Spectral fits with correlated errors}
\label{sec:correlatedFits}

For spectral fits to the extracted GCE spectra (as they are shown in
figures~\ref{fig:spectrumSysStat} and~\ref{fig:DMsplitSpectraSyst}), we make
use of a $\chi^2$ function with a non-diagonal covariance matrix.  This allows
to take into account the correlated empirical model systematics that we derived
in the previous section.\footnote{It is worth pointing out that summing
systematic and statistical errors in quadrature, which is common practice in
the DM-phenomenology literature, does \emph{not} lead to the weakest (or `most
conservative') constraints on model parameters in almost all of the cases.}

The $\chi^2$ function is given by
\begin{equation}
    \chi^2 = \sum_{ij} 
    \left(\frac{d\bar N}{dE_i}(\boldsymbol\theta)-\frac{dN}{dE_i} \right) 
    \Sigma^{-1}_{ij}
    \left(\frac{d\bar N}{dE_j}(\boldsymbol\theta)-\frac{dN}{dE_j} \right) \;,
\end{equation}
with the covariance matrix
\begin{equation}
    \Sigma_{ij} = (\sigma^\text{stat.}_i)^2 \delta_{ij} +
    \Sigma_{ij,\rm\,mod}^\text{trunc} + \Sigma_{ij,\rm\,res}\;.
\end{equation}
Here, $dN/dE_i$ ($d\bar N/dE_i$) denotes the measured (predicted) GCE flux in
the $i^\text{th}$ energy bin, $\boldsymbol\theta$ the model parameters,
$\sigma_i^\text{stat.}$ the corresponding statistical error,
$\Sigma_{ij,\rm\,mod}^\text{trunc}$ the truncated $(24\times24)$ covariance
matrix accounting for empirical model systematics, and $\Sigma_{ij,\rm\,res}$
the residual systematics at sub-GeV energies that we discussed in
subsection~\ref{sec:otherSystematics}.  For fits to the \emph{segmented} GCE
template fluxes, the corresponding $(240\times240)$ correlation matrix is taken
to be block diagonal in the different GCE segments (we neglect
segment-to-segment correlations), and we set $\Sigma_{ij,\rm\,res}=0$, as it is
not very relevant for morphology fits.

\medskip

Like above, all fits are performed using the minimizer \texttt{Minuit}.  For
the two-dimensional contour plots, we define the one, two and three sigma
contours (which we show in the plots if not otherwise stated) at $\Delta\chi^2=
2.3$, 6.2 and, 11.8, and derive them with the \texttt{minos} algorithm.  Note
that we will neglect the effects of the finite energy resolution of \Fermi-LAT,
which is below 15\% in the energy range of interest, but could be easily
incorporated.

\subsection{Dark Matter models}
\label{sec:DMfits}

The most exciting interpretation of the GCE is that it is caused by the
annihilation of DM particles, and indeed all of the previous studies analyzing
\Fermi-LAT data focus on this possibility~\cite{Goodenough:2009gk,
Hooper:2010mq, Boyarsky:2010dr, Hooper:2011ti, Abazajian:2012pn,
Macias:2013vya, Abazajian:2014fta, Daylan:2014rsa}.  Instead of presenting fits
to a large number of DM annihilation spectra, we will here simply concentrate
on the most common cases discussed in the literature.  We concentrate on the
hadronic annihilation channels $\bar{b}b$ and $\bar{c}c$ and on pure
$\tau^+\tau^-$ lepton final states.  The gamma-ray yields are taken from
\texttt{DarkSUSY 5.1.1}~\cite{Gondolo:2004sc}.  

\medskip

\begin{figure}[t]
    \begin{center}
        \includegraphics{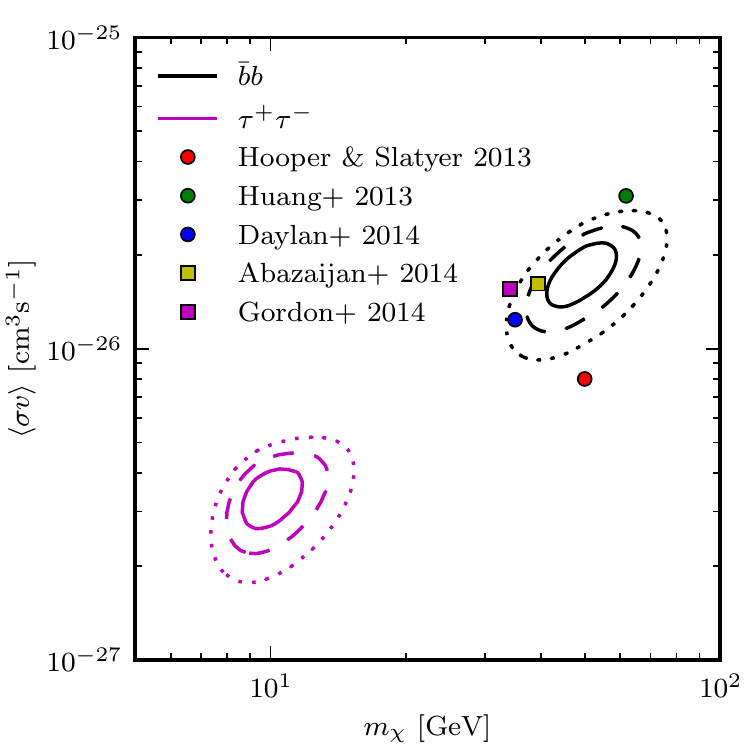}
        \includegraphics{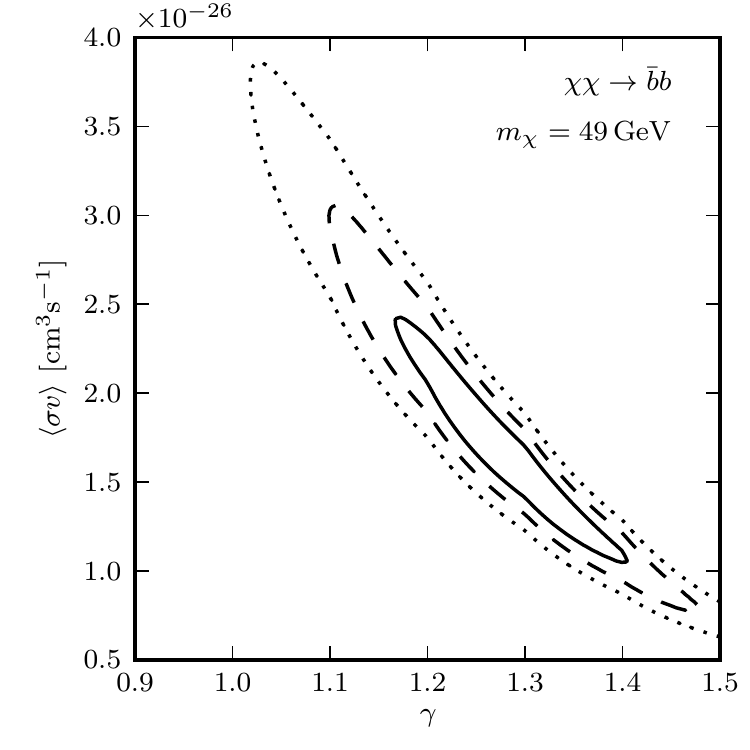}
    \end{center}
    \caption{ \emph{Left panel:} Constraints on the $\langle \sigma
    v\rangle$-vs-$m_\chi$ plane for three different DM annihilation channels,
    from a fit to the spectrum shown in figure~\ref{fig:spectrumSysStat} (\cf
    table~\ref{tab:fitResults}). \emph{Colored points} (\emph{squares}) refer
    to best-fit values from previous Inner Galaxy (Galactic center) analyses
    (see discussion in section~\ref{sec:DMdiscussion}).  \emph{Right panel:}
    Constraints on the $\langle\sigma v\rangle$-vs-$\gamma$ plane, based on the
    fits with the ten GCE segments.}
    \label{fig:DMfits1}
\end{figure}

\begin{figure}
    \begin{center}
        \includegraphics{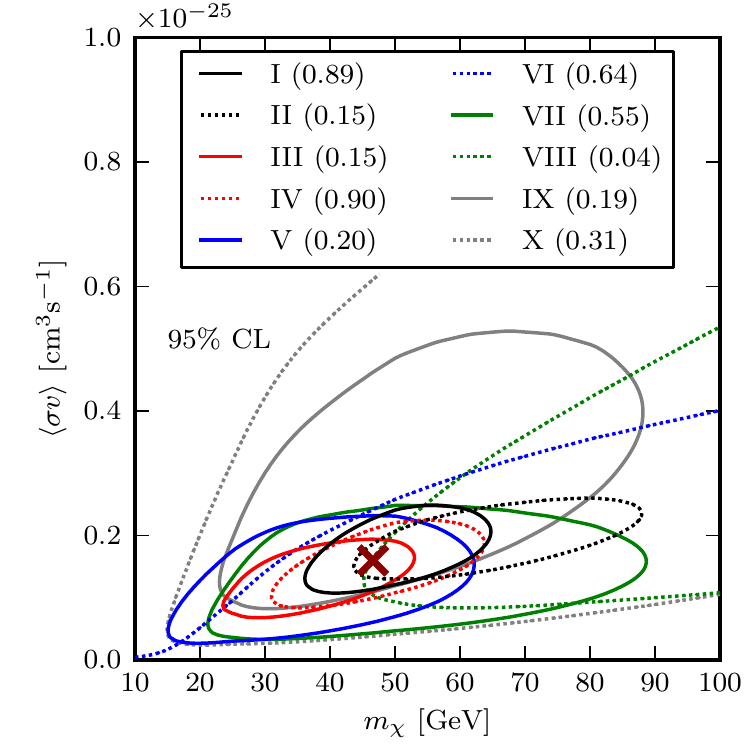}
    \end{center}
         \caption{Constraints on the $\langle \sigma v\rangle$-vs-$m_\chi$
         plane at 95\% CL, individually for the GCE template segments shown in
         figure \ref{fig:ROIsplits}, for the channel $\chi\chi\to\bar{b}b$.
         The cross indicates the best-fit value from a fit to all regions
         simultaneously ($m_\chi\simeq46.6\GeV$, $\langle\sigma
         v\rangle\simeq1.60\times10^{-26}\cm^3\s^{-1}$).  Note that we assume a
         NFW profile with an inner slope of $\gamma=1.28$.  The individual
         $p$-values are shown in the figure legend; the combined $p$-value is
         0.11.}
    \label{fig:DMfits2}
\end{figure}

In the left panel of figure \ref{fig:DMfits1} we show the constraints in the
$\langle\sigma v\rangle$-vs-$m_\chi$ plane that we obtain from a fit to the GCE
spectrum in figure \ref{fig:spectrumSysStat}.  Correlated model systematics are
taken into account as discussed above.  We find that both $\bar{b}b$ and
$\bar{c}c$ provide rather good fits to the data, with $p$-values around
0.4--0.5 (see table~\ref{tab:fitResults}).  For $\tau^+\tau^-$ final states,
the $p$-value is with 0.065 significantly lower, though it remains marginally
compatible with the data at $95\%$ CL.  We find that in the canonical case of
$\bar{b}b$ final states, DM masses around $m_\chi=49^{+6.4}_{-5.4} \rm\ GeV$
are favored by the data, and an annihilation cross-section of $\langle\sigma
v\rangle = 1.76^{+0.28}_{-0.27}\times10^{-26}\cm^3\s^{-1}$.\footnote{We remind
that we adopt a local DM density of $\rho_\odot = 0.4\rm\,GeV\,cm^{-3}$.}

\smallskip

In the right panel of figure \ref{fig:DMfits1} we show how the cross-section
$\langle\sigma v\rangle$ depends on the slope $\gamma$ of the adopted NFW
profile, for the case of annihilation into $\bar{b}b$ final states with the DM
mass fixed at 49\GeV.  This plot is based on the fluxes from the segmented GCE
template, see figure~\ref{fig:DMsplitSpectraSyst}.  As expected, the
cross-section is strongly correlated with the profile slope.  We find best-fit
values of $\gamma=1.28^{+0.8}_{-0.7}$ and $\langle\sigma
v\rangle=1.65^{+0.5}_{-0.4}\times10^{-26}\cm^3\s^{-1}$.

Note that in the case of $\gamma=1.2$, we obtain actually a somewhat larger
cross-section than in the left panel, though still marginally consistent within
one sigma. By using different values of $\gamma$ in the template analysis that
lead to figure~\ref{fig:spectrumSysStat}, we checked that the
$\gamma$-dependence of the cross-section shown in the left panel of
figure~\ref{fig:DMfits1} is practically identical to the one shown in the right
panel.

The constraints on the cross-section that we show in the right panel of
figure~\ref{fig:DMfits1} are, for a fixed value of $\gamma$, somewhat tighter
than in the left panel.  This is likely related to the neglect of
segment-to-segment correlations when determining the empirical model
systematics, which then tend to average out in fits to the spectrum.  This is a
caveat of spectral fits based on the segmented GCE templates.  For that reason,
we recommend using the spectral constraints derived from the non-segmented GCE
template instead, as shown in figure~\ref{fig:spectrumSysStat} and used in the
left panel of figure~\ref{fig:DMfits1}.

\medskip

In figure \ref{fig:DMfits2}, we show constraints at $95\%$ CL on the
$\langle\sigma v\rangle$-vs-$m_\chi$ plane that are obtained from fits to the
spectra from the individual GCE segments shown in
figure~\ref{fig:DMsplitSpectraSyst}.  We consider for definiteness only the
case of annihilation $\bar{b}b$ final states, and assume our contracted NFW
profile with an inner slope of $\gamma=1.28$, which gives the best fit in this
case.  Again, we fully take into account the empirical model systematics as
discussed above.  We find that the resulting constraints on the GCE spectrum
that we derive from the individual GCE segments are in mutual good agreement to
within $95\%$ CL.  There is no obvious bias for the GCE in the north/south or
east/west direction, though at higher latitudes the preferred DM masses are
slightly higher.  

The constraints in figure~\ref{fig:DMfits2} are also in good agreement with the
central value that we obtain from a fit in all of the ten segments
simultaneously (assuming $\gamma=1.28$; note that this is different from a fit
using a single GCE template), which yields $m_\chi\simeq 46.6\GeV$ and
$\langle\sigma v\rangle \simeq 1.60\times10^{-26}\cm^3\s^{-1}$.  We show the
corresponding $p$-values in the figure label.  For the hypothesis of a single
uniform excess spectrum (again using a $\bar{b}b$ spectrum with $m_\chi$ and
$\langle \sigma v\rangle$ as the two fitting parameters, keeping $\gamma=1.28$
fixed) we obtain a combined $p$-value of 0.11.

\medskip

\emph{We find that the hypothesis of a spherically symmetric excess emission
with a uniform energy spectrum is compatible with the \Fermi-LAT data from the
inner Galaxy to within $95\%$ CL.}  

\smallskip

Obviously, this is not a proof that the emission absorbed by the GCE template
in different parts of our ROI is caused by the \emph{same} physical mechanism,
but the result is rather suggestive.  We will below proceed under the
assumption that the entire GCE emission is generated by a single extended
source with uniform spectrum, and explore the consequences of this hypothesis.

\begin{table}
    \small
    \begin{center}
        \begin{tabular}{cccc}
            \toprule
            Spectrum & Parameters & $\chi^2$/dof & $p$-value\\\midrule
            broken PL &  $\alpha_1 =  1.42_{-0.31}^{+0.22} $,  $\alpha_2 =  2.63_{-0.095}^{+0.13} $,  $E_{\rm break} =  2.06_{-0.17}^{+0.23} {\rm\ GeV}$ & 1.06 & 0.47 \\[3pt]
            DM $\chi\chi\to\bar{b}b$ &  $\langle\sigma v\rangle =  1.76_{-0.27}^{+0.28} \times 10^{-26}{\rm\ cm^3\, s^{-1}}$,  $m_\chi =  49_{-5.4}^{+6.4} {\rm\ GeV}$ & 1.08 & 0.43 \\[3pt]
            DM $\chi\chi\to\bar{c}c$ &  $\langle\sigma v\rangle =  1.25_{-0.18}^{+0.2} \times 10^{-26}{\rm\ cm^3\, s^{-1}}$,  $m_\chi =  38.2_{-3.9}^{+4.6} {\rm\ GeV}$ & 1.07 & 0.44 \\[3pt]
            PL with exp. cutoff &  $E_{\rm cut} =  2.53_{-0.77}^{+1.1} {\rm\ GeV}$,  $\alpha =  0.945_{-0.5}^{+0.36} $ & 1.37 & 0.16 \\[3pt]
            DM $\chi\chi\to\tau^+\tau^-$ &  $\langle\sigma v\rangle =  0.337_{-0.048}^{+0.047} \times 10^{-26}{\rm\ cm^3\, s^{-1}}$,  $m_\chi =  9.96_{-0.91}^{+1.1} {\rm\ GeV}$ & 1.52 & 0.065 \\
            \bottomrule
        \end{tabular}
    \end{center}
    \caption{Results of spectral fits to the GCE emission as shown in figure
    \ref{fig:spectrumSysStat}, together with $\pm1\sigma$ errors (which include
    statistical as well as model uncertainties, see text).  We also show the
    reduced $\chi^2$, and the corresponding $p$-value.  The best-fit is given
    by a broken power-law, though annihilation into $\bar{b}b$ final states is
    completely compatible with the observed spectrum as well.  We find that
    even annihilation into $\tau^+\tau^-$ cannot be excluded with 95\% CL
    significance.}
    \label{tab:fitResults}
\end{table}

\subsection{Astrophysical models}
\label{sec:AstroFits}

We start this subsection with a simple broken power-law fit to the GCE
spectrum.  This spectrum is so generic that it is capable of approximately
describing a large number of scenarios, and it is given by
\begin{equation}
    \frac{dN}{dE} = \zeta \left(\frac{E}{E_\text{break}}\right)^{-\alpha}
    \quad \text{with} \quad \alpha=
    \left\{
        \begin{array}{ll}
            \alpha_1 & \mbox{if } E < E_\text{break} \\
            \alpha_2 & \mbox{if } E \geq E_\text{break}
        \end{array}
    \right.\;,
\end{equation}
where $0>\alpha_{1}>\alpha_2$ denote respectively the spectral indices above
and below the break energy $E_\text{break}$, and $\zeta$ is a normalization
parameter.  

\begin{figure}[t]
    \begin{center}
        \includegraphics{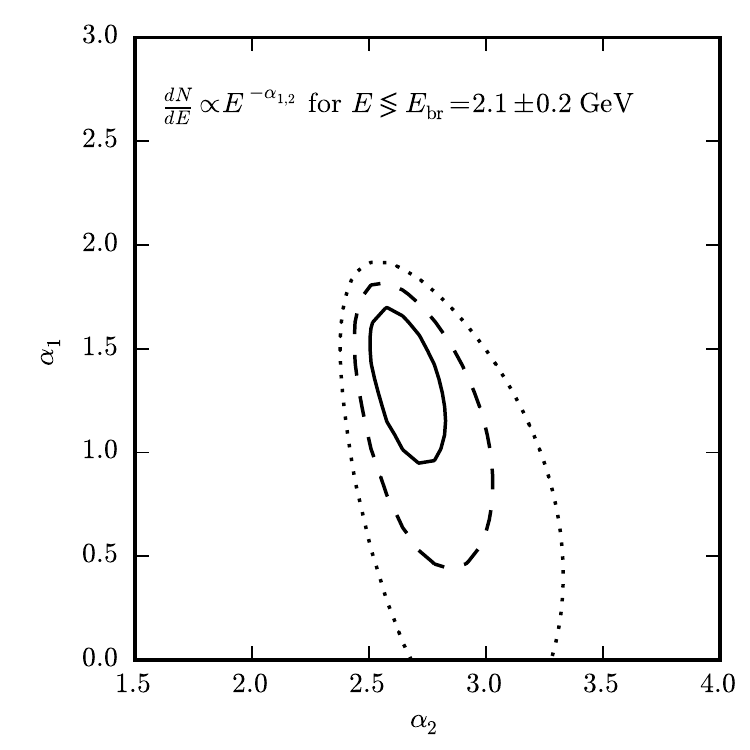}
        \includegraphics{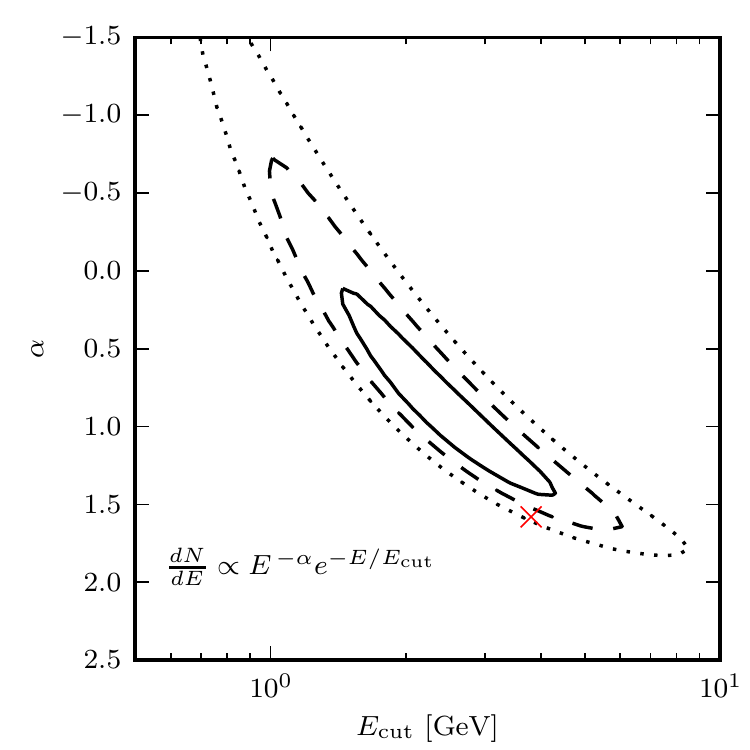}
        \vspace{-0.9cm}
    \end{center}
    \caption{ \emph{Left panel:} Constraints on the two spectral indices of a
    broken power-law, where we leave the break position free to float, from a
    fit to the spectrum in figure \ref{fig:spectrumSysStat}.  \emph{Right
    panel:} Constraints on a power-law with exponential cutoff.  For comparison
    we also show the values corresponding to observed MSPs from
    ref.~\cite{Cholis:2014noa}, where red point shows the best-fit value from a
    fit to a stacked MSP spectrum. We emphasize that, although the spectrum
    appears to be marginally compatible, the normalization of the observed GCE
    is too high by a factor of $\simeq 20-30$~\cite{Hooper:2013nhl,
    Cholis:2014lta} to be explained by MSPs.  }
    \label{fig:AstroFits}
\end{figure}

In the left panel of figure~\ref{fig:AstroFits}, we show constraints on the
broken PL spectrum in the $\alpha_1$-vs-$\alpha_2$ plane, obtained from a fit
to the data shown in figure \ref{fig:spectrumSysStat}.  The break position is
left free to vary in the fit.   As best-fit parameters for the slopes we find
$\alpha_1=1.42^{+0.22}_{-0.31}$ and $\alpha_2=2.63^{+0.13}_{-0.10}$ (\cf
table~\ref{tab:fitResults}); the position of the break is given by
$E_\text{cut}=2.06^{+0.23}_{-0.17}\GeV$.  We find that a simple broken PL
provides already a very good fit to the data, with a $p$-value of 0.47.  This
is marginally smaller than the $p$-values that we found for the DM annihilation
spectra.

\medskip

Another generic and simple spectrum is a power-law with an exponential cutoff,
as given by
\begin{equation}
    \frac{dN}{dE} = \zeta \left(\frac{E}{1\GeV}\right)^{-\alpha}
    e^{-E/E_\text{cut}}\;.
\end{equation}
Here, $\alpha$ is the spectral index, $E_\text{cut}$ denotes the cutoff energy,
and $\zeta$ is a normalization parameter.  Constraints on the
$\alpha$-vs-$E_\text{cut}$ plane that we found from a fit to the GCE spectrum
in figure~\ref{fig:spectrumSysStat} are shown in the right panel of figure
\ref{fig:AstroFits}.  In this figure, we also indicate the point in the
parameter space that corresponds to the stacked spectrum of MSPs that was
derived from a reanalysis of the \Fermi-LAT data in ref.~\cite{Cholis:2014noa}.
We find that the best-fit is obtained for a cutoff energy of
$E_\text{cut}=2.53^{+0.11}_{-0.77}$ and a spectral index of
$\alpha=0.945^{+0.36}_{-0.5}$.  However, the $p$-value for this fit is with
0.16 relatively poor.  Again, the best-fit parameters are summarized in table
\ref{tab:fitResults}.

\medskip

Finally, we explore the \emph{morphology} of the GCE with a simple parametric
model.  As a spatial template, we consider a generic spherically symmetric
volume emissivity with a radial dependence given by
\begin{equation}
    q \propto r^{-\Gamma} e^{-r/R_\text{cut}}\;,
    \label{eq:AltTempl}
\end{equation}
where $r$ denotes the Galacto-centric distance.  In contrast to the generalized
NFW profile, it features a well-defined spatial cutoff at the Galacto-centric
distance $R_\text{cut}$, which makes it possible to quantify the spatial extent
of the GCE emission in the sky (note that the spatial index $\Gamma \approx
2\gamma$ at distances close to the GC).

\smallskip

\begin{figure}
    \begin{center}
        \includegraphics{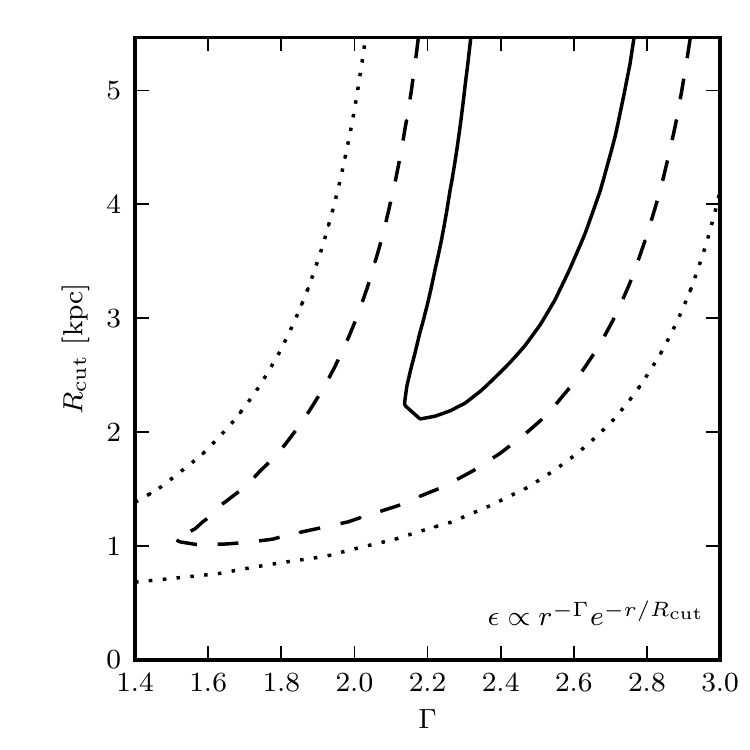}
        \includegraphics{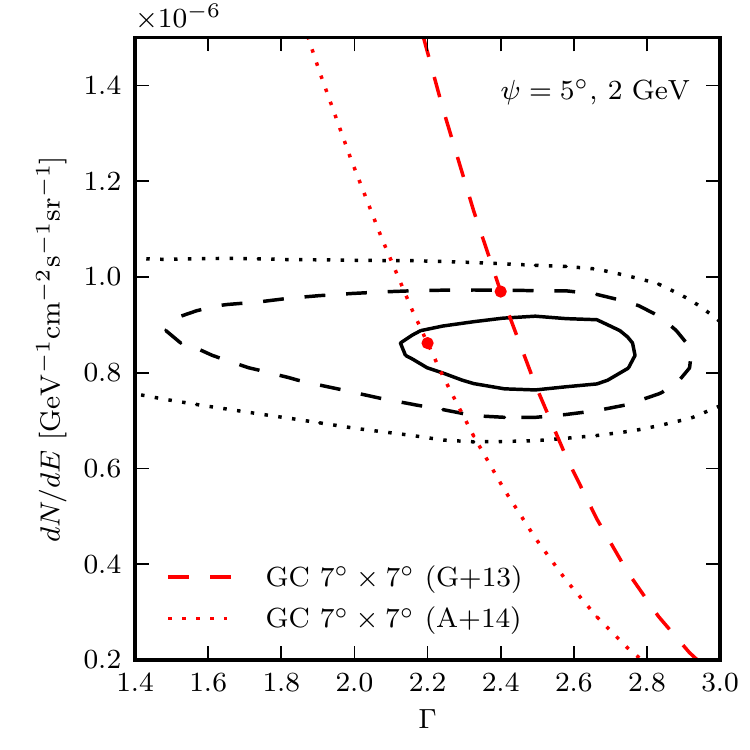}
        \vspace{-0.9cm}
    \end{center}
    \caption{ \emph{Left panel:} Constraints on the morphology of the GCE,
    assuming a spherically symmetric emission profile with a radial dependence
    that follows a power-law with index $\Gamma$ and exponential cutoff at
    radius $R_\text{cut}$ as indicated in the panel. A $\bar{b}b$ spectrum with
    $m_\chi=49\GeV$  is adopted as energy (results are similar for other
    spectra).  \emph{Right panel:} Confidence contours for the flux of the GCE
    at $5^\circ$ from the GC, and at an energy of 2\GeV, as function of the
    radial slope $\Gamma$ (we assume a $\bar{b}b$ spectrum with
    $m_\chi=49\GeV$, $R_\text{cut}$ is left free to float in the fit). The
    \emph{red lines} show an extrapolation to higher latitudes of the flux from
    the inner $7^\circ\times7^\circ$ of the Galaxy ,as determined in
    ref.~\cite{Abazajian:2014fta} (\emph{dotted line}) and
    ref.~\cite{Gordon:2013vta} (\emph{dashed line}), with the \emph{dots} being
    the actually measured values, and the lines being the expected scaling as
    function of $\Gamma$ (see text for details).}
    \label{fig:ExtensionFit}
\end{figure}

We perform a fit to the GCE fluxes as shown in
figure~\ref{fig:DMsplitSpectraSyst}, assuming as fiducial spectrum the
$\bar{b}b$ spectrum from table~\ref{tab:fitResults} with fixed mass and free
normalization (the precise form of the spectrum does not matter much).  We find
that the template can fit the data as well as the NFW profile (which is not a
surprise given that we allow $\Gamma$ and $R_{\rm cut}$ freely to vary).
Constraints on the parameters in the $\Gamma$-vs-$R_\text{cut}$ plane are shown
in the left panel of figure \ref{fig:ExtensionFit}.  We find, for a freely
varying $\Gamma$, as lower limit on the spatial extend of the excess
$R_\text{cut}>1.1\kpc$ at $95\%$ CL.  This corresponds to an angular distance
of about $7.4^\circ$ from the GC.  The constraint on $\Gamma$ is for large
values of $R_\text{cut}$ about $\Gamma \sim 2.2$--$2.9$ ($95\%$ CL), which is
compatible with what we found for the generalized NFW profile.  

\medskip

In the right panel of figure \ref{fig:ExtensionFit}, we show constraints on the
GCE intensity-vs-$\Gamma$ plane, leaving $R_\text{cut}$ free to vary.  The
intensity of the GCE emission is here determined at an energy of 2 GeV and at
an angular distance of $\psi = 5^\circ$ from the GC.  We find that $\psi =
5^\circ$ is a rather good pivot point for the determination of the signal
brightness in the adopted inner Galaxy ROI, since the flux at this point
is largely independent of the profile slope $\Gamma$.

It is instructive to compare the flux in the right panel of
figure~\ref{fig:ExtensionFit} to an extrapolation of the GeV excess emission
that is seen in the inner few degrees of the Galactic center.  We adopt here
results from refs.~\cite{Gordon:2013vta, Abazajian:2014fta}, which analyzed
gamma-ray emission from the inner $7^\circ\times7^\circ$, and found profile
slopes in the range $\Gamma \approx 2\gamma = 2.2$--$2.4$.  Taking their
best-fit values for $\bar{b}b$ final states, we calculate what flux would be
expected $5^\circ$ away from the GC at 2\GeV, and indicate these values in
figure~\ref{fig:ExtensionFit}.   We also show how this flux would approximately
vary when different values of $\Gamma$ are adopted (we require here that the
flux in the $7^\circ\times7^\circ$ region remains constant).  We conclude that
for values of roughly $\Gamma\sim2.3$--2.6 the excess emission in the inner
Galaxy can indeed be interpreted as the high-latitude counterpart of the GeV
excess seen in the inner few degrees.

\medskip

Finally, as shown in the right panel of figure~\ref{fig:ExtensionFit}, a
profile slope $\Gamma=2.2$ is about the smallest value for which one can
interpret the GCE in the inner Galaxy as extended counterpart of the GeV excess
that is seen in the inner few degrees of the Galaxy.  For this value of
$\Gamma$, we obtain a stronger lower limit on the extent of the GCE, namely
$R_\text{cut}>1.48\kpc$ at $95\%$ CL.  The corresponding angular distance from
the GC is $\psi=10.0^\circ$.

\section{Discussions}
\label{sec:discussions}

\subsection{Astrophysical interpretations}
\label{sec:Adiscussion}

In what follows we briefly discuss the astrophysical interpretations that have
been put forward in light of the results of section~\ref{sec:AstroFits}.
Astrophysical processes that might explain the GC excess fall in two main
categories: The emission from an unresolved population of sources concentrated
at the GC and diffuse emission related to the interaction of CRs with the ISM
and radiation fields in the extreme environment of the GC. 

\bigskip

As for the emission from an unresolved point source population, MSPs has been
for quite some time promising candidates for the explanation of the anomalous
excess.  Indeed, the gamma-ray MSPs source spectrum  is measured to be
compatible with a power-law with exponential cutoff usually at $E_{\rm cut}
\sim 2-3$ GeV, thus suggesting a compatibility with the GCE spectrum.
Recently, the best-fit (stacked) spectrum of the 61 MSPs detected by \Fermi-LAT
in 5.6 years has been parametrized by $E_{\rm cut} = 3.78^{+0.15}_{-0.08}$ GeV
and $\alpha = 1.57^{+0.02}_{-0.01}$ \cite{Cholis:2014noa}.  Those parameter
values lie inside our 99\% CL contour of fig.~\ref{fig:AstroFits}, where the
red cross represents the best fit result of ref.~\cite{Cholis:2014noa}.  In
section~\ref{sec:AstroFits}, we find that a power-law with exponential cutoff
prefers a cutoff energy of $E_\text{cut}=2.53^{+1.1}_{-0.77}$ GeV and a slope
of $\alpha=0.945^{+0.36}_{-0.5}$.  Our spectral parameters are in agreement
with previous results from analyses of the inner few degrees ($7^{\circ} \times
7^{\circ}$ around the GC) of the Galaxy.  Ref.~\cite{Abazajian:2014fta} found
$E_{\rm cut} = 1.65 \pm 0.20$ GeV and $\alpha = 0.45 \pm 0.21$ (full model),
while the values $E_{\rm cut} = 3^{+2}_{-1}$ GeV and $\alpha = 1.4 \pm 0.3$
were derived in ref.~\cite{Macias:2013vya} with the main difference being
different methods for background and point source modeling.  

In our analysis, the power-law with an exponential cutoff performs less well
than the $\bar{b}b$, $\bar{c}c$ spectra or the broken power-law, as shown in
table~\ref{tab:fitResults}.  Most interestingly, this is not so much due to the
steep rise in the spectrum at low energies as is visible in
figure~\ref{fig:spectrumSysStat_fits}, but to the sizable amount of excess
emission at high energies above 10 GeV, which is difficult to accommodate with
a power-law with an exponential cutoff.  A sub-exponential cutoff (as indicated
from stacked MSP spectra in ref.~\cite{Cholis:2014noa}) would here certainly
help to improve the quality of the fit.

\smallskip

Additionally, it has been shown that the spatial distribution of MSPs is not
easily able to account for the spatial extension of the
GCE~\cite{Hooper:2013nhl, Calore:2014oga}.  Indeed, observed gamma-ray MSPs are
well described by a disk-like distribution. In that case, predictions of the
unresolved flux originating from MSPs generally estimate that this population
might contribute up to 5\%--10\% of the excess emission in both GC and inner
Galactic regions (see refs.~\cite{Hooper:2013nhl, Calore:2014oga}). 

The MSPs spatial profile has to steeply increase towards the center like
approximately $\propto r^{-2.5}$ to be compatible with the morphology of the
GCE.  Hence, a population associated with the \emph{bulge} is required (and to
certain extent expected), yet a bulge component can not explain the extension
of the signal up to $\sim 15^{\circ}$ ~\cite{Hooper:2013nhl}.  To derive the
spatial distribution of bulge MSPs, one can use low-mass X-ray binaries (LMXB)
that are believed to be MSP progenitors as proxies.  Measurements of the
INTErnational Gamma-Ray Astrophysics Laboratory (INTEGRAL) suggest that  the
LMXB profile in the centre of the Milky Way could be steeper than $\propto
r^{-2}$. One measure is available for LMXB in M31 \cite{Voss:2006az}. Using
that, ref.~\cite{2012PhRvD..86h3511A} claimed  that MSPs may account for the
emission in the inner $2^{\circ}$ about the GC.  Yet, making a connection
between observed MSPs and LMXBs at globular clusters and with bright LMXBs
observations towards the GC \cite{Cholis:2014lta} found that only $\sim 5\%$ of
the  excess emission can be accounted for by MSPs laying at the inner
$5^{\circ}$ ($\simeq$ 0.8 kpc). 

In conclusion, our current knowledge of MSPs makes it questionable whether they
are the main source of the GCE.  Nonetheless, the excess emission may still be
interpreted in terms of an unresolved diffuse emission from a yet unidentified
source population. Those would be concentrated at the GC, without obvious
members sitting close to us, which would help derive information on their
spectra, spatial distribution, luminosity function or multi-wavelength
properties. 

\medskip

Non-equilibrium processes involving burst-like events during an active past of
our GC have been explored as well,  see refs.~\cite{Carlson:2014cwa,
Petrovic:2014uda}.  Although they deserve a careful study in light of the data,
we will only mention them briefly at this point. In the case of one or more
proton bursts injected from a few kilo-years up to a few mega-years ago
\cite{Carlson:2014cwa}, the induced morphology would be in general more
disk-like than the GCE, trace the gas distribution and extend up from few
degrees to tens of degrees depending on the age of the event. On the other
hand, the injection of an additional electron population with total energy
$\sim 10^{53}$~erg on the timescale of few mega-years ago
\cite{Petrovic:2014uda} is expected to lead to a more symmetric emission around
the GC because the ICS traces the smooth radiation field distribution, but no
attempts of performing template fits based on this scenario have been made yet.
In this case, the induced spectrum would need to reproduce, depending in the
initial conditions and properties of the event, our measured broken power-law
spectrum (with  $\alpha_1 \sim 1.42$, $\alpha_2 \sim  2.63$,  $E_{\rm break}
\sim  2.06 {\rm\ GeV}$ as quoted in table~\ref{tab:fitResults}). 

\subsection{A signal from dark matter annihilation?}
\label{sec:DMdiscussion}

As demonstrated in section~\ref{sec:DMfits}, the spectrum and morphology of the
GCE are compatible with a signal from WIMP DM annihilation in the halo of the
Milky Way.  Fits with typical annihilation spectra are found to give good
$p$-values, despite the fact that, at first sight, the peaked DM spectra do not
appear to be a good representation of the data shown \fex in
figure~\ref{fig:spectrumSysStat_fits}.  This is due to the strong correlations
of the empirical model systematics in energy,  which to some degree allow a
further subtraction of ICS-like and $\pi^0$-like background spectra.  When
performing fits to the data this results especially at low and high energies,
where the GCE flux is already small, in an additional suppression of the GCE
spectrum, leading to a more pronounced peak at 1--3 GeV.

\medskip 

The results quoted in table~\ref{tab:fitResults} prefer annihilation into
quarks ($\bar{b} b, \bar{c} c$), while heavy leptonic final states are almost
excluded at 95\% CL. For a generalized NFW DM spatial distribution with slope
$\gamma$=1.2 and the representative case of annihilation into $\bar{b} b$, our
fits prefer a DM mass in the range $m_\chi=43.6$--55.4\,GeV and a
velocity-averaged annihilation cross-section of $\langle\sigma v\rangle =
1.76_{-0.27}^{+0.28} \times 10^{-26}{\rm\ cm^3\, s^{-1}}$.  Compared to
findings in the previous literature, our results  are in general in good
agreement, although we find typically somewhat larger masses and
cross-sections.  We will present a more detailed comparison in the following.
In figure~\ref{fig:DMfits1} (left panel), we display the points corresponding
to the best-fit results from previous works we compare with.  Notice that the
$\langle\sigma v\rangle$ values have been opportunely rescaled for a fair
comparison.\footnote{To fairly compare previous findings with our results, we
firstly rescale the DM profiles used in the previous literature.  In
particular, the generalised NWF profile can be reparametrized as:
\begin{equation}
    \rho(r) = \rho_\odot \left( \frac{r}{r_\odot} \right) ^{-\gamma} \left(\frac{(1 + r_\odot/r_s)}{(1 + r/r_s)} \right)^{3 - \gamma} \,,
\end{equation}
In the limit of small radii, we rescale the cross-sections in the literature
such to match our parameters choice for $\rho_\odot, r_\odot, r_s$.  We then
further rescale the cross-sections in order to take into account the different
mass and $\gamma$ best-fit values found in the work we are comparing with.  To
this end, we derive the corresponding  $\langle\sigma v\rangle$-vs-$\gamma$
contours with our fitting procedure (for $m_\chi$ fixed).  In the text, we
quote the final  \emph{rescaled} cross-section we compare with in the
$\langle\sigma v\rangle$-vs-$m_\chi$ plane of figure~\ref{fig:DMfits1}.  } 

\smallskip

The first paper about the \emph{inner Galaxy} found values $m_\chi \sim 50$ GeV
and $\langle\sigma v\rangle \sim 0.8 \times 10^{-26}\rm\ cm^3
s^{-1}$,~\cite{Hooper:2013rwa}, which fall largely outside our 99\% CL contour.
The results of the follow-up work in ref.~\cite{Huang:2013pda} at latitudes
$10^{\circ} < |b|  < 20^{\circ}$ suggest a somewhat larger value for the WIMP
mass $m_\chi \sim 61.8$ GeV and cross-section $\langle\sigma v\rangle \sim 3.1
\times 10^{-26}{\rm\ cm^3 s^{-1}}$ (for \eg $\bar{b} b$), which lies slightly
outside our 99\% CL contour.  The most recent analysis of the \emph{inner
Galaxy} was presented in ref.~\cite{Daylan:2014rsa}, and pointed towards a DM
mass $m_\chi  \sim 35$ GeV (lower than our best-fit value) and cross-section
$\langle\sigma v\rangle \sim 1.24 \times 10^{-26}{\rm\ cm^3 s^{-1}}$ (for
$\bar{b} b$), that agree with our results at 95\% CL.  Compared to this most
recent analysis, there are a number of differences in the analysis set up.
Ref.~\cite{Daylan:2014rsa} uses \fex the \Fermi-LAT model \texttt{P6V11} as
their baseline GDE model to subtract diffuse backgrounds, which likely absorbs
more ICS emission than any of our GDE models.  Moreover,
ref.~\cite{Daylan:2014rsa} performs a full-sky fit, while we concentrate on a
smaller region.  In general, we find that with respect to previous results from
the \emph{inner Galaxy} our spectral fits do not show a strong trend and are
broadly consistent. 

\smallskip

There is a number of analyses concentrating on studying the GCE in the
\emph{inner few degrees} of the Galaxy.  Comparing our results with the results
of the analyses of the GC is a quantitative way to test the consistency of the
excesses observed at the GC and in the inner Galaxy in light of the DM
interpretation.  In ref.~\cite{Abazajian:2014fta}, the preferred mass and
cross-section values for $\bar{b} b$ annihilation are found to be $m_{\chi}
\sim 40$ GeV and $\langle\sigma v\rangle \sim 1.62 \times 10^{-26}{\rm\ cm^3\,
s^{-1}}$, for an inner slope of the generalized NFW profile of $\gamma=1.1$. We
find that this results is in well in agreement with our analysis, lying inside
the 95\% CL contour.  Instead, the results from ref.~\cite{Gordon:2013vta},
$m_{\chi} \sim 34.1$ GeV and $\langle\sigma v\rangle \sim 1.56 \times
10^{-26}{\rm\ cm^3\, s^{-1}}$ (where $\gamma = 1.2$), agree with our findings
at 99\% CL.  Lastly, we show in the right panel of
figure~\ref{fig:ExtensionFit} an extrapolation of the GeV excess emission
measured in the inner $7^\circ\times7^\circ$ of the Galaxy to higher latitudes,
based on the results from refs.~\cite{Gordon:2013vta, Abazajian:2014fta} (see
discussion above).  For profile slopes of $\gamma\approx 1.15$--$1.3$ we indeed
find good agreement between these and our results from the inner Galaxy.

\bigskip

We briefly mention the constraints on DM annihilation that come from other
targets and messengers (see ref.~\cite{Bringmann:2014lpa} for a recent
discussion in context of the GCE).  
Some of these constraints are already in mild tension
with or show mild support for the DM interpretation of the GCE, and in the near
future these probes will further help to support or exclude this possibility.

For gamma rays, the most robust limits on DM self-annihilation in hadronic
final states (including also $\tau^+ \tau^-$) come currently from observations
of dwarf spheroidal galaxies~\cite{Ackermann:2013yva, Cholis:2012am,
GeringerSameth:2011iw}, and probe DM annihilation down to the thermal
cross-section for masses less than about 10 GeV (for $\bar{b}b$ final states).
Interestingly, the most recent combined analysis of 15 dwarfs might already
indicate first signs for a DM signal compatible with the GCE (at the level of
$2.3\sigma$)~\cite{Ackermann:2013yva}.  Antiprotons provide also a strong probe
for DM \cite{Donato:2008jk, Cholis:2010xb, Cirelli:2013hv}, though
uncertainties in the propagation assumptions remain
relevant~\cite{Evoli:2011id}.  For the most common CR propagation scenarios
they are already in mild tension with the DM interpretation of the
GCE~\cite{Bringmann:2014lpa, Cirelli:2014lwa}. For leptonic two-body final
states the strongest limits come from a spectral analysis of the positron
fraction measured by AMS-02~\cite{Bergstrom:2013jra, Ibarra:2013zia}, which
excludes a thermal rate for $\chi\chi\to e^+ e^-$ ($\chi\chi\to \mu^+\mu^-)$ up
to DM masses of $m_\text{DM}\simeq 200\rm \ GeV$ (100\GeV), so that this
channel can only play a very subdominant role in models explaining the
GCE~\cite{Bringmann:2014lpa}.  Limits from radio observations of the GC are in
some tension with the GCE DM interpretation, unless the DM profile at the GC is
somewhat cored out~\cite{Bringmann:2014lpa}, or strong convective winds or ICS
energy losses are important~\cite{Cholis:2014fja}.  Finally, limits from
observations of the cosmic microwave background constrain leptonic models for
DM annihilation~\cite{Galli:2009zc, Slatyer:2009yq, Galli:2011rz, Evoli:2012qh,
Lopez-Honorez:2013cua}, though not at a level that is relevant for the GCE
right now.

\subsection{The energy spectrum at sub-GeV energies}

Results below 1 GeV are potentially subject to large systematics related to the
relatively large PSF of \Fermi-LAT at those low energies.  At the same time,
the low-energy part of the GCE spectrum is of utmost importance when aiming at
a discrimination of DM signals from astrophysical processes, see
ref.~\cite{Daylan:2014rsa}.  Indeed, in the prototypical case of $\bar{b}b$
final states, DM spectra are predicted to follow a rising power-law with a
spectral index of roughly $\sim 1.5$ and with a prounced cutoff at energies
above $\sim 0.05\,m_\chi$.  The observation of such a spectrum with high
accuracy would be a strong support for a DM interpretation of the GCE.

\smallskip

We think our results below 1 GeV are reliable for a number of reasons.  As
described in section~\ref{sec:otherSystematics}, we estimated the impact of a
number of analysis choices that we make throughout our analysis.  These include
details about the point source mask, the treatment of the \Fermi-LAT PSF, and
the impact of fixed IGRB and \Fermi\ bubble contributions.  As shown in
figure~\ref{fig:otherSyst}.  Some of these effects indeed start to dominate
over the statistical error at energies below 1 GeV.  We account for these
uncertainties by additional terms in the covariance matrix for the spectral
fits.  In appendix~\ref{sec:otherModeling} we furthermore show that our results
do not change much when neglecting these above systematics in the fit, or when
instead restricting the fits to ranges above 600 MeV.  Lastly, the absence of a
stronger logitudinal variation at latitudes $4^\circ \leq|b|\leq 6^\circ$ as
shown in the lower left panel of figure~\ref{fig:profiles} is a visible
indicator for the GCE spectrum featuring a spectral index harder than $\sim2$
at energies below 2 GeV.

\medskip

Our analysis supports a rise at energies below 1 GeV, with a spectral index
harder than $\sim2$ as can be seen \fex in the bottom left panel of
figure~\ref{fig:profiles}.  Moreover, figure~\ref{fig:DMsplitSpectraSyst}
suggests that such a low-energy trend extends up to $5^{\circ}$--$10^{\circ}$
in latitude and higher.  In the case of the broken power-law fit, the
low-energy slope is $\alpha_1 = 1.42^{+0.22}_{-0.31}$, while in case of the
power-law with exponential cutoff we obtain $\alpha = 0.945^{+0.36}_{-0.5}$,
with the harder value being related to the GCE excess emission above 10 GeV.
As discussed above, the spectrum from $\bar{b}b$ is perfectly consistent with
the data.

\section{Conclusions}
\label{sec:conclusions}

During the last few years the presence of a gamma-ray excess at GeV energies
and towards the GC has been suggested by a sequence of studies performed using
\Fermi-LAT gamma-ray data \cite{Goodenough:2009gk, Hooper:2010mq,
2011PhRvD..84l3005H, Boyarsky:2010dr, 2012PhRvD..86h3511A, Abazajian:2014fta,
Gordon:2013vta, Macias:2013vya, Daylan:2014rsa, Zhou:2014lva}. Given the
sphericity of the excess, its amplitude and the fact that it is centered within
$\sim 0.05^{\circ}$ of the GC~\cite{Daylan:2014rsa}, a large number of
follow-up studies have been interpreted this excess as a signal from DM
self-annihilation. Additionally, less exotic possibilities, such as populations
of dim point sources or diffuse emission from recently injected CRs, have also
been proposed.  Important ingredients for understanding the characteristics and
the origin of this excess are first its morphology, namely its extension to
Galactic latitudes above the inner few degrees, and second its energy spectrum.  

\smallskip

Any description of the excess emission necessitates the adoption of several
assumptions on all of the relevant Galactic diffuse and point-souce
backgrounds.  One of the major limitations of previous high latitude analyses
is the lack of a thorough study of background model systematics, which  often
causes overly constraining and sometimes biased results.  As an example,
previous analyses often adopted the \PV model, which provides a prediction for
only the \textit{total} Galactic diffuse emission ($\pi^{0}$ $+$ bremsstrahlung
$+$ ICS).  We demonstrated that the \PV model features an extremely hard ICS
emission at energies above $10\GeV$ (figure~\ref{fig:P6V11}), which -- when
used in a template analysis -- easily over-subtracts components in regions
relevant for DM searches, potentially leading to pronounced drops in the
spectrum at energies around 10\GeV, as \eg found in refs.~\cite{Hooper:2013rwa,
Daylan:2014rsa}.  

\medskip

In this paper, we reanalyzed the \Fermi-LAT data in the \emph{inner Galaxy}.
More specifically, we concentrated on high Galactic latitudes $|b| \ge
2^{\circ}$ in a Galacto-centric box of $40^{\circ} \times 40^{\circ}$.  Given
the often large variations in the predicted spectra and morphologies of the
Galactic diffuse emission components (figures~\ref{fig:TemplateComparisons1},
\ref{fig:spectrumDiff} and \ref{fig:TemplateComparisons2}), we modeled the
three contributions $\pi^{0}$, bremsstrahlung and ICS as separate templates.
To probe the associated uncertainties in the Galactic diffuse emission, we used
existing models in the literature \cite{FermiLAT:2012aa} as well as our own
models. The latter account for even extreme variations in the CR source
distribution and injection, the gas distribution, the diffusion of CRs,
convection and re-acceleration, the interstellar radiation field distribution
and the distribution and amplitude of the Galactic magnetic field.  Equipped
with these templates, and additional templates for the \Fermi\ bubbles, the
IGRB and known point sources, we performed a multi-linear regression analysis
of the \Fermi-LAT data in 24 energy bins from 300 MeV to 500 GeV.  We repeated
that analysis in several test regions along the Galactic disk.  Our main
results can be summarized as follows.
\begin{itemize}
    \item We confirmed the existence of a diffuse Galacto-centric excess
        emission (``Galactic center excess") in the inner Galaxy, above the
        modeled astrophysical backgrounds.  We showed that the spectral
        properties are remarkably stable against theoretical model systematics,
        which we bracketed by exploring a large range of Galactic diffuse
        emission models (figure~\ref{fig:spectrum60models}).  The excess
        emission shows a clear peak at 1--3 GeV, which rises steeply at lower
        energies, and follows a power-law with slope $\sim-2.7$ above.

    \item We found residuals above the modeled astrophysical backgrounds in
        various test regions along the Galactic plane, which are of almost the
        same size as the Galactic center excess (figure~\ref{fig:sliding}).  By
        means of a principal component analysis we traced these residuals back
        to non-uniform variations in the normalizations and spectral indices of
        the primary Galactic diffuse emission components
        (figure~\ref{fig:variance}).  We folded these uncertainties back into
        the analysis of the Galactic center excess as an estimate for the
        empirical model systematics (figure~\ref{fig:spectrumSysStat}).

    \item In order to explore the morphology of the excess emission, we split
        the adopted Galactic center excess template in ten segments
        (figure~\ref{fig:ROIsplits}), and repeated the above estimate of
        theoretical and empirical model systematics individually for each of
        them, finding consistent results (figure~\ref{fig:DMsplitSpectraSyst}).

    \item We showed that the hypothesis of a spherically symmetric and
        spectrally uniform excess emission is compatible with the data from the
        inner Galaxy at 95\% CL (right panel of figure~\ref{fig:DMfits1}).
        Under that hypothesis, and assuming a radial profile of the volume
        emissivity with a floating index $\Gamma$ and an exponential cutoff
        $R_\text{cut}$, we found a robust lower limit on the radial extension
        of the excess of $R_\text{cut}>1.1\kpc$ at 95\% CL.  This corresponds
        to an angular distance from the GC of $7.4^\circ$ (left panel of
        figure~\ref{fig:ExtensionFit}). For a radial index of $\Gamma=2.2$,
        which is compatible with previous results from the inner few degrees of
        the Galaxy (right panel of figure~\ref{fig:ExtensionFit}), we obtained
        an even stronger limit of $R_\text{cut}>1.48\kpc$ at $95\%$ CL, which
        corresponds to an angular distance of $10.0^\circ$.
        
    \item We do \emph{not} confirm previous results that indicated that the
        Galactic center excess spectrum would drop to zero at $E\gtrsim10\GeV$
        energies (figure~\ref{fig:spectrumSysStat}).  However, when we included
        the existing large model systematics as correlated errors into the
        spectral fits, we found that the excess spectrum is well described both
        by a broken power law with spectral indices
        $\alpha_1=1.42^{+0.22}_{-0.31}$ and $\alpha_2=2.63^{+0.13}_{-0.10}$ and
        break energy $E_{\rm break}=2.06^{+0.23}_{-0.17}\GeV$
        (figure~\ref{fig:AstroFits}), and by the gamma-ray spectrum produced by
        DM particles annihilating into $\bar{b}b$ final states.  In the latter
        case, we obtained as best-fit values $\langle \sigma
        v\rangle=1.76^{+0.28}_{-0.27}{\rm\ cm^3\, s^{-1}}$ and
        $m_\chi=49^{+6.4}_{-5.4} \GeV$ (figure~\ref{fig:DMfits1}).  In both
        cases, the $p$-values of the spectral fit are close to 0.5
        (table~\ref{tab:fitResults}).
\end{itemize}

Although we find that the hypothesis of a spherical and spectrally uniform
excess emission is in good agreement with the data, this does not exclude a
more complicated morphology,  a spatial dependence of the energy spectrum, or
simultaneous contributions from different physical mechanisms.  The results
shown in figure~\ref{fig:DMfits2} suggest that an excess emission that becomes
harder towards higher latitudes could easily be accommodated by the data.  This
would be relevant for explanations of the Galactic center excess in terms of
the ICS emission (see section~\ref{sec:Adiscussion}) from leptonic burst-like
events.

Our results concerning the DM interpretation of the Galactic center excess are
in reasonable agreement (at 95\% CL) with most of the previous results from the
inner Galaxy~\cite{Daylan:2014rsa, Hooper:2013rwa} and Galactic center
\cite{Gordon:2013vta}  analyses, however they are in slight tension (at 99\%
CL) with some result from the inner few degrees around the
GC~\cite{Abazajian:2014fta} (see section~\ref{sec:DMdiscussion}).  Concerning
the interpretation in terms of MSPs, the typically expected power-law with
exponential cutoff spectrum is slightly disfavoured by the data, with a
$p$-value of 0.16 (see discussion in section~\ref{sec:Adiscussion}).
Additionally, the magnitude of the excess emission and its extension up to
latitudes of $\sim15^\circ$ make the potential contribution of an unresolved
MSP population towards the Galactic center even less significant than the
quoted 5--10$\%$ upper limits discussed in refs.~\cite{Hooper:2013nhl,
Calore:2014oga, Cholis:2014lta}. 

\medskip

As mentioned above, we find that when we folded  the strongly correlated
residuals that are present along the Galactic disk back onto the GC, spectral
fits in the inner Galaxy do not significantly discriminate between broken
power-laws and DM related spectra for the Galactic center excess.  Reducing
these empirical model systematics, by improving our understanding of the
residuals along the disk, is an important first step towards more reliable
Galactic diffuse emission models for the inner Galaxy and a benchmark for
future work,  which might finally lead to a more precise determination of the
Galactic center excess spectrum and morphology at high latitudes.

\paragraph*{Acknowledgments.}
We thank Markus Ackermann, Luca Baldini, Sam McDermott, Douglas Finkbeiner, Dan
Hooper, Tim Linden, Luigi Tibaldi, Tracy Slatyer, Meng Su, Piero Ullio and Neal
Weiner for useful discussions.  We thank Mark Lovell for the careful reading of
the manuscript. This work makes use of SciPy~\cite{SciPy},
PyFITS\footnote{\url{http://www.stsci.edu/resources/software_hardware/pyfits}},
PyMinuit\footnote{\url{http://code.google.com/p/pyminuit}},
IPython~\cite{IPython} and HEALPix \cite{Gorski:2004by}.  This work has been
supported by the US Department of Energy. F.C. acknowledges support from the
European Research Council through the ERC starting grant WIMPs Kairos, P.I. G.
Bertone.

\clearpage
\appendix


\section{Description of Galactic diffuse models}
\label{app:60models}
In this appendix we give in tables~\ref{tab:ModelsParameters1}
and~\ref{tab:ModelsParameters2} all the parameters for the 60 Galactic diffuse
models that we use to probe the range of uncertainties under the physical
assumptions that could be relevant in the identification of the GC ``GeV
excess''.

The CR sources, as described in section~\ref{sec:Uncertainties}, can have
different radial distributions. We include this possibility by using the four
source distributions (``SNR", ``Pls$_{L}$'', ``Pls$_{Y}$'' and ``OB'') modeled
from ref.~\cite{FermiLAT:2012aa}.  The CR electron and proton injection spectra
are assumed to be power-laws with indices $\alpha_{e}$ and $\alpha_{p}$ above
rigidity breaks $R_{0}^{e}$ and $R_{0}^{p}$ respectively. The rigidity breaks
are chosen to be in the narrow ranges of 2.18$< R_{0}^{e} <$3.05 GV and 11.3$<
R_{0}^{p} <$11.7 GV.  Below the rigidity breaks, CR electrons and protons are
injected with power-laws of $\alpha_{e}^{\rm low E} = $1.6 and 1.89$<
\alpha_{p}^{\rm low E} < $1.96.  These injection spectra are normalized
($N_{e}$ and $N_{p}$) at 34.5 and 100 GeV kinetic energies. For the diffusion
scale heights and radii we use models from ref.~\cite{FermiLAT:2012aa}, which
considers ranges of $4 \leq z_{D} \leq10$ kpc and for $20 \leq r_{D} \leq 30$
kpc. For the gasses, our spin temperature $T_{S}$ choices are 150 K and
$10^{5}$ K, and E(B-V) magnitude cuts of 2 or 5 \cite{FermiLAT:2012aa}.  In
addition, in converting the observed CO map to a H2 map for our own GDE models,
we used the conversion factor $X_{CO}$ profile of ref.~\cite{Strong:2004td}.
Given that the 13 models from ref.~\cite{FermiLAT:2012aa} were compared to the
full-sky data and thus could also probe the $X_{CO}$ profile, each of these
models has its own $X_{CO}$ factor that we take as it is.  The $X_{CO}$ profile
assumptions can have some impact on the overall GDE models, but given that they
affect only the molecular hydrogen gas component, which is highly non-spherical
and very concentrated towards the disk, they cannot compensate in any way for a
spherically symmetric excess.  For the 13 models that we used from
ref.~\cite{FermiLAT:2012aa} and the re-naming of them see
tables~\ref{tab:ModelsParameters1} and~\ref{tab:ModelsNames}.

Using the webrun version of \texttt{Galprop v54}, we build GDE models that test
the remaining uncertainties, i.e diffusion coefficient (for $z_{D} = 4$ kpc,
$r_{D} = 20$ kpc), re-acceleration, convection, ISRF and $B$-field
distributions.  For the diffusion coefficient described in
eq.~\ref{eqn:Diffusion}, we vary the value of $D_{0}$ in the range of
2--60$\times 10^{28}$ $\cm^{2} \s^{-1}$.  For the diffusive re-acceleration
described in eq.~\ref{eqn:Reacceleration}, we vary the value of the
Alfv$\acute{\textrm{e}}$n speed within 0 and 100 $\km \s^{-1} $.  For the
gradient of convection velocity $dv/dz$, we assume values between 0 and 500
$\km \s^{-1} \kpc^{-1}$.  The Galactic magnetic field is described in
eq.~\ref{eqn:B-field}.  In the ``$B$-field'' column, the first three out of the
nine digits refer to the $B_{0} \times 10$ value in $\mu$G (thus 090 is $B_{0}
=$ 9 $\mu$G), the next three to the $r_{c} \times 10$ in kpc and the last three
to $z_{c} \times 10$ in kpc.  We take combinations of $B_{0}$, $r_{c}$ and
$z_{c}$ that result in $5.8 \leq B(r=0,z=0) \leq 117$ $\mu$G, with the large
range scaling of the $B$-field to be $5 \leq r_{c} \leq 10$ kpc and $1 \leq
z_{c} \leq 2$ kpc.  Finally, the ISRF model in the \texttt{Galprop} webrun has
three multiplication factors, for the ``optical'', ``IR'' and CMB components.
We take the ``optical'' and ``IR'' factors to have values between 0.5--1.5 as
our extreme options.

\begin{table}
    \centering
    \footnotesize
    \begin{tabular}{cccccccccc}
     \toprule
        Name & $z_{D}$ & $D_{0}$ & $v_{A}$ & $dv/dz$ & Source & $\alpha_{e}$($\alpha_{p}$) & $N_{e}$($N_{p}$)  & $B$-field  & ISRF \\
     \midrule
        A        &  4  &  5.0  &   32.7    &   50  &  SNR  &  2.43(2.47)  & 2.00(5.8) & 090050020 & 1.36,1.36,1.0  \\  
        B        &  4  &  28.0  &   31.0   &   0  &  SNR  &  2.43(2.39)  & 1.00(4.9) & 105050015 & 1.4,1.4,1.0  \\  
        C        &  4  &  5.0  &   32.7    &   0  &  SNR  &  2.43(2.39) & 0.40(4.9) & 250100020 & 1.0,1.0,1.0  \\  
        D        &  4  &  5.2  &   32.7    &   0  &  SNR  &  2.43(2.39)  & 0.40(4.9) & 050100020 & 0.5,0.5,1.0  \\  
        E        &  4  &  2.0  &  32.7    &   0  &  SNR  &  2.43(2.39)  & 0.40(4.9) & 050100020 & 1.0,1.0,1.0  \\  
        F        &  6  &  8.3  &   32.7    &   0  &  Pls$_{L}$  &  2.42(2.39)  & 0.49(4.8) & 050100020 & 1.0,1.0,1.0  \\  
        G        &  6  &  7.9  &  35.4    &   0  &  Pls$_{L}$  &  2.42(2.39)  & 0.49(4.8) & 050100020 & 1.0,1.0,1.0  \\  
        H        &  6  &  7.4  &   33.0    &   0  &  Pls$_{L}$  &  2.42(2.39)  & 0.49(4.8) & 050100020 & 1.0,1.0,1.0  \\  
        I        &  6  &  8.8  &  39.6    &   0  &  Pls$_{L}$  &  2.42(2.39)  & 0.49(4.8) & 050100020 & 1.0,1.0,1.0  \\  
        J        &  10  &  10.3  &  32.2    &   0  &  Pls$_{L}$   &  2.44(2.39)  & 0.49(4.8) & 050100020 & 1.0,1.0,1.0  \\  
     \midrule[0pt]
        K        &  10  &  9.0  &  28.6    &   0  &  OB  &  2..47(2.40)  & 0.53(5.1) & 050100020 & 1.0,1.0,1.0  \\  
        L        &  10  &  10.1  &  31.6    &   0  &  Pls$_{Y}$  &  2.45(2.39)  & 0.50(5.0) & 050100020 & 1.0,1.0,1.0  \\  
        M        &  10  &  9.8  &  30.8    &   0  &  SNR  &  2.46(2.39)  & 0.50(4.9) & 050100020 & 1.0,1.0,1.0  \\  
        N        &  8  &  8.5  &  30.9    &   0  &  SNR  &  2.45(2.39)  & 0.50(4.9) & 050100020 & 1.0,1.0,1.0  \\  
        O        &  6  &  7.1  &  31.9    &   0  &  SNR  &  2.44(2.39)  & 0.50(4.9) & 050100020 & 1.0,1.0,1.0  \\  
        P        &  4  &  5.2  &  32.7    &   0  &  SNR  &  2.43(2.39)  & 0.49(4.9) & 050100020 & 1.0,1.0,1.0  \\  
        Q        &  10  &  10.0  &  29.1    &   0  &  OB  &  2.46(2.39)  & 0.48(4.9) & 050100020 & 1.0,1.0,1.0  \\  
        R        &  10  &  10.6  &  31.1    &   0  &  Pls$_{L}$  &  2.44(2.39)  & 0.49(4.8) & 050100020 & 1.0,1.0,1.0  \\  
        S        &  4  &  5.2  &  32.7    &   0  &  SNR  &  2.43(2.39)  & 0.40(4.9) & 050100020 & 1.0,1.0,1.0  \\  
        T        &  4  &  8.0  &  32.7    &   0  &  SNR  &  2.43(2.39)  & 0.40(4.9) & 050100020 & 1.0,1.0,1.0  \\  
     \midrule[0pt]
        U        &  4  &  12.0  &   32.7    &   0  &  SNR  &  2.43(2.39)  & 0.40(4.9) & 050100020 & 1.0,1.0,1.0  \\  
        V        &  4  &  20.0  &  32.7    &   0  &  SNR  &  2.43(2.39)  & 0.40(4.9) & 050100020 & 1.0,1.0,1.0  \\  
        W       &  4  &  40.0  &  32.7    &   0  &  SNR  &  2.43(2.39)  & 0.40(4.9) & 050100020 & 1.0,1.0,1.0  \\  
        X        &  4  &  60.0  &  32.7    &   0  &  SNR  &  2.43(2.39)  & 0.40(4.9) & 050100020 & 1.0,1.0,1.0  \\  
        Y        &  4  &  4.0  &  32.7    &   0  &  SNR  &  2.43(2.39)  & 0.40(4.9) & 050100020 & 1.0,1.0,1.0  \\  
        Z        &  4  &  3.0  &  32.7    &   0  &  SNR  &  2.43(2.39)  & 0.40(4.9) & 050100020 & 1.0,1.0,1.0  \\  
        GI        &  4  &  5.0  &   32.7    &   50  &  SNR  &  2.43(2.39)  & 0.40(4.9) & 050100020 & 1.0,1.0,1.0  \\  
        GII     &  4  &  5.0  &   32.7    &   100  &  SNR  &  2.43(2.39)  & 0.40(4.9) & 050100020 & 1.0,1.0,1.0  \\  
        GIII    &  4  &  5.0  &  32.7    &   200  &  SNR  &  2.43(2.39)  & 0.40(4.9) & 050100020 & 1.0,1.0,1.0  \\  
        GIV    &  4  &  5.0  &  32.7    &   500  &  SNR  &  2.43(2.39)  & 0.40(4.9) & 050100020 & 1.0,1.0,1.0  \\  
     \midrule[0pt]
        GV    &  4  &  5.0  &  32.7    &   0  &  SNR  &  2.43(2.39)  & 0.40(4.9) & 025100020 & 1.0,1.0,1.0  \\  
        GVI    &  4  &  5.0  &  32.7    &   0  &  SNR  &  2.43(2.39)  & 0.40(4.9) & 100100020 & 1.0,1.0,1.0  \\  
        GVII   &  4  &  5.0  &  32.7    &   0  &  SNR  &  2.43(2.39)  & 0.40(4.9) & 500100020 & 1.0,1.0,1.0  \\  
        GVIII  &  4  &  5.0  &  0.0    &   0  &  SNR  &  2.43(2.39)  & 0.40(4.9) & 050100020 & 1.0,1.0,1.0  \\  
        GIX   &  4  &  5.0  &  16.0    &   0  &  SNR  &  2.43(2.39)  & 0.40(4.9) & 050100020 & 1.0,1.0,1.0  \\  
        GX    &  4  &  5.0  &  50.0    &   0  &  SNR  &  2.43(2.39)  &  0.40(4.9) & 050100020  & 1.0,1.0,1.0  \\  
        GXI    &  4  &  5.0  &  100.0    &   0  &  SNR  &  2.43(2.39)  & 0.40(4.9) & 050100020 & 1.0,1.0,1.0  \\  
        GXII   &  4  &  5.2  &  32.7    &   0  &  SNR  &  2.43(2.39)  & 0.40(4.9) & 050100020 & 0.8,0.8,1.0  \\  
        GXIII  &  4  &  5.2  &  32.7    &   0  &  SNR  &  2.43(2.39)  & 0.40(4.9) & 050100020 & 1.2,1.2,1.0  \\  
        GXIV    &  4  &  5.2  &  32.7    &   0  &  SNR  &  2.43(2.39)  & 0.40(4.9) & 050100020 & 1.5,1.5,1.0  \\  
        \bottomrule
    \end{tabular}
    \caption{Parameters of the 60 Galactic diffuse models. $z_{D}$ is in kpc,
    $D_{0}$ is in units of $\times10^{28}$cm$^{3}$s$^{-1}$, $v_{A}$  in $\km
    \s^{-1} $, $dv/dz$ in $\km \s^{-1} \kpc^{-1}$, $N_{e}$($N_{p}$) is in units
    of $\times 10^{-9}$ in $\cm^{-2} \sr^{-1} \s^{-1} \MeV^{-1}$ at $E_{\rm
    kin}$ of 34.5 (100) GeV.  $\alpha_{e}$($\alpha_{p}$) is the electron
    (proton) injection index above rigidity of 2.18 (11.3) GV.  All models
    assume $r_{D}=20$ kpc apart from models ``Q'' and ``R'', for which we take
    $r_{D}=30$ kpc. In addition, all models have a spin temperature $T_{S}=$150
    K apart from ``F'' and ``I'' that have $T_{S}=10^{5}$ K. Finally, all
    models have a magnitude cut in the E(B-V) of 5 apart from models ``G'' and
    ``I'' which have a magnitude cut of 2.}
    \label{tab:ModelsParameters1}
\end{table}

\begin{table}
    \centering
    \footnotesize
    \begin{tabular}{cccccccccccc}
     \toprule
        Name & $z_{D}$ & $D_{0}$ & $v_{A}$ & $dv/dz$ & Source & $\alpha_{e}$(p) & $N_{e}(p) $ & $B$-field  & ISRF \\
     \midrule
        GXV   &  4  &  5.8  &  33.0    &   0  &  SNR  &  2.43(2.39)  & 3.20(4.9) & 140060020 & 1.2,1.2,1.0  \\  
        GXVI  &  4  &  5.8  &  33.0    &   0  &  SNR  &  2.43(2.39)  & 3.20(4.9) & 105050020 & 1.2,1.2,1.0  \\  
        GXVII &  4  &  5.8  &  33.0    &   0  &  SNR  &  2.43(2.39)  & 3.20(4.9) & 170070020 & 1.2,1.2,1.0  \\  
        GXVIII &  4  &  5.8  &  33.0    &   0  &  SNR  &  2.43(2.39)  & 3.20(4.9) & 105050030 & 1.2,1.2,1.0  \\  
        GXIX  &  4  &  5.8  &  33.0    &   0  &  SNR  &  2.43(2.39)  & 3.20(4.9) & 105050015 & 1.2,1.2,1.0  \\  
        GXX   &  4  &  5.8  &   33.0    &   0  &  SNR  &  2.43(2.39)  & 3.20(4.9) & 105050010 & 1.2,1.2,1.0  \\ 
        GXXI   &  4  &  10.0  &  33.0    &   0  &  SNR  &  2.43(2.39)  & 2.00(4.9) & 105050015 & 1.2,1.2,1.0  \\  
        GXXII   &  4  &  20.0  &  31.0    &   0  &  SNR  &  2.43(2.39)  & 1.00(4.9) & 105050015 & 1.4,1.4,1.0  \\  
        GXXIII   &  4  &  28.0  &  25.0    &   0  &  SNR  &  2.59(2.39)  & 1.20(3.4) & 105050015 & 1.3,1.3,1.0  \\  
        GXXIV    &  4  &  28.0  &  100.0    &   0  &  SNR  &  2.59(2.39)   & 1.20(3.4) & 105050015 & 1.3,1.3,1.0  \\  
     \midrule[0pt]
        GXXV &  4  &  28.0  &   25.0   &   200  &  SNR  &  2.59(2.39)  & 1.20(3.4) & 105050015 & 1.3,1.3,1.0  \\  
        GXXVI &  4  &  2.0  &   25.0    &   0  &  SNR  &  2.59(2.39) & 1.20(3.4) & 105050015 & 1.3,1.3,1.0  \\  
        GXXVII &  4  &  28.0  &   25.0    &   0  &  SNR  &  2.59(2.39)  & 1.20(3.4) & 105050015 & 0.5,0.5,1.0  \\  
        GXXVIII &  4  &  28.0  &  25.0    &   0  &  SNR  &  2.59(2.39)  & 1.20(3.4) & 210050015 & 1.3,1.3,1.0  \\  
        GXXIX  &  4  &  5.0  &   32.7    &   50  &  SNR  &  2.43(2.39)  & 2.0(4.9) & 050100020 & 0.8,0.8,1.0  \\  
        GXXX &  4  &  5.0  &  32.7    &  50  &  SNR  &  2.43(2.39)  & 2.0(4.9) & 060070020 & 1.0,1.0,1.0  \\  
        GXXXI &  4  &  5.0  &   32.7    &   50  &  SNR  &  2.43(2.39)  & 2.0(4.9) & 090050020 & 1.0,1.0,1.0  \\  
        GXXXII  &  4  &  5.0  &  32.7    &   50  &  SNR  &  2.43(2.39)  & 2.0(4.9) & 090050020 & 1.36,1.36,1.0  \\  
        GXXXIII &  4  &  5.0  &  50.0    &   50  &  SNR   &  2.43(2.39)  & 2.0(4.9) & 090050020 & 1.36,1.36,1.0  \\  
        GXXXIV &  4  &  5.0  &  80.0    &   50  &  SNR  &  2..43(2.39)  & 2.0(4.7) & 090050020 & 1.36,1.36,1.0  \\  
        \bottomrule
    \end{tabular}
    \caption{Continuing Table~\ref{tab:ModelsParameters1}. $z_{D}$ is in kpc,
    $D_{0}$ is in units of $\times10^{28}$cm$^{3}$s$^{-1}$, $v_{A}$  in $\km
    \s^{-1} $, $dv/dz$ in $\km \s^{-1} \kpc^{-1}$, $N_{e}$($N_{p}$) in units of
    $\times 10^{-9}$ in $\cm^{-2} \sr^{-1} \s^{-1} \MeV^{-1}$ at $E_{\rm kin}$
    of 34.5 (100) GeV.  $\alpha_{e}$($\alpha_{p}$) is the injection index above
    rigidity of 2.18 (11.3) GV. All models shown in this table have $r_{D}=20$
    kpc, $T_{S} = 150$ K and a E(B-V) magnitude cut of 5.}
    \label{tab:ModelsParameters2}
\end{table}

\begin{table}
    \centering
    \footnotesize
    \begin{tabular}{cc}
     \toprule
        Name in this paper & Name in \cite{FermiLAT:2012aa}\\
     \midrule
        F &  $^{S}$L$^{Z}$6$^{R}$20$^{T}$100000$^{C}$5  \\  
        G & $^{S}$L$^{Z}$6$^{R}$20$^{T}$150$^{C}$2  \\  
        H &  $^{S}$L$^{Z}$6$^{R}$20$^{T}$150$^{C}$5  \\  
        I  &  $^{S}$L$^{Z}$6$^{R}$20$^{T}$100000$^{C}$2 \\  
        J  &  $^{S}$L$^{Z}$10$^{R}$20$^{T}$150$^{C}$5  \\  
        K &  $^{S}$O$^{Z}$10$^{R}$20$^{T}$150$^{C}$5  \\  
        L  &  $^{S}$Y$^{Z}$10$^{R}$20$^{T}$150$^{C}$5  \\  
        \bottomrule
    \end{tabular}
    \begin{tabular}{cc}
     \toprule
        Name in this paper & Name in \cite{FermiLAT:2012aa}\\
     \midrule
        M & $^{S}$S$^{Z}$10$^{R}$20$^{T}$150$^{C}$5  \\  
        N &  $^{S}$S$^{Z}$8$^{R}$20$^{T}$150$^{C}$5  \\  
        O &  $^{S}$S$^{Z}$6$^{R}$20$^{T}$150$^{C}$5  \\  
        P &  $^{S}$S$^{Z}$4$^{R}$20$^{T}$150$^{C}$5  \\  
        Q &  $^{S}$O$^{Z}$10$^{R}$30$^{T}$150$^{C}$5  \\ 
        R &  $^{S}$L$^{Z}$10$^{R}$30$^{T}$150$^{C}$5  \\\\
        \bottomrule
    \end{tabular}
    \caption{The 13 models that we used from ref.~\cite{FermiLAT:2012aa}.}
    \label{tab:ModelsNames}
\end{table}

\begin{figure}
    \begin{center}
        \includegraphics{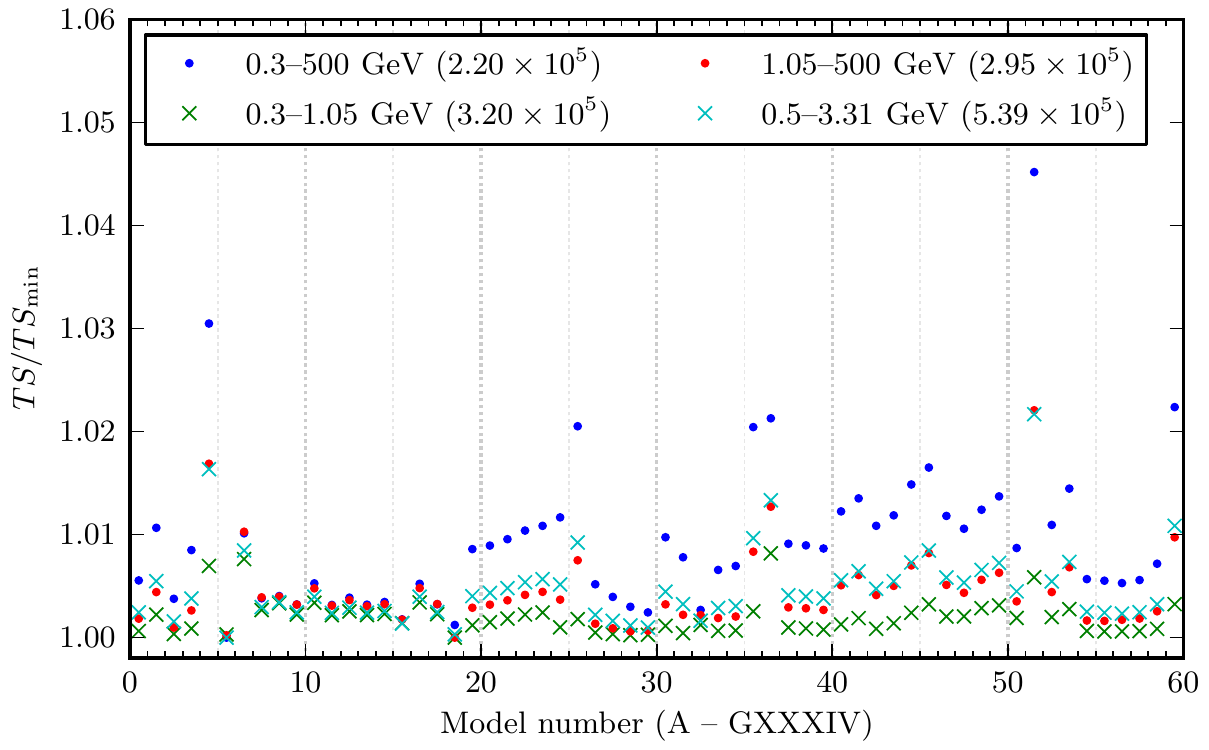}
    \end{center}
    \caption{$TS$ values of various GDE models used in this analysis, in
    different energy ranges, divided by the smallest $TS$ value in that energy
    range (and indicated in the legend in parenthesis). The ordering is the
    same as in tables~\ref{tab:ModelsParameters1} and
    \ref{tab:ModelsParameters2}. Model F performs well in all energy ranges,
    and is used as a baseline best-fit model throughout this paper.}
    \label{fig:TSvalues}
\end{figure}

In figure~\ref{fig:TSvalues}, we show for all our 60 models the ratio of
$TS$(model)/$TS_{\rm min}$, where $TS_{\rm min}$ refers to the $TS$ of model F
(which gives the best-fit to the data at all energies). We show that ratio for
four different energy ranges that span the entire energy interval that we
analyze. There are clear patterns in that ratio vs model, which are associated
to the fact that GDE models with consecutive names may differ only by a
specific subset of assumptions\footnote{The six models that are referred in the
main text were chosen for simplicity to have names between A-F.}.  As for the
quality of the fit, our results depend at different levels on the various
physical assumptions. More specifically:
\begin{itemize}
    \item The results depend slightly on the assumptions regarding the CR
        source distribution, with the OB-star distribution being the least
        preferred.

    \item The results depend significantly on the gas distribution, not
        favoring a magnitude cut of 2 in E(B-V) map.

    \item The diffusion scale heights of $\sim$4 kpc and diffusion coefficients
        $D_{0}$ close to the relevant standard assumptions $\sim 5 \times
        10^{28}$ $\cm^{2} \s^{-1}$ are preferred with values of much slower
        diffusion than that clearly disfavored.  

    \item The $TS$ ratio  is insensitive to assumptions on convection with a
        very weak preference towards non zero values for $dv/dz$

    \item High values for the Alfv$\acute{\textrm{e}}$n speed give very poor
        $TS$ ratio (and as a result $\Delta TS$).

    \item ISRF model assumptions have a small impact on the $TS$ ratio. 

    \item A preference towards values of $\sim$50--100 $\mu$G in the amplitude
        of the $B$-field at the GC and with larger values for parameters
        $r_{c}$ and $z_{c}$ is observed. 
\end{itemize}
Yet, we remind that these results come only from the study of the reference
ROI: $2^{\circ} \leq | b | \leq 20^{\circ}$ and $| l | \leq 20^{\circ}$ and
other constraints would need to be taken into account, together with the fact
that there are choices on the set of assumptions that lead to almost degenerate
results.  

\begin{figure}
    \begin{center}
        \includegraphics{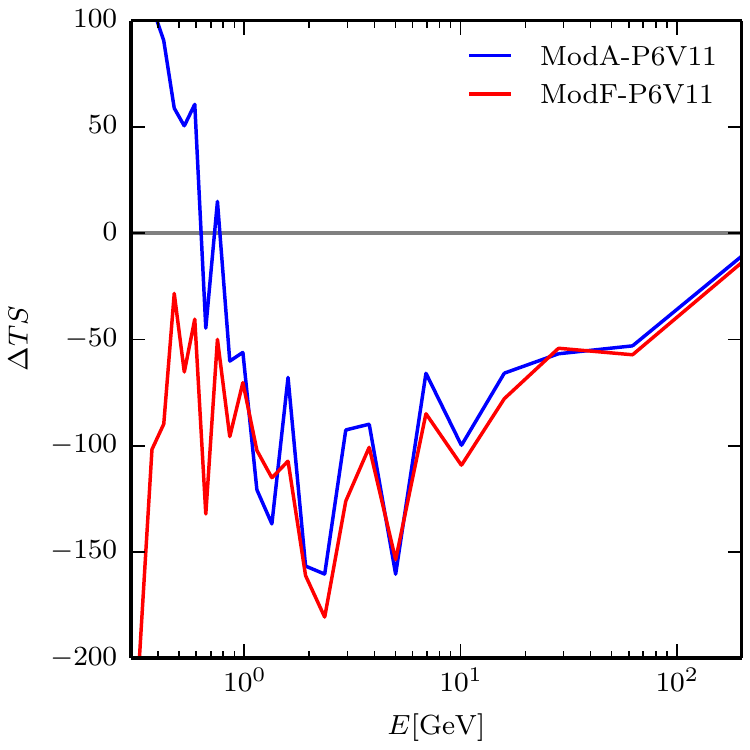}
    \end{center}
    \caption{$\Delta TS$ value as function of energy, comparing model A and
    model F with the often adopted \texttt{P6V11}. At energies above 1 GeV,
    both of our models perform significantly better than \PV.}
    \label{fig:DeltaTSvaluesAF}
\end{figure}

Finally, in figure \ref{fig:DeltaTSvaluesAF} we show the variation of the $TS$
in function of the energy for model A and F when compared to the \PV diffuse
model.  The resulting  $\Delta TS$ indicate that model A performs better than
\PV at all energies above 1 GeV, while model F performs better than \PV in the
entire energy range of interest. 

\clearpage
\section{Further properties of the Galactic center excess}
\label{app:properties}

\subsection{Characterization of the morphology}

\begin{figure}
    \begin{center}
        \includegraphics[width=0.45\linewidth]{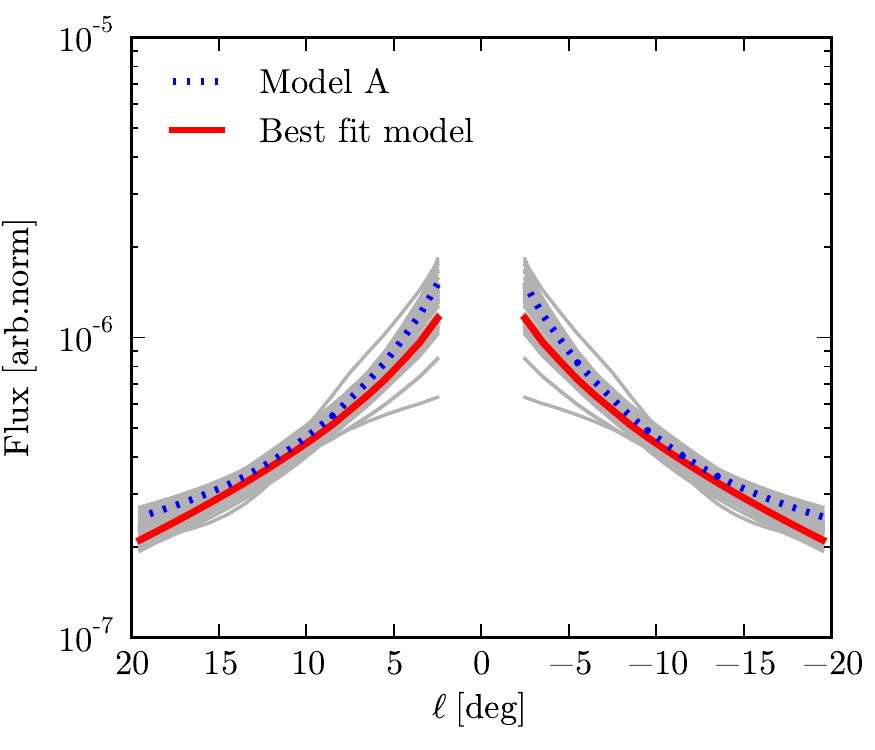}
        \includegraphics[width=0.45\linewidth]{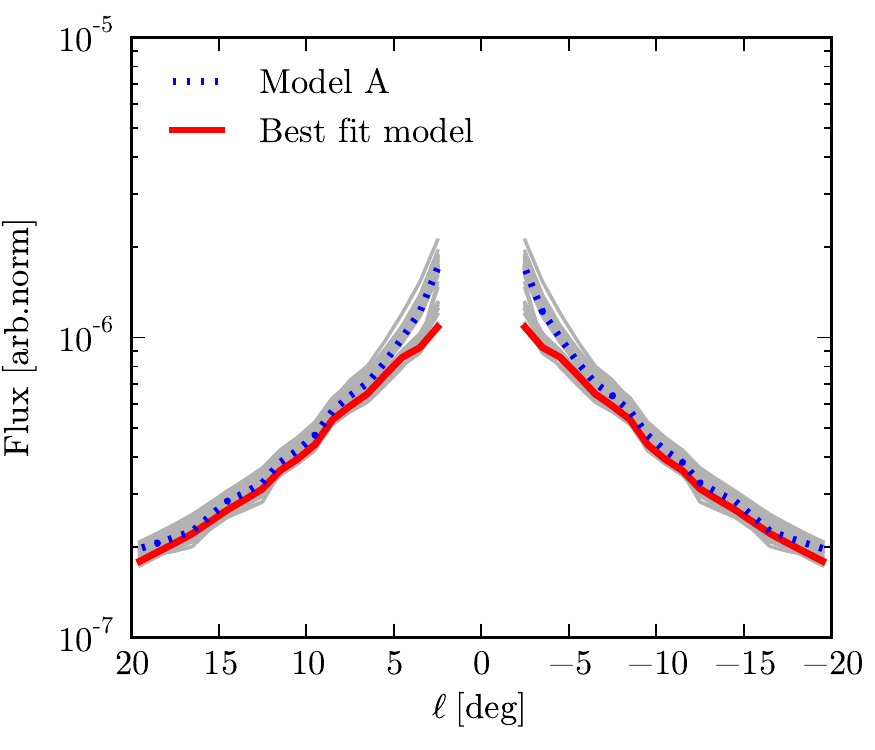}
    \end{center}
    \caption{\emph{Left panel:} Spread of latitude profile of ICS component for
    the 60 GDE models. The \emph{red line} indicates the best-fit GDE model F.
    The \emph{blue line} indicates the self-consistent GDE model A.
    \emph{Right panel:} Same as left panel, but when allowing the ICS component
    at different latitudes to float freely (\emph{ICS slicing}).}
    \label{fig:ICSprofile}
\end{figure}

We here further characterize the properties of the emission associated with the
GCE template.  We have seen in section~\ref{sec:morphology} that introducing
the ICS slicing helps in the determination of the morphology of the excess
emission, that is expected to be much more susceptible to variations in the
morphology of the GDE components.  For all the results presented below the ICS
slicing is applied. These additional degrees of freedom are indeed highly
important in determining the profile slope.  Nevertheless, we point out that
this new freedom does not spoil the physical properties of the ICS emission
associated to the GDE model.  Figure~\ref{fig:ICSprofile} indeed confirms that
allowing the ICS slices' normalizations to vary freely shrinks the dispersion
due to the different GDE models but does not corrupt the latitude dependence of
the ICS total flux.  In figure \ref{fig:ICSprofile} we show the measured
latitude profile of the ICS component of all 60 GDE models without (left panel)
and with (right panel) the slicing of the ICS component applied.

\begin{figure}
    \begin{center}
        \includegraphics[width=0.32\linewidth]{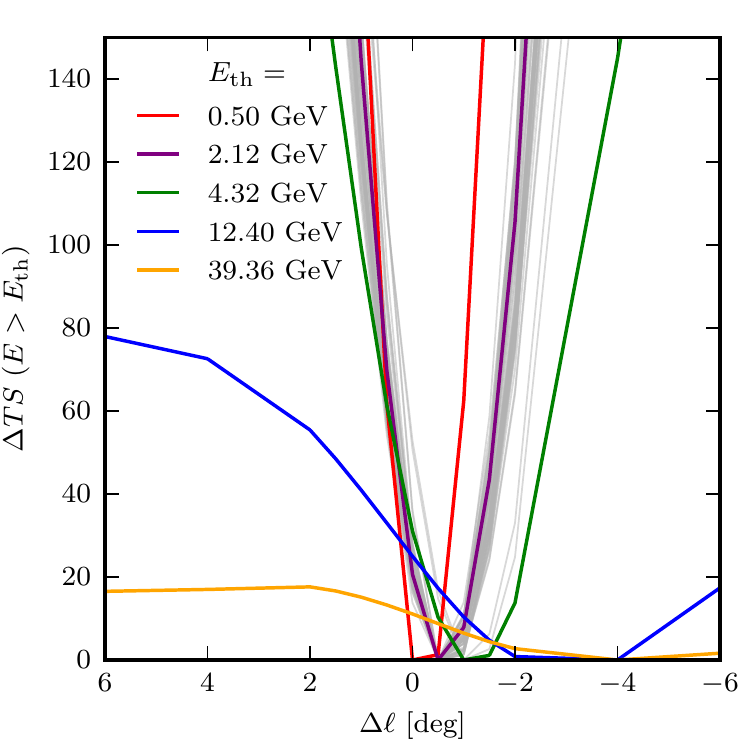}
        \includegraphics[width=0.32\linewidth]{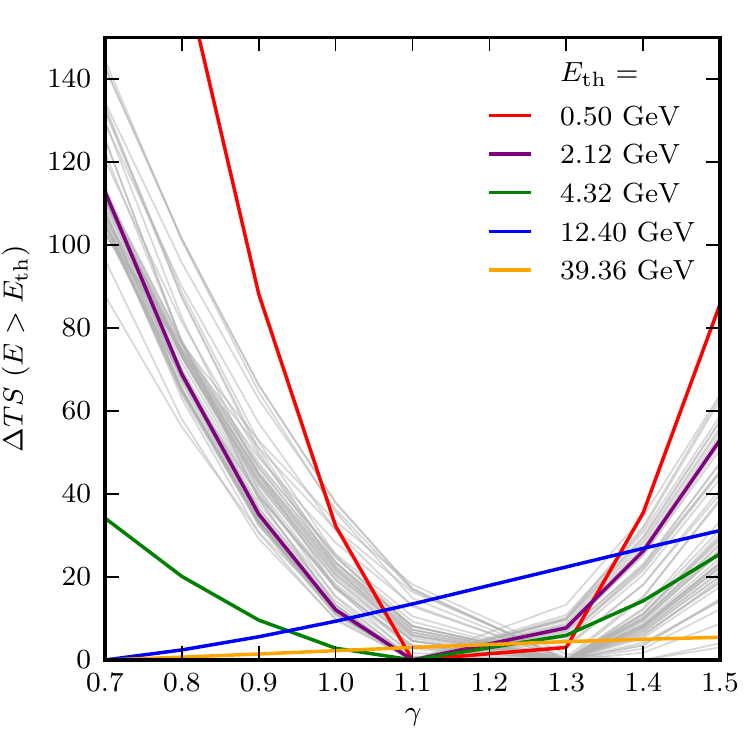}
        \includegraphics[width=0.32\linewidth]{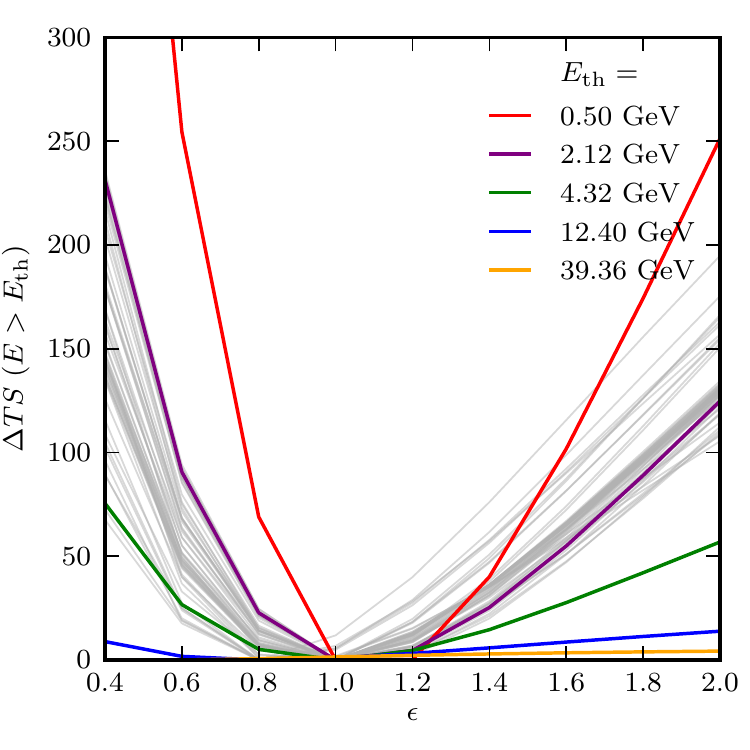}
    \end{center}
    \caption{Using ICS latitude cuts, for the 60 GDE models (\emph{gray lines})
    and for model A (\emph{colored lines}) at different energies we show in the
    \emph{left panel} the variation of $TS$ ($\Delta TS$) as function of
    displacement of GC excess template. In the \emph{central panel} the $\Delta
    TS$ as a function of slope of the GC excess template, $\gamma$.  In the
    \emph{right panel} the $\Delta TS$ as function of the elongation parameter
    $\epsilon$.}
    \label{fig:shift}
\end{figure}

In order to investigate whether the GC excess is centered at the GC, we perform
fits with a GCE template displaced along Galactic longitudes, with a
displacement parameter $\Delta \ell$.  The left panel of figure \ref{fig:shift}
shows the variation of the $TS$ when moving the centre of the GCE template
along $\ell$ for all of the 60 GDE models at 2 GeV in function of the
displacement $\Delta \ell$. All 60 models indicate that the best value of the
displacement parameter $\Delta \ell$ is about 0$^{\circ}$, meaning that a GCE
template centered at the GC (or slightly shifted at $\ell = 1^{\circ}$) is
preferred. In the same panel the $TS$ variation for different energies of the
self-consistent GDE model is shown. At low energies the excess is centered at
about the GC, while at  high energies the morphology changes and the preferred
displacement moves to $\ell \sim 4^{\circ}$. This general behavior holds for
all analyzed models.  Nevertheless, although the position with the minimum $TS$
value moves to the west, the observed flux within the systematic uncertainties
is consistent with being maximal at the GC at all energies.

In order study the fall-off behavior the of the GC excess emission, we examine
how the quality of our fits changes when allowing the slope of the generalized
NFW profile, $\gamma$, to vary in the range 0.7--1.5.  In the central panel of
figure \ref{fig:shift}, we show the resulting $\Delta TS$ as a function of the
profile slope $\gamma$ for all 60 GDE models at 2 GeV, when allowing the ICS
component to vary as function of latitude as discussed above.  In this case it
is possible to constrain $\gamma$ between 1.1 and 1.3 according to the results
from all models. Overlaid in the same panel is the result for the
self-consistent GDE model at different energies. As in the case of the
displacement, at low energy (consistently for 0.5, 2, and 4 GeV) the fits
prefers $\gamma = 1.1$, while the curves corresponding to higher energies do
not constrain $\gamma$ (at least in the range explored) mainly because of the
lack of enough events.

Finally, in order to test the sphericity of the GC excess, we build templates
that are elongated in a direction parallel to the disk or perpendicular to it.
We parametrize the angle $\xi$ of eq.~\ref{eqn:fluxADM} as $\cos \xi \equiv
\cos b \cos (\ell /\epsilon)$, instead of the standard definition $\cos \xi
\equiv \cos b \cos \ell$.  Here, $\epsilon$ can be thought of as an effective
``elongation scale factor": $\epsilon < 1$ leads to an GC excess that is
elongated perpendicularly to the Galactic disk, $\epsilon = 1$ corresponds to a
spherical symmetric profile, and $\epsilon > 1$ stretches the profile along the
direction of the disk. The right panel of  figure \ref{fig:shift} shows the
variation of the $TS$ value as function of the elongation scale factor
$\epsilon$ for the different GDE models and for model A at different energies.
Values of $\epsilon =1$ are preferred at all energies and for all models.

\medskip

We can thus conclude that we find the excess emission associated to the GCE
template to be:
\begin{itemize}
    \item Centered at the GC ($\ell \sim 0^{\circ}$);

    \item With a fall-off behavior in radius consistent with a generalized NFW
        profile with $\gamma$=1.1--1.3;

    \item Not elongated in a direction parallel or perpendicular to the
        Galactic disk and, thus, most likely spherically symmetric.
\end{itemize}

\subsection{Correlation with \Fermi~bubbles}
\label{sec:Bubbles}

\begin{figure}
    \begin{center}
        \raisebox{8.5mm}{\includegraphics[width=0.34\linewidth]{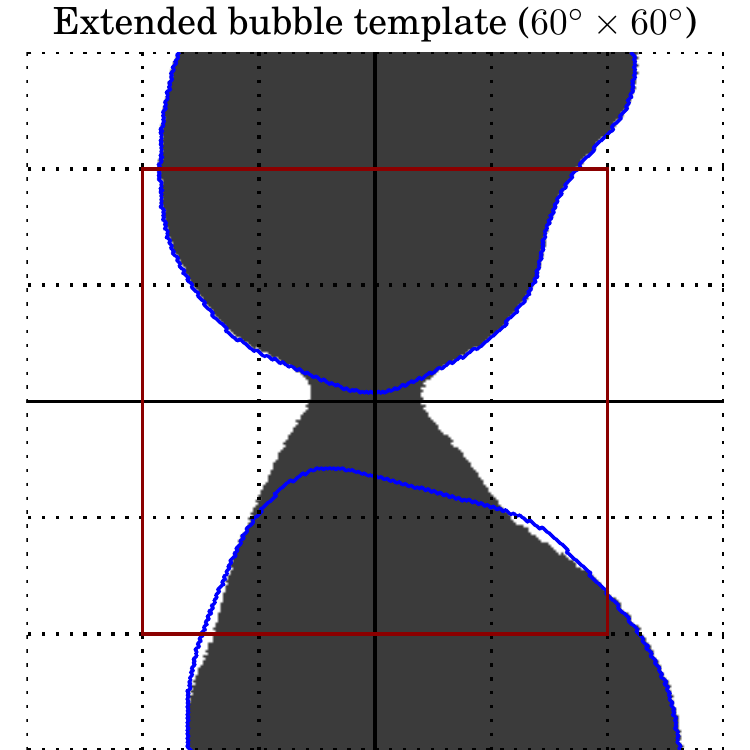}}
        \includegraphics[width=0.65\linewidth]{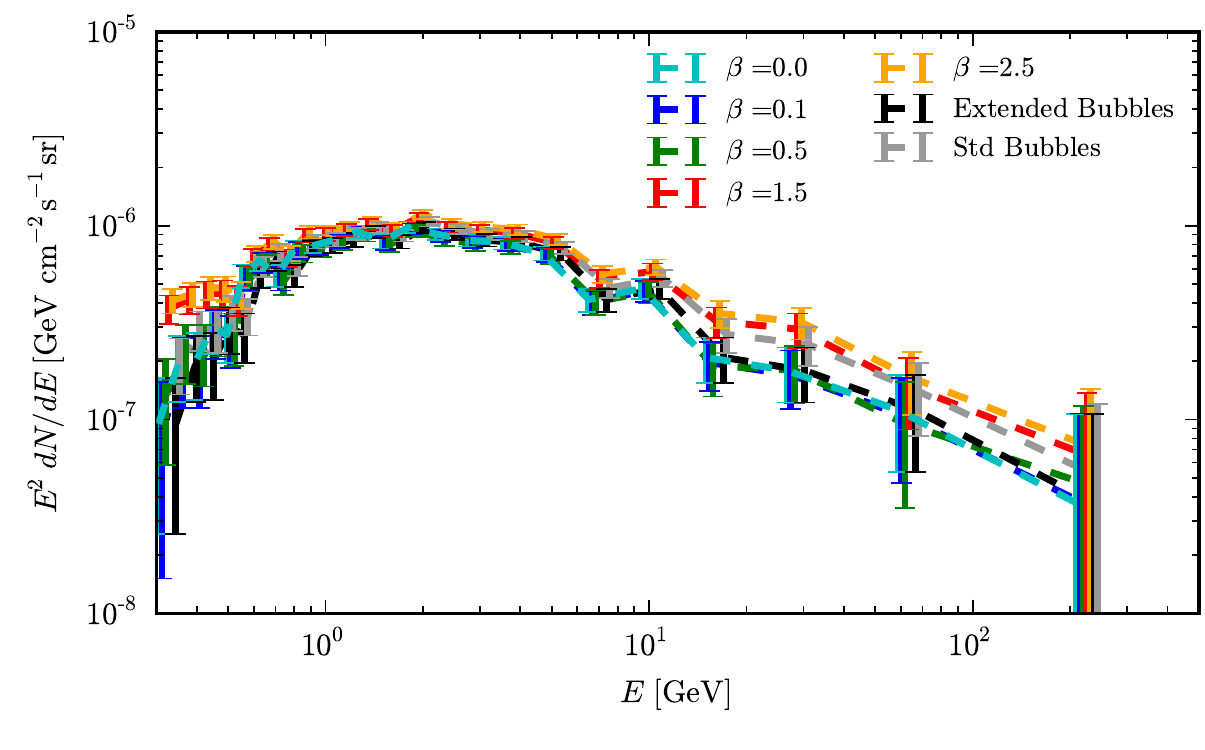}
    \end{center}
    \caption{\emph{Left panel:} Extended (\emph{shaded area}) and standard
    (\emph{blue line}, from ref.~\cite{Su:2010qj}) \Fermi~bubbles templates
    used in this work.  The \emph{red square} indicates our ROI.  \emph{Right
    panel:} Spectrum of GC excess when a latitude-dependent bubbles template is
    included in the fit (results are given for model A).}
    \label{fig:spectra_Bub}
\end{figure}

The template associated to the \Fermi~bubbles is, as explained in
section~\ref{sec:simpleTemplates}, a brightness-uniform template that covers
the region where the bubbles are defined accordingly to ref.~\cite{Su:2010qj}.
Latest studies of the \Fermi~bubbles reanalyzed this region and the emission
associated to it \cite{Fermi-LAT:2014sfa}.  The LAT Collaboration analysis cuts
the disk at $|b|>10^{\circ}$. Nevertheless, it is likely that the region of the
bubbles extends down to the GC, also in the light of the possible astrophysical
processes that have been attempted to explain this yet unknown emission, \eg
emission of a jet from the central black hole, wind from SN explosion close to
the GC, 1$^{\rm st}$ or 2$^{\rm nd}$ order Fermi acceleration of CR electrons
\cite{Guo:2011eg, Guo:2011ip, Crocker:2010dg, Cheng:2011xd, Mertsch:2011es}.
Therefore we consider also a new definition of the \Fermi~bubbles
edges\footnote{Meng Su, Private Communication.}, that extends the bubbles down
to the GC.  This new region is displayed in figure~\ref{fig:spectra_Bub}, left
panel, together with the contour of the bubbles region as defined in
ref.~\cite{Su:2010qj}.  The spectrum associated with the GCE template when the
``extended'' bubbles template is adopted is displayed in figure
\ref{fig:spectra_Bub} (right panel).  The new bubbles template is consistent
with the ``standard" one -- \ie used as standard working hypothesis -- at low
energies, $E \leq 5$ GeV, while it differs at high energies of about 30\% at
100 GeV. 

We then investigate the possibility that part of the emission absorbed by the
GCE template, in particular the high energy end of the spectrum in figure
\ref{fig:spectrumSysStat}, might be correlated spatially with the
\Fermi~bubbles.  To this end, we adopt an additional template for the bubbles
(built from the new extended bubbles template  and in addition to this one),
allowing its intensity to be dependent on the latitude $\propto |b|^{-\beta}$,
with $\beta <$ 1 or  $\beta >$ 1.  The emission absorbed by the GCE template
when adding a latitude dependent template for the \Fermi~bubbles is shown in
figure \ref{fig:spectra_Bub} for different values of the index $\beta$. By
comparing the $TS$ values of the different templates, it emerges that the fit
prefers a template with a very mild latitude dependence, $\beta$ = 0.1, as well
as $\beta$ = 0 and the extended bubbles template. The standard template adopted
in the analysis is only slightly worse than the best-fit one ($\Delta TS =32$),
while $\beta \geq 0.5$ is highly disfavored by the fit ($\Delta TS \geq
\mathcal{O}(100)$).  This means that the GCE template does not lose too much
power when using the latitude-dependent bubbles template, thus confirming the
robustness of our results derived under the assumption of uniform-brightness
template.

\section{Systematic uncertainties}

\subsection{An analytical model for the empirical model uncertainty}
\label{sec:systModeling}

In subsection~\ref{sec:ModelingUncertainties} we estimated the model
uncertainties from a fit to 22 test regions along the Galactic disk.  The
resulting fluctuations where shown in figure \ref{fig:variance} in terms of
mean and standard deviation (left panel), as well decomposed into their
principal components (right panel).  As can be seen from this plot, the first
three principal components are sizeable and larger than or comparable to the
statistical error, and appear to have a relatively well defined functional
dependence on energy.  We will show here that this functional dependence is
actually expected as the result of uncertainties in the normalization and slope
of the main background components ICS and $\pi^0$+Bremss, and can be well
fitted with a simple model.

\medskip

We will start by assuming that the true background flux associated with
component $k$ in a given ROI is related to the measured background flux as
determined by our template analysis by
\begin{equation}
    \frac{dN_k^\text{true}}{dE} = (1+\alpha_k)
    \left(\frac{E}{E_\text{ref}}\right)^{\gamma_k} \frac{dN_k}{dE}\;,
\end{equation}
where, $\alpha_k$ and $\gamma_k$ are parameters that account for small
corrections in the normalization and slope of the flux.  These two parameters
are unknown, but assumed to be normal distributed with mean zero and standard
deviation $\Delta\alpha_k$ and $\Delta \gamma_k$.\footnote{Since the errors
that we will discuss are typically very small, a more precise modeling of these
uncertainties in terms of a \fex log-normal distribution is not of relevance
here.}

We further assume that any mis-modeling of the background will be absorbed by
the GCE template, hence
\begin{equation}
    \frac{dN_{\rm GCE}^\text{true}}{dE} = \left(1-\sum_k\left(\alpha_k -
    \gamma_k\ln \frac{E}{E_\text{ref}}\right)\right)
    \frac{dN_{\rm GCE}}{dE}\;,
\end{equation}
where we neglect second and higher order terms in $\alpha_k$ and $\gamma_k$.
If only a fraction of mis-modeling is absorbed by the GCE template, this can be
accounted for by a rescaling of $\Delta\alpha_k$ and $\Delta\gamma_k$.

\medskip

The above uncertainty in the GCE spectrum corresponds to a correlation matrix
with the simple form
\begin{equation}
    \Sigma_{ij,\rm\,mod} \simeq \sum_k \left( \Delta \alpha^2_k + \Delta \gamma^2_k
        \ln \frac{E_i}{E_\text{ref}}
    \ln \frac{E_j}{E_\text{ref}} \right)
    \frac{dN_k}{dE_i}
    \frac{dN_k}{dE_j}\;.
    \label{eqn:SigmaModel}
\end{equation}
This is true as long as the background variations can be modeled as a
multivariate normal distribution, which requires here roughly
$\Delta\alpha_k\lesssim 0.3$ and $\Delta\gamma_k\lesssim0.1$, which we will
find to be the case in the present situation.  

The background fluxes are derived from the data in our baseline ROI.  We will
only take into account the dominant $\pi^0$+Bremss and ICS components, and
parametrize them analytically in the simple approximate form
\begin{equation}
    \frac{dN_{\pi^{0}+B}}{dE}=
    1.0\times10^{-5} \frac{x^{-2}}{\sqrt{x^{1.2}+x^{-0.4}}} \rm GeV^{-1}\,cm^{-2}\,s^{-1}\,sr^{-1}
\end{equation}
and
\begin{equation}
    \frac{dN_{\rm ICS}}{dE} = 4.0\times10^{-6} x^{-2.3} \rm\, GeV^{-1}\,cm^{-2}\,s^{-1}\,sr^{-1}\;,
\end{equation}
where $x\equiv E/1\GeV$.  This parametrization is sufficiently accurate for the
present purpose.  As pivot point for the variations we adopt $E_\text{ref} =
3\rm\,GeV$, which is again not critical.

\medskip

We then perform a simple $\chi^2$ fit to the three principal components shown
in figure \ref{fig:variance}, giving equal weight to every energy bin and all
three components.  As a result, we find the best-fit values $\Delta
\alpha_{\pi^{0}+B} \simeq 0.031$, $\Delta \gamma_{\pi^{0}+B} \lesssim 10^{-4}$,
$\Delta \alpha_{\rm ICS} \simeq 0.025$ and $\Delta \gamma_{\rm ICS} \simeq
0.0093$.  The corresponding model predictions are again shown in figure
\ref{fig:variance}, and provide overall a good description of the observed
fluctuations.

Taking the above results at face value and referring to fluxes averaged over
our baseline ROI as \fex shown in figure \ref{fig:consistency}, the impact of
typical mis-modelling of the ICS or $\pi^0$+Bremss components as observed along
the Galactic plane is expected to affect the GCE spectrum only at the level of
a few percent of the background components.  Yet, since the background are one
order of magnitude stronger, this effect can be still sizeable.  In fact, it
constitutes one of the major uncertainties  when discussing spectral and
morphological properties of the GCE excess.

\subsection{Various method uncertainties}
\label{sec:otherModeling}

\begin{figure}
    \begin{center}
        \includegraphics{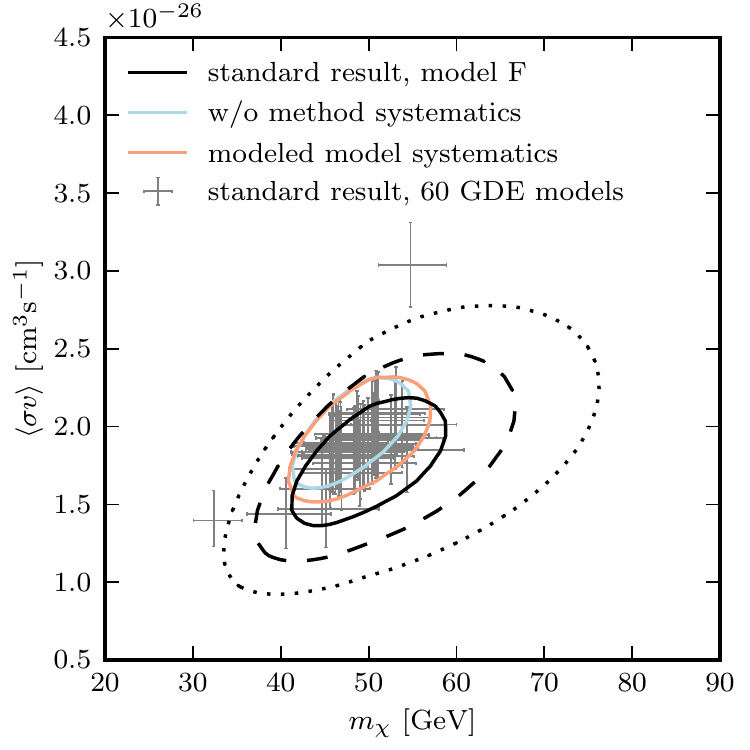}
        \includegraphics{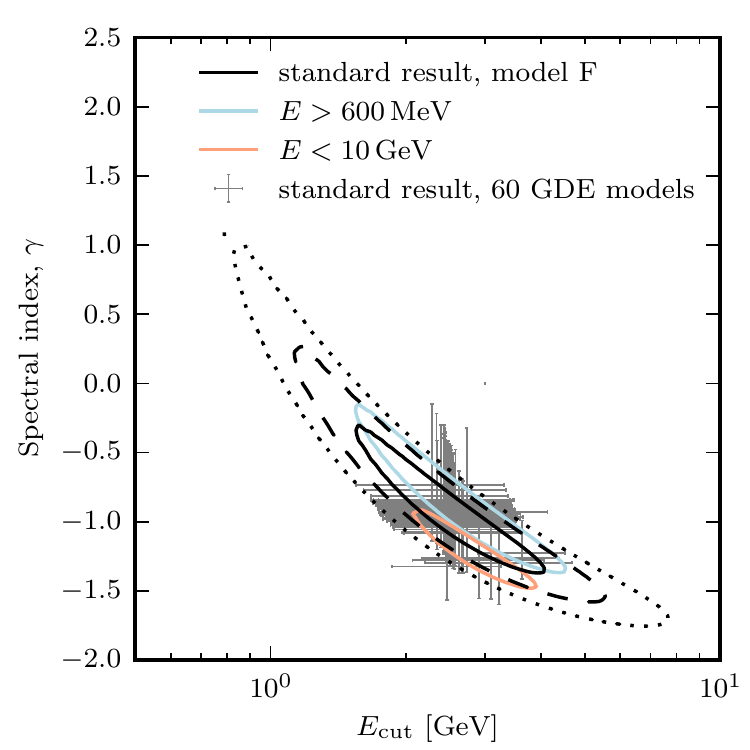}
    \end{center}
    \caption{\emph{Left panel:}  Like left panel of figure \ref{fig:DMfits1}
    for $\bar{b}b$ final states (\emph{black contours}, based on model F), but
    showing in addition the effect of modeling the background model systematics
    (\emph{light pink contour}), of neglecting the method systematics
    (\emph{light blue contour}), and the $\pm1\sigma$ errors that one obtains
    for all of the 60 GDE (\emph{gray crosses}).  \emph{Right panel:}  Like
    right panel of figure \ref{fig:AstroFits}, but showing in addition the
    results in the case of restricted energy ranges in the fit (\emph{light
    blue and red curves}; only 68\% CL contours for clarity), and using the 60
    GDE models instead (\emph{gray crosses}).}
    \label{fig:fitValidation}
\end{figure}

In figure \ref{fig:fitValidation} we show how our results are affected by the
most critical analysis choices.  In the left panel, we show a fit to the flux
in figure \ref{fig:spectrumSysStat} with a $\bar{b}b$ spectrum.  We compare the
results that we find for our baseline analysis choices with the results
obtained when neglecting the method uncertainty (\cf figure
\ref{fig:otherSyst}), or when adopting one of the other 59 GDE scenarios to
model the backgrounds.  We find that our results are largely consistent;
outliers correspond to GDE models which give a particularly bad TS value.

In the right panel, we show how the fit of a power-law with exponential cutoff
is affected by various analysis choices (\cf figure \ref{fig:AstroFits}).  We
show the contours that we obtain when excluding data below 600 MeV or above 10
GeV (only 68\% CL contours for clarity), as well as the results obtained when
using one of the other 59 GDE models.  Again, our results remain consistent.
Note that excluding data above 10 GeV has the effect of lowering the value of
the spectral index $\gamma$.  This can be understood by realizing that the
power-law like behaviour of the GCE (see figure \ref{fig:spectrumSysStat}) at
energies above 10 GeV prefers a larger subtraction of a smooth background
(encoded in the correlated model uncertainties, see
appendix.~\ref{sec:systModeling}) when fitted with an exponential cutoff.  This
subtraction necessarily affects the flux below 1 GeV and makes the spectrum
harder.

\section{Miscellaneous}

\begin{figure}
    \begin{center}
        \includegraphics[width=0.94\linewidth]{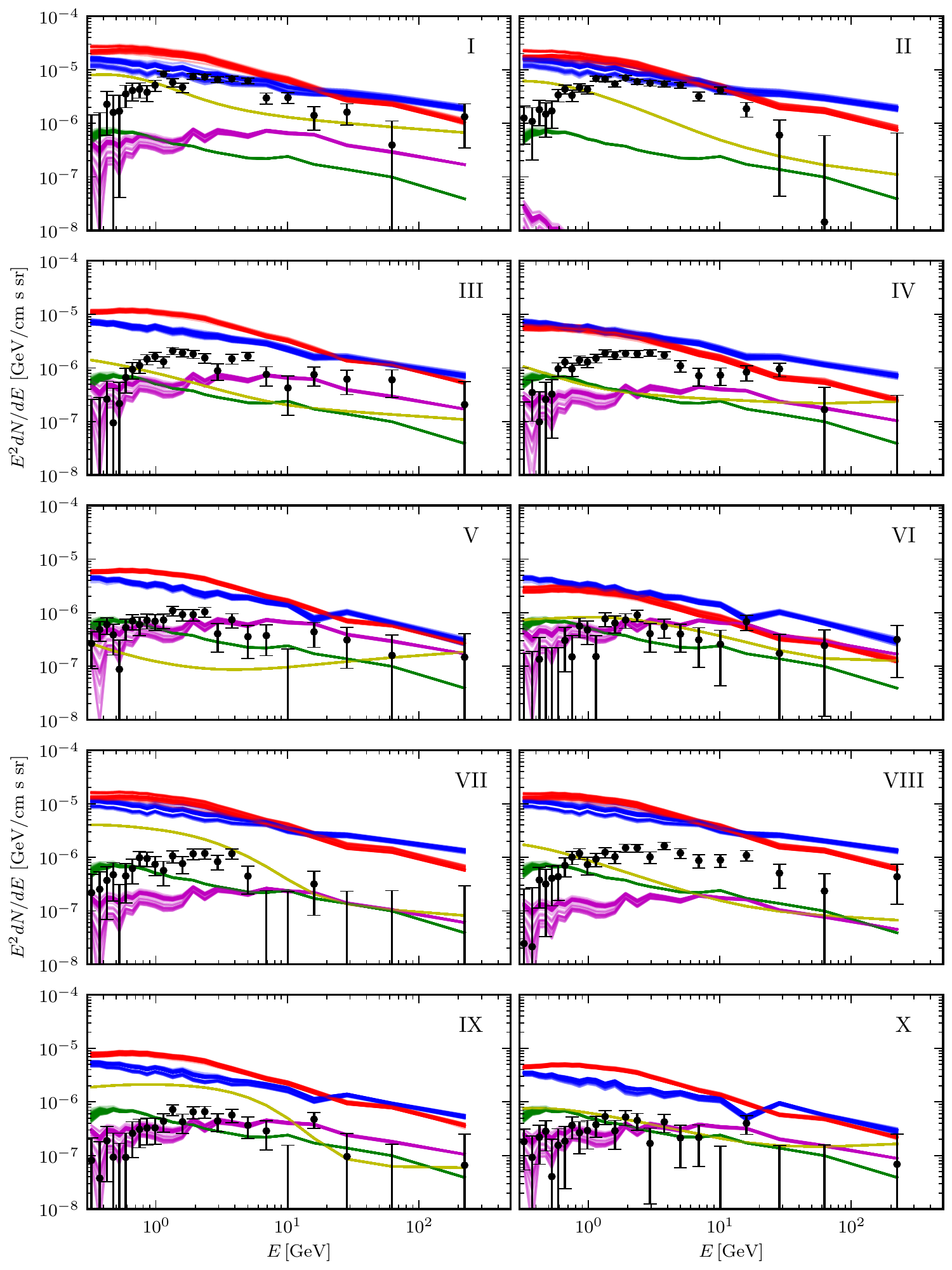}
    \end{center}
    \caption{Energy spectra of the different background components
    ($\pi^0$+Bremss, ICS, the isotropic, bubbles, PSCs) in the template fit,
    for \emph{all} 60 GDE models, averaged over the segments shown in figure
    \ref{fig:ROIsplits}.  The color coding is the same as in figure
    \ref{fig:consistency}.  The black points show the results for the GC excess
    for model F, as in figure \ref{fig:DMsplitSpectraSyst}.}
    \label{fig:DMsplitBG}
\end{figure}

In figure \ref{fig:DMsplitBG}, we show the spectra of the individual background
components, as observed in each of the ten segments, for all 60 GDE models.  We
observe only small variations in the main GDE components, with the largest
variations at low energies in the subdominant \Fermi~bubbles and the isotropic
emissions.  For reference, we also show the GC excess spectrum for \emph{one}
of the GDE models (model F); the envelope for all other models is shown in
figure \ref{fig:DMsplitSpectraSyst}.

\clearpage
\bibliography{GCdiff.bib}
\bibliographystyle{JHEP}

\end{document}